\documentclass[aps,prb,twocolumn,groupedaddress]{revtex4-1}

\usepackage{graphicx}
\usepackage{amsmath}        
\usepackage{mathtools, bm}
\usepackage{amssymb, bm}
\usepackage{times}      
\usepackage{setspace}
\usepackage{hyperref}
\usepackage{amsthm}
\usepackage{titlesec}

\newcommand{\bra}[1]{\langle #1|}
\newcommand{\ket}[1]{| #1\rangle}
\newcommand{\mn}[1]{\langle #1 \rangle}
\newcommand{\bk}[2]{\langle #1 | #2 \rangle}

\begin{document}


\title{Energy of fermionic ground states with low-entanglement single-reference expansions, and tensor-based strong-coupling extensions of the coupled-cluster method}


\author{Dominic Bergeron}
\email[]{dominic.bergeron@usherbrooke.ca}
\affiliation{D\'epartement de physique, Universit\'e de Sherbrooke, Qu\'ebec, Canada}


\date{\today}

\begin{abstract}

We consider a fermionic system for which there exist a single-reference configuration-interaction (CI) expansion of the ground state wave function that converges, albeit not necessarily rapidly, with the increasing number of particle-hole excitations. For such systems, we show that, whenever the coefficients of Slater determinants (SD) with $l\leq k$ excitations can be defined with a number of free parameters $N_{\leq k}$ bounded polynomially in $k$, the ground state energy $E$ only depends on a small fraction of all the wave function parameters, and is the solution of equations of the coupled-cluster (CC) form. This generalizes the standard CC method, for which $N_{\leq k}$ is bounded by a constant. Based on that result and low-rank tensor decompositions (LRTD), we discuss two possible extensions of the CC approach for wave functions with general polynomial bound for $N_{\leq k}$. The most straightforward of those extensions uses the LRTD to represent the amplitudes of the CC cluster operator $T$ which, unlike in the CC case, is not truncated with respect to number of excitations, and the energy and tensor parameters are given by a LRTD-adapted version of standard CC equations. The LRTD can also be used to directly parametrize the wave function coefficients, which involves different equations of the CC form. We derive those equations for the coefficients of SD's with $l\leq 4$ excitations, using the CC exponential wave function ansatz with a different type of excitation operator, and a representation of the Hamiltonian in terms of excited particle and hole operators. To complete the proposed computational methods, we construct compact tensor representations of coefficients, or $T$-amplitudes, using superpositions of tree tensor networks which take into account different possible types of entanglement between excited particles and holes. Finally, we discuss why the proposed CC extensions are theoretically applicable at larger coupling strengths than those treatable by the standard CC method.

\end{abstract}


\maketitle

\section{Introduction}

The accurate prediction of low temperature properties of many particles system is the main goal of theoretical condensed matter and quantum chemistry research. The quantum nature of those systems however poses a seemingly unsurmountable challenge because of the exponential growth of the problem's complexity with the number of particles. For bosonic systems, quantum Monte Carlo (QMC) methods allow to overcome the exponential complexity. At a fundamental level however, matter is made of fermions, and the computation remains exponentially hard even with QMC methods because of the well known \textit{fermionic sign problem} due to the anti-commutation of fermion operators \cite{Troyer_2005}.

Although no general recipe exists to solve the many-fermions problem, many different methods have been developed to successfully study specific types of molecules or phases of matter using adapted approximations. Those approximations typically become exact either in the weak- or the strong-coupling (interaction) limits. One weak-coupling approach that has been very successful in quantum chemistry is the \textit{coupled-cluster} (CC) method \cite{Coester_Kummel_1960, Cizek_1966}  (review in Ref. \onlinecite{Bartlett_Musial_2007}), which uses the wave function ansatz $\exp(T)\ket{\phi}$, where $\ket{\phi}$ is a reference Slater determinant (SD) playing the role of a vacuum, and $T$ is an excitation operator creating states with different numbers of particle-holes excitations on that vacuum. Unlike the closely related \textit{configuration-interaction} (CI) method, the CC method is size-extensive, i.e., the energy scales correctly with system size, which makes it suitable as well for weakly correlated condensed matter systems \cite{Bartlett_1981}. Although exact only in the weak-coupling limit, it can however include a reasonable amount of quantum correlations, as compared to density functional theory for instance, but it eventually fails dramatically at strong coupling \cite{Bulik_2015}. There are also several extensions of the CC method aimed at treating strong correlations, among which are the multireference CC methods (reviewed in Ref. \onlinecite{Lyakh_2012}), and various single reference extensions that use correlation or projection operators \cite{Krotscheck_1980, Xian_2008, Neuscamman_2013, Bulik_2015, Yiheng_2017,Gomez_2019}.

The ability to treat strong correlations in fermionic systems is necessary to model some of the most interesting materials, for instance, Mott insulators, Heavy Fermions systems and high-temperature cuprate superconductors. For spin systems, i.e. very strongly interacting half-filled systems, tensor networks (TN) states have proven very effective (reviews in Refs. \onlinecite{Orus_2014,Ran_2019}). TN methods have also been applied to fermionic systems \cite{Corboz_2009, Changlani_2009, Marti_2010, Chan_2011, Poilblanc_2014, Czarnik_2014, Zohar_2015, Szalay_Pfeffer_2015, Zhao_2017}. The main idea behind a TN, which is a type of low-rank tensor decomposition (LRTD), is to take advantage of the small entanglement actually present in the ground state of systems with local Hamiltonians to drastically reduce the number of free parameters required to represent the ground state. TN design therefore relies on the locality of the Hamiltonian ($H$). Other types of LRTD, which are not TN and do not rely on the locality of $H$, have also been used in quantum chemistry, and CC calculation in particular, to reduce computational complexity \cite{Benedikt_2013, Hohenstein_2012, Hohenstein_2013, Hohenstein_2013a, Parrish_2014, Schutski_2017, Tichai_2019, Parrish_2019, Lesiuk_2019}. In those cases however, the LRTD were used as purely mathematical tools, without reference to the system's entanglement, and without modifying the nature of the CC approximation.


In the present work, we explicitly assume a ground state with low entanglement as in TN methods, but with a single-reference wave function expansion, as in the CI and CC methods, and a Hamiltonian assumed to be non-local. We then discuss the use of LRTD not to reduce computational complexity, but to construct different size-extensive approximations than the standard CC ones, in order to treat stronger correlations, without using projection or correlation operators. 

More specifically, given a single-reference expansion of the ground state of a fermionic system, we first show that the energy of that state is the solution of generalized CC equations whenever the coefficients of SD's with $l\leq k$ particle-hole excitations can be defined with a total number of parameters $N_{\leq k}$ bounded polynomially in $k$. This also implies that the energy depends only on a small fraction of all the wave function parameters and coefficients. We will see that the standard CC approximation corresponds to the simplest of this type of parametrization, where $N_{\leq k}$ is bounded by a constant. We then discuss two other types of parametrization with arbitrary polynomial bound for $N_{\leq k}$ based on LRTD: one in which the LRTD parametrize the amplitudes in the CC cluster operator $T$ and allow to close the CC equations without truncating $T$ on the number of excitations axis, and another in which the LRTD parametrize the wave function coefficients directly. For the latter, we derive generalized CC equations for coefficients of SD's with up to four excitations. To do so, we use a $T$ operator such that $T^l\ket{\phi}$ is proportional to the $l$ excitation part of the wave function and $\exp(T)\ket{\phi}$ is the formally exact full CI wave function, and a representation of the Hamiltonian in terms of excited particle and hole operators involving only standard second quantization, instead of the usual CC particle-hole normal ordering notation. For the two proposed approaches, we have to consider higher-order tensors than in the tensor implementations of CC, and for which more complex tensor representations are required to obtain a globally compact low-entanglement representation of the involved coefficients or $T$-amplitudes. For that purpose, we construct representations which can be described as superpositions of tree tensor networks (STTN), and are designed to maximize compactness by taking into account the different possible types of entanglement between excited particles and holes and by sharing tensors between different sets of coefficient. The STTN also allow to estimate the computational complexity of the proposed approaches. The justification for the low-entanglement assumption, despite the non-locality of the Hamiltonian, and the conditions of validity of the CC extensions are also discussed. 

The paper is organized as follows: A brief review of the CI and CC methods is given in section \ref{sec:CI_CC}. Then, the ground state energy computation result is obtained in section \ref{sec:comput_GS_E}. The two types of LRTD parametrization and the choices of orbital basis and reference are discussed in section \ref{sec:CC_LRTD}. Section \ref{sec:H_ph} describes the particle-hole representation of the Hamiltonian used to derive the generalized CC equations for the CI coefficients. Those equations are then presented in section \ref{sec:eqns_coeffs}, with their derivation provided in appendix and as \hyperlink{suppl_mat}{supplemental material}. Finally, the STTN representation is described in section \ref{sec:coeffs_tensor_decomp}, followed by a discussion and conclusion in sections \ref{sec:discussion} and \ref{sec:conclusion}, respectively.

\section{The configuration-interaction and coupled-cluster methods}\label{sec:CI_CC}

Let us begin by briefly reviewing the basis of the configuration-interaction and coupled-cluster methods to which we will refer later. More details on those methods can be found in Refs. \onlinecite{Lowdin_1955, Szabo_Ostlund_2012,Bartlett_Musial_2007}.  First, we define
\begin{equation}\label{eq:def_phi}
\ket{\phi}= a_{i_1}^\dagger a_{i_2}^\dagger \ldots a_{i_N}^\dagger \ket{0}\,,
\end{equation}
where $a_{i}^\dagger$ creates an electron on the spin-orbital $i$. $\ket{\phi}$ will be called the \textit{reference} Slater determinant (SD) for the $N$ electrons system and will be used as an approximate vacuum state for the system. We also define
\begin{equation}\label{eq:phi_i_j}
\ket{\phi^{i_1,i_2,\ldots, i_k}_{j_1,j_2,\ldots,j_k}}=a_{i_k}^\dagger  a_{j_k} a_{i_{k-1}}^\dagger  a_{j_{k-1}} \ldots a_{i_1}^\dagger a_{j_1} \ket{\phi}\,,
\end{equation}
a SD with $k$ particle-hole excitations with respect to $\ket{\phi}$. Now, let us write the system's ground state wave function in the Born-Oppenheimer approximation as
\begin{equation}\label{eq:psi_sum}
\begin{split}
\ket{\psi}&=\ket{\phi}+\sum_{i,j} c_j^i \ket{\phi^i_j}+\sum_{\substack{\langle i_1,i_2\rangle \\\langle j_1,j_2\rangle}} c_{j_1j_2}^{i_1i_2}  \ket{\phi_{j_1j_2}^{i_1i_2}}\\
&+\ldots+ \sum_{\substack{\langle i_1i_2\ldots i_n\rangle,\\\langle j_1,j_2\ldots j_n\rangle}} c_{j_1,j_2\ldots j_n}^{i_1i_2\ldots i_n} \ket{\phi_{j_1,j_2\ldots j_n}^{i_1i_2\ldots i_n}}
\end{split}
\end{equation}
where the $i$ and $j$ indices refer to empty and occupied spin-orbitals in $\ket{\phi}$, respectively, $\langle i_1i_2\ldots i_n\rangle$ is a combination of distinct indices such that $i_m<i_l$ for $m>l$, and $n\leq n_{max}=min(N_e,N_o)$, where $N_e$ and $N_o$ are the number of empty and occupied spin-orbitals, respectively, in $\ket{\phi}$. Expression \eqref{eq:psi_sum} is the well known configuration-interaction (CI) form of the wave function. In CI calculations, the coefficients are obtained by minimizing the energy $\bra{\psi}H\ket{\psi}/\bk{\psi}{\psi}$, where $H$ is the system's Hamiltonian, which is usually done by diagonalizing the matrix representation of $H$ in the $\ket{\phi_{j_1,j_2\ldots j_k}^{i_1i_2\ldots i_k}}$ basis, where $0\leq k \leq n$. If $n=n_{max}$, the method is called \textit{full} CI or \textit{exact diagonalization}. Usually $n< n_{max}$ in the CI method because the entire Hilbert space is too large for the full CI calculation to be tractable. 

The single particle orbitals in \eqref{eq:def_phi} are typically Hartree-Fock spin-orbitals. However, the optimal orbitals for CI are called \textit{natural} spin-orbitals and are the orbitals that diagonalize the single particle density matrix $\bra{\psi_{FCI}}a_i^\dagger a_j\ket{\psi_{FCI}}$, where $\ket{\psi_{FCI}}$ is the full CI ground state. They are also the orbitals that minimize the variance $\sum_i \left(\mn{\hat{n}_i^2}-\mn{\hat{n}_i}^2\right)=\sum_i\mn{\hat{n}_i}(1-\mn{\hat{n}_i})$ ($\hat{n}_i^2=\hat{n}_i$ for fermions), of the spin-orbital occupation number $\hat{n}_i=a_i^\dagger a_i$ \cite{Lowdin_1955}, which minimizes the size of the active space, i.e. the number of partially filled spin-orbitals, and therefore reduces the size of the Hilbert space for the many-particle problem. However, since $\ket{\psi_{FCI}}$ is unknown, an approximate $\ket{\psi}$ must necessarily be used in practice, which yields only approximate natural orbitals.

The justification for expanding the wave function in number of excitations with respect to $\ket{\phi}$ is that, because the Hamiltonian has only one- and two-particle terms, it has a block-band-matrix form in such a basis. It is therefore expected that the coefficients of the ground state will decrease rapidly as a function of the number of excitations if the reference SD is well chosen. 

The main flaw of the CI method is that it is not size-extensive, i.e., the energy of the system does not behave correctly as the system becomes large. For instance, for a homogeneous system of size $V$, instead of being proportional to $V$ as $V\rightarrow \infty$, the CI energy is proportional to $\sqrt{V}$ \cite{Bartlett_1981}. This is a consequence of the truncation of Hilbert space, which is equivalent to the inclusion of unlinked diagrams from a perturbation theory perspective, while any size-extensive approximation must only include linked diagrams \cite{Brueckner_1955, Goldstone_1957,Bartlett_1981,Bartlett_Musial_2007}. The CI method is however a practical method for small molecules and impurity problems such a the Anderson model, one of the cornerstone of Dynamical Mean Field Theory \cite{Zgid_Gull_Chan_2012,Go_Millis_2017}.

Instead of truncating Hilbert space at a given number of particle-hole excitations, one can also define a wave function that spans the whole Hilbert space, by using ``composite'' coefficients above a certain number of excitations. This is what the coupled-cluster (CC) method does by postulating a ground state of the form
\begin{equation}\label{eq:CC_ansatz}
\ket{\psi}=e^{T}\ket{\phi}\,,
\end{equation}
where $T$ is called the \textit{cluster operator} and is defined as
\begin{subequations}\label{eq:def_T_CC}
\begin{equation}\label{eq:def_T_CC_a}
T=T_1+T_2+\ldots+T_{n}
\end{equation}
where the $T_l$'s are excitation operators defined as
\begin{equation}\label{eq:T_N_CC}
T_l=\sum_{\substack{\langle i_1i_2\ldots i_l\rangle,\\\langle j_1,j_2\ldots j_l\rangle}} d_{j_1,j_2\ldots j_l}^{i_1i_2\ldots i_l} a_{i_l}^\dagger a_{j_l} a_{i_{l-1}}^\dagger a_{j_{l-1}} \ldots a_{i_1}^\dagger a_{j_1}\,.
\end{equation}
\end{subequations}
When the exponential is expended as a Maclaurin series, the SD's with any number of excitations are present, up to $n_{max}$. In addition, while the $l\leq n$ excitation coefficients have an irreducible (connected) part, namely an amplitude in $T$, the coefficients of higher order SD's are fully expressed as combinations of products of lower order amplitudes generated by the form $e^{T}$, i.e., they are reducible. For instance, if $n=2$, the single excitation SD's are generated by $T_1$ and double excitations SD's are generated by $T_2+T_1^2/2$, while the three excitations SD's are generated by $T_1T_2+T_1^3/6$. The series \eqref{eq:def_T_CC} is also a type of cumulant expansion, but in excitation operator space with respect to $\ket{\phi}$, i.e., it is the connected part of the operator relating $\ket{\psi}$ to $\ket{\phi}$. For large systems, the CC ansatz \eqref{eq:CC_ansatz} produces much better results than discarding completely the SD's with more than $n$ excitations as in the CI method \cite{Bartlett_Musial_2007}. On the other hand, the CC method is exact only in the weak coupling limit and fails quite dramatically at strong coupling \cite{Bulik_2015}, which implies that the decoupled expressions for the high order coefficients are bad approximations for the true coefficients in that regime.


If \eqref{eq:CC_ansatz} is an eigenstate of the Hamiltonian $H$, we have $H\ket{\psi}=E\ket{\psi}=e^{T}E\ket{\phi}$. Therefore, assuming $\ket{\phi}$ is normalized, since any state $\ket{\phi_{j_1,j_2\ldots j_l}^{i_1i_2\ldots i_l}}$ is orthogonal to $\ket{\phi}$ for $l\neq 0$, if we project $H\ket{\psi}$ on $\bra{\phi}e^{-T}$ or $\bra{\phi_{j_1,j_2\ldots j_l}^{i_1i_2\ldots i_l}}e^{-T}$, we obtain the equations
\begin{equation}\label{eq:CC_equations}
\begin{split}
\bra{\phi}e^{-T}He^{T}\ket{\phi}&=E\,,\\
\bra{\phi_{j_1,j_2\ldots j_l}^{i_1i_2\ldots i_l}}e^{-T}He^{T}\ket{\phi}&=0\,,\quad 1\leq l\leq n\,,
\end{split}
\end{equation}
which, after substitution of $T$ by Eq. \eqref{eq:def_T_CC} and $H$ by its second quantization expression, yields a set of nonlinear equations defining the $T$ amplitudes $d_{j_1,j_2\ldots j_l}^{i_1i_2\ldots i_l}$ for $1\leq l\leq n$.

By expanding the exponentials in the similarity-transformed Hamiltonian $e^{-T}He^{T}$, we obtain
\begin{equation}\label{eq:emT_H_eT}
\begin{split}
&e^{-T}He^{T}= H+[H,T]+\frac{1}{2}\left[[H,T],T\right]\\
&\qquad +\frac{1}{3!}\left[\left[[H,T],T\right],T\right]+\frac{1}{4!}\left[\left[\left[[H,T],T\right],T\right],T\right]+\ldots
\end{split}
\end{equation}
For $T$ given by \eqref{eq:def_T_CC}, this series ends after the five terms included above. Very importantly, \eqref{eq:emT_H_eT} shows that $e^{-T}He^{T}$ only contains connected terms because the commutators cancel any disconnected terms. Because of this connectedness of the CC equations, the CC energy is size-extensive \cite{Brueckner_1955, Goldstone_1957,Bartlett_1981,Bartlett_Musial_2007}.
Note also that, because $T$ contains only excitation terms, any term in which it appears on the left of $H$ is disconnected. Consequently,
\begin{equation}\label{eq:simil_H_CC}
e^{-T}He^{T}=(He^{T})_c
\end{equation}
where the subscript $c$ indicates that only connected terms are kept.

The size-extensivity is a crucial aspect of the CC method, which makes it suitable for large systems, and thus applicable to condensed matter. However, because particle correlations are directly accounted for in the CC wave function up to two and sometimes three excitations in practice \cite{Bartlett_Musial_2007}, while higher order coefficients are decoupled, important strong-coupling physics cannot be properly treated. For example, for a SD with more than $n$ excitations and two excited particles or holes with opposite spins occupying an orbital where the Coulomb repulsion is strong, the coefficient should be small in order to minimize the energy, but in most terms defining the $l>n$ excitations coefficients of the CC wave function, the strongly interacting particles or holes are decoupled, and thus do not include the effect of that repulsion. If all the higher order coefficients were to be negligibly small, this effect would be small. However, on the contrary, the convergence rate of the single-reference expansion \eqref{eq:psi_sum} decreases as the coupling strength increases because many different SD's become nearly degenerate, and thus have comparable contributions in the ground state. At larger coupling, it therefore becomes more important to control higher order coefficients. Unfortunately, the computational complexity increases exponentially with the truncation order $n$ in the CC method. Other strategies are thus required to treat more strongly correlated systems. 

\section{Energy of ground states with low entanglement CI expansions}\label{sec:comput_GS_E}


Let us consider a system of $N=N_{\uparrow}+N_{\downarrow}$ fermions, for which the CI form of the wave function, Eq. \eqref{eq:psi_sum}, converges, though not necessarily rapidly. In the following, we will show that, when the $l\leq k$ excitation coefficients can be defined with a number of free parameters $N_{\leq k}^{TD}$ bounded above by a polynomial in $k$, then the grounds state energy $E$ is the solution of a set of equations of the CC form, Eqs. \eqref{eq:CC_equations}, but where $T$ is generally different from \eqref{eq:def_T_CC}. Consequently, $E$ also depends only on a small fraction of all the wave function parameters and wave function coefficients. Note that the convergence condition only serves to ensure that $E$ is the true ground state energy, hence the assumption that the convergence can be slow. This is an important aspect of that result that will be discussed further below.


Before we begin with the proof, let us give a few definitions about tensors. First, the term ``order'' will be used to specify the number of indices of a tensor. Then, the term ``rank'' will be used in a similar sense as for matrices, where it is the number of linearly independent columns, or rows, i.e., the number of non-zero singular values of the matrix. However, for tensors, there are more than one definition of rank. First, if we define a simple tensor as the tensor product of vectors, we can represent any tensor as a sum of simple tensors. Then, the simplest definition of a tensor rank is the minimal number of simple tensor in that representation. Another definition is the multilinear rank, or multirank. For a tensor of order $k$, the multirank is the vector $(r_1,r_2,\ldots,r_k)$  corresponding to the dimensions of the core tensor in the higher order singular value decomposition (HOSVD) \cite{Bergqvist_2010}. In the following, we use the expression ``full rank'' in the sense that a tensor does not have a more compact representation than as a multidimensional array of the same dimensions as the original, and ``low rank'' in the general sense that it can be represented by a tensor product involving less free parameters than the number of elements in the tensor.
To prove the result stated in the first paragraph in the most general case, we first define the following excitation operator $T$
\begin{subequations}\label{eq:T_SCC}
\begin{equation}\label{eq:T_SCC_D_op}
T=\sum_{i,j} a_i^\dagger a_j \hat{D}_j^i\\
\end{equation}
where $\hat{D}_j^i$ is an operator defined as
\begin{equation}\label{eq:eig_Eqn_D_op}
\begin{split}
\hat{D}_{j_k}^{i_k}\, \ket{\phi_{j_1j_2\ldots j_{k-1}}^{i_1i_2\ldots i_{k-1}}}=\frac{1}{k}\frac{c_{j_1j_2\ldots j_k}^{i_1i_2\ldots i_k}}{c_{j_1j_2\ldots j_{k-1}}^{i_1i_2\ldots i_{k-1}}}\,  \ket{\phi_{j_1j_2\ldots j_{k-1}}^{i_1i_2\ldots i_{k-1}}}\,.
\end{split}
\end{equation}
\end{subequations}
where the coefficients are assumed as those of the full CI wave function, Eq. \eqref{eq:psi_sum}. We could write explicitly the operator $\hat{D}_j^i$ in terms of products of particle number operators, but the definition \eqref{eq:eig_Eqn_D_op} is sufficient. With this definition of $T$, we have
\begin{equation}\label{eq:Tk_cfs}
\begin{split}
T\ket{\phi}&=\sum_{i,j} a_i^\dagger a_j \hat{D}_j^i\ket{\phi}\\
&=\sum_{i,j} c_{j}^{i}\ket{\phi_{j}^{i}}\\
T^2\ket{\phi}&=\sum_{i_2j_2} a_{i_2}^\dagger a_{j_2} \hat{D}_{j_2}^{i_2}\sum_{i_1j_1} c_{j_1}^{i_1}\ket{\phi_{j_1}^{i_1}}\\
&=\frac{1}{2}\sum_{\substack{i_1,i_2\\j_1,j_2}}   c_{j_1j_2}^{i_1i_2} \ket{\phi_{j_1j_2}^{i_1i_2}}\\
&\;\vdots\\
T^k\ket{\phi}&=\frac{1}{k!}\sum_{\substack{i_1,i_2\ldots i_k\\j_1,j_2,\ldots,j_k}} c_{j_1,j_2,\ldots,j_k}^{i_1,i_2\ldots i_k} \ket{\phi_{j_1,j_2,\ldots,j_k}^{i_1,i_2\ldots i_k}}\,.
\end{split}
\end{equation}
The definition \eqref{eq:T_SCC} assumes that, if a coefficient $c_{j_1,j_2,\ldots,j_k}^{i_1,i_2\ldots i_k}$ is finite, all the coefficients which upper and lower indices are subsets of $\{i_1,i_2\ldots i_k\}$ and $\{j_1,j_2,\ldots,j_k\}$, respectively, are also finite. They can however be arbitrarily small.

Using \eqref{eq:T_SCC} in the CC ansatz \eqref{eq:CC_ansatz} yields
\begin{equation}\label{eq:psi_CC_exact}
\begin{split}
e^T\ket{\phi}&=\left(1+T+\frac{T^2}{2}+\frac{T^3}{3!}+\ldots\right)\ket{\phi}\\
&=\ket{\phi}+\sum_{i,j} c_{j}^{i}\ket{\phi_{j}^{i}}+\frac{1}{4}\sum_{\substack{i_1,i_2\\j_1,j_2}}   c_{j_1j_2}^{i_1i_2} \ket{\phi_{j_1j_2}^{i_1i_2}}+\ldots\\
&+\left(\frac{1}{n_{max}!}\right)^2\sum_{\substack{ i_1i_2\ldots i_{n_{max}},\\ j_1,j_2\ldots j_{n_{max}}}} c_{j_1,j_2\ldots j_{n_{max}}}^{i_1i_2\ldots i_{n_{max}}} \ket{\phi_{j_1,j_2\ldots j_{n_{max}}}^{i_1i_2\ldots i_{n_{max}}}}\\
&=\ket{\phi}+\sum_{i,j} c_j^i \ket{\phi^i_j}+\sum_{\substack{\langle i_1,i_2\rangle \\\langle j_1,j_2\rangle}} c_{j_1j_2}^{i_1i_2}  \ket{\phi_{j_1j_2}^{i_1i_2}}\\
&\quad+\ldots+ \sum_{\substack{\langle i_1i_2\ldots i_{n_{max}}\rangle,\\\langle j_1,j_2\ldots j_{n_{max}}\rangle}} c_{j_1,j_2\ldots j_{n_{max}}}^{i_1i_2\ldots i_{n_{max}}} \ket{\phi_{j_1,j_2\ldots j_{n_{max}}}^{i_1i_2\ldots i_{n_{max}}}}\\
&=\ket{\psi_{FCI}}
\end{split}
\end{equation}
where we have used the fact that $c_{j_1,j_2\ldots j_n}^{i_1i_2\ldots i_n} \ket{\phi_{j_1,j_2\ldots j_n}^{i_1i_2\ldots i_n}}$ is invariant under the $(n!)^2$ different permutations of the indices. The operator \eqref{eq:T_SCC} therefore allows to put the full CI form of the wave function in the CC form. The bijective relation between $T^k$ and the $k$ excitations part of the wave function will also be useful below.

Now, with $T$ given by \eqref{eq:T_SCC}, the series \eqref{eq:emT_H_eT} has $n_{max}+1$ terms, unlike the CC case for which the number of terms is 5. Thus, if we expand the exponentials as Maclaurin series in the CC equations \eqref{eq:CC_equations}, we obtain
\begin{equation}\label{eq:SCC_gen_equation_lhs}
\begin{split}
E\delta_{l0}&=\bra{\phi_{j_1,j_2\ldots j_l}^{i_1i_2\ldots i_l}}e^{-T}He^{T}\ket{\phi}\\
&=\bra{\phi_{j_1,j_2\ldots j_l}^{i_1i_2\ldots i_l}}\left(1-T+\frac{T^2}{2}-\frac{T^3}{3!}+\ldots+(-1)^l\frac{T^l}{l!}\right)\\
&\quad\times H\left(1+T+\frac{T^2}{2}+\frac{T^3}{3!}+\ldots+\frac{T^{l+2}}{(l+2)!}\right)\ket{\phi}\,,\\
&\hspace{2.4in}0\leq l\leq n\,,
\end{split}
\end{equation}
where $\delta_{l0}$ is the Kronecker delta function and $\bra{\phi_{j_1,j_2\ldots j_0}^{i_1i_2\ldots i_0}}=\bra{\phi}$. From Eq. \eqref{eq:Tk_cfs}, those equations depend on the coefficients of the $1\leq l\leq n+2$ excitation SD's in \eqref{eq:psi_CC_exact}. The maximum power of $T$ on the right-hand side of $H$ is $l+2$ because $H$ can annihilate two particle-hole pairs at most and the maximum power of $T$ on the left-hand side of $H$ is $l$ since the smallest number of excitations is zero. Equations \eqref{eq:SCC_gen_equation_lhs} are exact since $e^{T}\ket{\phi}$ is the full CI wave function and thus $He^{T}\ket{\phi}=Ee^{T}\ket{\phi}$ holds exactly.





The number of SD's with $l$ excitations increases nearly exponentially with $l$. Therefore, since equations \eqref{eq:SCC_gen_equation_lhs} depend on the $1\leq l\leq n+2$ excitation coefficients, while the number of equations is equal to the number of SD's with $0\leq l\leq n$ excitations, the number of parameters is much larger than the number of equations if the equations are truncated at a given $n<n_{max}$ and the coefficients have full rank. As described in section \ref{sec:CI_CC}, the CC strategy to close the equations amounts to expressing the coefficients of SD's with $l>n$ excitations using only cluster operator amplitudes for $l\leq n$ excitations. However, this is only one of many possible ways of closing the equations. Indeed, Eqs. \eqref{eq:SCC_gen_equation_lhs} can be closed at a value $n<n_{max}$ if the number of free parameters defining the coefficients increases at a slower rate with the number of excitations than the number of SD's itself. This might seem like a strange assumption, but it is possible if there exist low-rank decompositions of each set of coefficients $\left\{c_{j_1,j_2\ldots j_{l}}^{i_1i_2\ldots i_{l}}\right\}$, when the latter is interpreted as a tensor. In practice, the increasing rate of the number of free parameters with $l$ must be slow enough that the equations are closed at $n$ small, so that the number of equations and free parameters are computationally tractable. This is the case if we assume an increasing rate smaller than some low order polynomial. Let us see this in more details. 

If the number of up and down spin-orbitals are both equal to $L$, the number of coefficients of the form $c_{j_1,j_2\ldots j_{k}}^{i_1i_2\ldots i_{k}}$, assuming that indices $i_1$ to $i_p$ have spins up and indices $i_{p+1}$ to $i_k$ have spins down, is
\begin{equation}
\begin{split}
N_{k,p}^{CI}&=\binom{L-N_\uparrow}{p}\binom{L-N_\downarrow}{k-p}\binom{N_\uparrow}{p}\binom{N_\downarrow}{k-p}\\
&=\frac{(L-N_\uparrow)!}{p!(L-N_\uparrow-p)!}\frac{(L-N_\uparrow)!}{(k-p)!(L-N_\uparrow-k+p)!}\\
&\quad \times\frac{N_\uparrow!}{p!(N_\uparrow-p)!}\frac{N_\uparrow!}{(k-p)!(N_\uparrow-k+p)!}\,.
\end{split}
\end{equation}
the number of coefficients with $k$ excitations is
\begin{equation}\label{eq:Nk_CI}
N_{k}^{CI}=\sum_{p=max(0,k-N_{\downarrow})}^{min(k,N_{\uparrow})} N_{k,p}^{CI}
\end{equation}
and the total number of coefficients, or CC equations, with up to $k$ excitations is
\begin{equation}
N^{CI}_{\leq k}=\sum_{l=0}^{k} N_{l}^{CI}\,.
\end{equation}
For $k$ fixed, $N^{CI}_{\leq k}$ is polynomial in $N=N_\uparrow+N_\downarrow$. 

Now, using the fact that
\begin{equation}
\binom{l}{m}\geq \left(\frac{l}{m}\right)^m\,,\quad 1\leq m\leq l\,,
\end{equation}
we have that
\begin{equation}
N_{k,p}^{CI}\geq \left(\frac{L-N_\uparrow}{p}\right)^p\left(\frac{L-N_\downarrow}{k-p}\right)^{k-p}\left(\frac{N_\uparrow}{p}\right)^p\left(\frac{N_\downarrow}{k-p}\right)^{k-p}
\end{equation}
If we consider only the cases $p=k$,
\begin{equation}
N_{k,k}^{CI}\geq \left(\frac{(L-N_\uparrow)N_\uparrow}{k^2}\right)^k\,.
\end{equation}
Now, if the number of parameters $N_{\leq k}^{TD}$ defining the $l\leq k$ excitation coefficients is bounded above polynomially in $k$, we have $N_{\leq k+2}^{TD}\leq ak^r$, where $r\in \mathbb{N}$ and $a>0$. We then have
\begin{equation}
\begin{split}
\frac{N_{\leq k}^{CI}}{N_{\leq k+2}^{TD}}> \frac{N_{k,k}^{CI}}{N_{\leq k+2}^{TD}}&\geq \frac{\left(\frac{(L-N_\uparrow)N_\uparrow}{k^2}\right)^k}{ak^r}\\
\frac{N_{\leq k}^{CI}}{N_{\leq k+2}^{TD}}&> \frac{1}{a}e^{k\ln \frac{(L-N_\uparrow)N_\uparrow}{k^2}-r\ln k}
\end{split}
\end{equation}
For $k^2\ll (L-N_\uparrow)N_\uparrow$, this ratio grows rapidly with $k$ and thus, for $r$ small, there is a small value of $k$ for which $N_{\leq k}^{CI}>N_{\leq k+2}^{TD}$. Therefore, given the polynomially bounded parametrization of the coefficients, the equations \eqref{eq:SCC_gen_equation_lhs} are closed at a small value of $n$ and the ground state energy $E$ depends only on the $k\leq n+2$ excitation coefficients, a fraction of order $\left(\frac{n}{N}\right)^r$ or smaller of the total number of free parameters defining the wave function, and $E$ can be computed in polynomial time by solving a set of equations of the CC form, or generalized CC equations. This concludes the proof of the result stated in the first paragraph of the present section.

A corollary to that computation result is that equal-time correlation functions at zero temperature can also be computed if the ground state energy can. This is the Hellmann-Feynman theorem, which is proved in Appendix \ref{sec:eq_time_corr}. In particular, this fact can be used to determine an optimal spin-orbital basis. This is discussed further in section \ref{sec:orb_basis}.






The standard CC method corresponds to the simplest application of the computation result we have just proved. Indeed, since the $l>n$ excitation coefficients depend only on low order amplitudes in CC, $N_{\leq k}^{CC}$ is constant for $k\geq n$, i.e., it is the simplest possible polynomial.


The fact that there is no lower bound on the convergence rate of the CI expansion in the above result implies that it also applies to strongly correlated systems. Indeed, as discussed at the end of section \ref{sec:CI_CC}, at strong, but finite, coupling strength, the ground state of such systems also possess a converging single-reference CI expansion. The difference with the weak coupling regime is that the expansion converges only slowly as the number of excitations increases, hence the failure of approximations that truncate the cluster operator with respect to that parameter. As discussed in section \ref{sec:CC_LRTD} that follows, the above result allows to use CC equations with other types of approximations much better suited to the strongly correlated regime.

As mentioned above, a polynomially bounded parametrization of the sets of coefficients $\left\{c_{j_1,j_2\ldots j_{l}}^{i_1i_2\ldots i_{l}}\right\}$ for $1\leq l\leq n+2$ can be obtained using low-rank tensor decompositions, which are based on the existence of basis transformations, in single and multiple particle and hole spaces, such that a superposition of $l$ excitations SD's can be written with a much smaller number of components in the transformed basis. Since the rank can be used as a measure of entanglement, this type of parametrization is also a low-entanglement representation of the coefficients. In section \ref{sec:CC_LRTD} that follows, we will discuss two types of LRTD-based parametrization which are applications of the ground state energy computation result given above, and are good candidates for strong-coupling extensions of the CC method. Adapted LRTD-based representations applicable in both cases will also be described in section \ref{sec:coeffs_tensor_decomp}.


Note that equations \eqref{eq:SCC_gen_equation_lhs} are also valid if $T$ is the CC cluster operator, Eq. \eqref{eq:def_T_CC}, but without any truncation. Then, the maximum power of $T$ is 4, as discussed in section \ref{sec:CI_CC}, and only the terms of $T$ up to $T_{l}$ contribute on the left-hand side of $H$ and up to $T_{l+2}$ on the right-hand side. However, as shall be discussed in section \ref{sec:CC_LRTD} that follows, that choice produces a particular type of approximation when combined with the LRTD-based representations. On the other hand, thanks to the operator \eqref{eq:T_SCC}, we obtain a result applicable to any converging single-reference expansion with a polynomially bounded parametrization of the wave function coefficients, which includes the CC approximation and the two types of approximations discussed in the section \ref{sec:CC_LRTD}.

\section{Extending the coupled-cluster method with low-rank tensor decompositions}\label{sec:CC_LRTD}

In this section we discuss two possible generalizations of the CC method based the result of section \ref{sec:comput_GS_E}. The first one is a straightforward extension of CC and the second approach uses a parametrization more directly connected to the CI form of the wave function. We will also discuss the choice of spin-orbital basis and reference SD.



\subsection{A coupled-cluster approach without truncation of the cluster operator}\label{sec:CC_no_trunc}

The simplest tensor-based extension of CC, or TCC for ``tensor-CC'', that uses a general polynomially bounded parametrization of the wave function coefficients applies the LRTD-based parametrization to the amplitudes $d_{j_1,j_2\ldots j_k}^{i_1i_2\ldots i_k}$ in the CC cluster operator, Eq. \eqref{eq:def_T_CC}, for all the terms with $1\leq k\leq n+2$ involved in the equations \eqref{eq:SCC_gen_equation_lhs} with $0\leq l \leq n$. The explicit form of those equations in terms of cluster operator amplitudes and Hamilitonian parameters are the usual CC equations, except that $T$ is not truncated with respect to excitation number, or equivalently, truncated at $n+2$ excitations. Therefore, if $n=2$, we use only the CC equations with $l\leq 2$ of the CC approximation with truncation of $T$ at four excitations (CCSDTQ). To complete the equations, the LRTD-based representation for sets of coefficients with $1\leq k\leq 4$ excitations provided in section \ref{sec:coeffs_tensor_decomp}, or other similar representations, can be used to parametrize the amplitudes.

The most crucial aspect of that approximation, as compared with standard CC, is the fact that there is no truncation of $T$ with respect to excitation number. Indeed, if we consider the wave function coefficients generated by the CC exponential ansatz with such a $T$ operator, the reducible parts then have irreducible corrections at \textit{all orders}, except that those corrections are expressed with LRTD. For instance, this type of parametrization of the $T$ amplitudes can take into account local correlations, which can have a similar effect as a Gutzwiller operator \cite{Gutzwiller:1963} that modulates the coefficients of SD's with strongly correlated excited quasiparticles (i.e. particles or holes) occupying the same spatial orbital. This is only the simplest possibility allowed by such a parametrization, since non-local correlations can be treated as well. Therefore, in principle, this type of tensor-based approximation yields results valid at stronger coupling than standard CC. In fact, while the CC approximation is exact in the weak coupling limit, corresponding to a fast converging cluster operator, TCC would be exact in the low-entanglement limit of the cluster operator $T$, which includes the weak coupling limit, but is more general. It would therefore be a very compact approximation at weak coupling, but also at moderately strong coupling, as long as the wave function coefficients can be represented by the reducible parts generated by the exponential ansatz, corrected by irreducible LRTD-based terms. On the other hand, as the coupling strength increases, the reducible parts become worse approximations of the actual wave function coefficients, which requires larger and more complex irreducible parts that can become difficult to represent using LRTD. The type of approximation discussed in section \ref{sec:CI_from_CC} that follows could then become useful.

\subsection{Computing CI coefficients using generalized CC equations}\label{sec:CI_from_CC}

Another possible tensor-based extension of the CC method uses the definition \eqref{eq:T_SCC} for $T$ and equations \eqref{eq:SCC_gen_equation_lhs}. This produces equations similar to the CC ones, though involving the CI coefficients instead of cluster operator amplitudes, which are closed by parametrizing the coefficients with LRTD. This approach could thus be called TCICC, for ``tensor-CI using CC equations''. In that case, unlike the approximation described in section \ref{sec:CC_no_trunc} above, the wave function coefficients are completely defined using LRTD and do not have reducible parts. Although such an approximation is probably not optimal at weak coupling, it is in principle applicable at very strong coupling, as long as there exist a single-reference CI representation of the wave function that converges at least slowly with the number of particle-hole excitation, ensuring that a good approximation to the actual ground state energy can be found. Such an approximation is exact in the low-entanglement limit of the CI coefficients, which is very different from the limit of validity of the CC approximation, and is more general than the TCC approach, which is a special case of TCICC.

The explicit form of equations \eqref{eq:SCC_gen_equation_lhs} in terms of CI coefficients and Hamiltonian parameters, obtained using the definition \eqref{eq:T_SCC} for $T$, are provided for the $l\leq 4$ excitation coefficients in section \ref{sec:eqns_coeffs}. To derive those equations, instead of the standard approach of CC that use a special notation to express particle-hole normal-ordered operators, we have used a representation of $H$ in terms of excited particles and holes operators which is described in section \ref{sec:H_ph}. To parametrize the CI coefficients, the same type of LRTD-based representation as proposed for the cluster amplitudes, and described in section \ref{sec:coeffs_tensor_decomp}, could be used. 

\subsection{Choice of spin-orbital basis and reference}\label{sec:orb_basis}

When working with a single-reference expansion of the wave function such as Eq. \eqref{eq:psi_CC_exact}, the spin-orbital basis that maximizes convergence is the natural spin-orbital basis. Since these orbitals diagonalize the single-particle density matrix $\bra{\psi_{FCI}}a_i^\dagger a_j\ket{\psi_{FCI}}$, which does not have any time dependance, they can be computed using the source-field method described in Appendix \ref{sec:eq_time_corr}. This implies that the exact wave function $\ket{\psi_{FCI}}$ is approximated by $\ket{\psi}$. For instance, one can start the calculation using Hartree-Fock spin-orbitals, then, after a first approximation of the ground state energy and wave function parameters are computed, a first approximation of the natural spin-orbitals can be computed with the source-field method, with which a new energy and wave function can be obtained. The procedure can then be repeated until the energy converges. If the number of spin-orbitals is $2L$, one diagonalization of $\bra{\psi}a_i^\dagger a_j\ket{\psi}$ requires $2L^2$ different solutions of the CC equations for a Hamiltonian perturbed by a small source field. Although that number of different CC calculations might seem daunting, each solution is only slightly different from the unperturbed one and is therefore much faster to compute.



\section{Particle-hole representation of the Hamiltonian for single reference-based calculations}\label{sec:H_ph}

To derive the CC equations, it is convenient to use a representation of the Hamiltonian $H$ where the operators are normal-ordered for both particles and holes. The standard method is to express the Hamiltonian operators using the notation $\{a_i^\dagger a_j\}$ and $\{a_i^\dagger a_j^\dagger a_k a_l\}$, which puts the annihilation operators for empty orbitals (in $\ket{\phi}$) \textit{and} the creation operators for occupied orbitals at the rightmost positions ($\{\ldots \}$ is not an anticommutator in that notation) \cite{Kucharski_Bartlett_1986, Bartlett_Musial_2007}. Here, we will also use such a normal-ordered Hamiltonian, but instead of using the $\{\ldots \}$ notation, we will define excited particle and hole operators $p_i^\dagger$, for the unoccupied spin-orbitals, and $h_i^\dagger$, for the occupied spin-orbitals, respectively, and express $H$ using those operators. The resulting representation is less compact than with the $\{\ldots \}$ notation, but it accomplishes a part of the work that is otherwise required during the algebraic derivation of the equations, and thus it actually simplifies it since that part is done only once. In addition, it only uses second quantization notation and is physically intuitive, as it explicitly takes the form of a Hamiltonian acting on a vacuum and the particles and anti-particles created from it, which is how the reference $\ket{\phi}$ is treated. Finally, no distinction has to be made between empty and occupied spin-orbitals indices during the derivation with that representation because that information is included in the definition of the excited particle and hole operators. 

First, let us define
\begin{equation}\label{eq:def_t_V}
\begin{split}
t_{ij}&=\delta_{\sigma_i\sigma_j}\int d^3r\, \eta_i^*(\mathbf{r}) \left(\frac{-\hbar^2\nabla^2}{2m_e}+V^c_{ei}(\mathbf{r})\right) \eta_j(\mathbf{r})\\
V_{ijkl}^c&=\delta_{\sigma_i\sigma_l}\delta_{\sigma_j\sigma_k}\int d^3r_1 d^3r_2\, \eta_i^*(\mathbf{r}_1) \eta_j^*(\mathbf{r}_2) V^c_{ee}(\mathbf{r}_1-\mathbf{r}_2)\\
&\qquad\times \eta_k(\mathbf{r}_2)\eta_l(\mathbf{r}_1)\,,
\end{split}
\end{equation}
where the indices are spin-orbital indices, $m_e$ is the electron mass, $\eta_i(\mathbf{r})$ are spatial orbitals, $\sigma_i=\uparrow,\downarrow$ are spin indices, $V^c_{ei}(\mathbf{r})$ is the static Coulomb potential experienced by an electron due to ions treated in the Born-Oppenheimer approximation, and $V^c_{ee}(\mathbf{r}_1-\mathbf{r}_2)$ is the electron-electron Coulomb potential. Then, the Hamiltonian in the second quantized form is
\begin{equation}
\hat{H}=\sum_{ij} t_{ij} a_i^\dagger a_j + \frac{1}{2}\sum_{ijkl} V_{ijkl}^c a_i^\dagger a_j^\dagger a_k a_l\\
\end{equation}
or, using the anti-symmetrized Coulomb interaction $V_{ijkl}=V_{ijkl}^c-V_{ijlk}^c$,
\begin{equation}\label{eq:H_K_V}
\begin{split}
\hat{H}&=\sum_{ij} t_{ij} a_i^\dagger a_j + \frac{1}{4}\sum_{ijkl} V_{ijkl} a_i^\dagger a_j^\dagger a_k a_l\\
\hat{H}&=\hat{K}+\hat{V}\,,
\end{split}
\end{equation}
where $\hat{K}$ is the one body term and $\hat{V}$, the two-body term. Note that, from \eqref{eq:def_t_V} and the hermicity of the potential energy term, we have that $V_{ijkl}^c=V_{lkji}^c$ and from the inversion symmetry of $V^c_{ee}(\mathbf{r}_1-\mathbf{r}_2)$, $V_{ijkl}^c=V_{jilk}^c$, and thus $V_{ijkl}^c=V_{klij}^c$. Therefore, $V_{ijkl}$ also has all those symmetries, in addition to $V_{ijkl}=-V_{ijlk}=-V_{jikl}$.

We now define excited particle and hole operators associated with the reference $\ket{\phi}$:
\begin{subequations}\label{eq:def_p_i_h_i}
\begin{equation}\label{eq:def_p_i}
p_i^\dagger =(1-n_i^{\phi})a_i^\dagger\,,
\end{equation}
\begin{equation}\label{eq:def_h_i}
h_i^\dagger=n_i^{\phi} a_i\,,
\end{equation}
\end{subequations}
where $n_i^{\phi}=\bra{\phi}a_i^\dagger a_i\ket{\phi}$ is the number of particles occupying spin-orbital $i$ in $\ket{\phi}$ and is thus constant. Therefore, $p_i^\dagger$ and $p_i$ act only on empty spin-orbitals of $\ket{\phi}$, while $h_i^\dagger$ and $h_i$ act only on occupied spin-orbitals of $\ket{\phi}$. From those definitions we have
\begin{equation}\label{eq:subst_ph}
\begin{split}
a_i^\dagger&=p_i^\dagger+h_i\\
\end{split}
\end{equation}
and the anticommutation relations
\begin{equation}
\begin{split}
\{p_i,p_j\}&=0,\\
\{p_i^\dagger,p_j\}&=\delta_{ij}(1-n_i^{\phi}),
\end{split}
\end{equation}
\begin{equation}
\begin{split}
\{h_i,h_j\}&=0,\\
\{h_i^\dagger,h_j\}&=\delta_{ij}n_i^{\phi},\\
\end{split}
\end{equation}
\begin{equation}
\begin{split}
\{p_i,h_j\}&=0,\\
\{p_i,h_j^\dagger\}&=0.\\
\end{split}
\end{equation}
Using the definitions \eqref{eq:def_p_i_h_i}, the SD \eqref{eq:phi_i_j} is written
\begin{equation}\label{eq:phi_i_j_p_h}
\ket{\phi^{i_1,i_2,\ldots, i_k}_{j_1,j_2,\ldots,j_k}}=p_{i_k}^\dagger  h_{j_k}^\dagger p_{i_{k-1}}^\dagger  h_{j_{k-1}}^\dagger \ldots p_{i_1}^\dagger h_{j_1}^\dagger \ket{\phi}\,.
\end{equation}

If we substitute \eqref{eq:subst_ph} in $\hat{K}$ and put each term in the usual normal order of second quantization, i.e., with all the annihilation operators on the right, we obtain
\begin{equation}\label{eq:K_phr}
\begin{split}
\hat{K}&=\sum_{ij} t_{ij}p_{i}^\dagger p_{j} -\sum_{ij} t_{ij}h_{i}^\dagger h_{j}+\sum_{ ij} t_{ij}\left(p_{i}^\dagger h_{j}^\dagger+H.c.\right)\\
&\qquad+\sum_{i}t_{ii}n_i^{\phi}\,,
\end{split}
\end{equation}
where $H.c.$ stands for Hermitian conjugate. We also have assumed that $t_{ij}$ is real and thus $t_{ij}=t_{ji}$. Here the indices can run over all spin-orbitals because of the prefactors in the definitions \eqref{eq:def_p_i_h_i}. Note how the energy of holes is formally the negative of the particles' and how the constant term $\bra{\phi}\hat{K}\ket{\phi}$ appears explicitly.

Now, if we substitute \eqref{eq:subst_ph} in $\hat{V}$, put the terms in normal order and use the symmetries of $V_{ijkl}$, we obtain
\begin{equation}\label{eq:def_V_ph}
\begin{split}
\hat{V}&=\sum_{ijk} V_{ikkj}n_k^{\phi}p_i^\dagger p_j-\sum_{ijk} V_{ikkj}n_k^{\phi}h_i^\dagger h_j\\
&\quad+\sum_{ ijk} V_{ikkj}n_k^{\phi}(p_i^\dagger h_j^\dagger + H.c.)+\frac{1}{2}\sum_{ij} V_{ijji}n_i^{\phi}n_j^{\phi}\\
&\quad+\frac{1}{4}\sum_{ijkl} V_{ijkl} p_i^\dagger p_j^\dagger p_k p_l +\frac{1}{4}\sum_{ijkl} V_{ijkl}h_i^\dagger h_j^\dagger h_k h_l\\
&\quad+\frac{1}{4}\sum_{ ijkl} V_{ijkl}(p_i^\dagger p_j^\dagger h_k^\dagger h_l^\dagger+H.c.)\\
&\quad+\frac{1}{2}\sum_{ ijkl} V_{ijkl}(p_i^\dagger p_j^\dagger h_k^\dagger p_l+H.c.)\\
&\quad +\frac{1}{2}\sum_{ ijkl} V_{ijkl}(h_i^\dagger h_j^\dagger p_k^\dagger h_l+H.c.)- \sum_{ijkl} V_{ikjl}p_i^\dagger h_j^\dagger h_k p_l\,.
\end{split}
\end{equation}

Here, it is interesting to consider only the two-body terms involving particle and hole number operators $n^p_{i}=p_i^\dagger p_i$ and $n^h_i=h_i^\dagger h_i$ in \eqref{eq:def_V_ph}, we then obtain
\begin{equation}\label{eq:dens_coulomb_int}
\hat{V}_{n_p,n_h}=\frac{1}{2}\sum_{ij} V_{ijji} n^p_{i}n^p_{j}+\frac{1}{2}\sum_{ij} V_{ijji} n^h_{i}n^h_{j} -\sum_{ij} V_{ijji} n^p_{i}n^h_{j}\,.
\end{equation}
Therefore, assuming $V_{ijji}>0$, holes repel each other in the same way as particles and particles and holes attract each other. From that expression and the expression for $\hat{K}$, Eq. \eqref{eq:K_phr}, we see  how particles and holes in an insulating condensed matter system and electrons and positrons in the vacuum are mathematically equivalent in the long wavelength limit.

Now, if we define
\begin{equation}
t^{\phi}_{ij}=t_{ij}+\sum_k V_{ikkj}n_k^{\phi}\,,
\end{equation}
which is called the Fock matrix, we finally obtain the particle-hole representation of the Hamiltonian:
\begin{equation}\label{eq:H_phi_r_fin}
\begin{split}
\hat{H}^{\phi}&=\hat{H}-\bra{\phi}\hat{H}\ket{\phi}\\
&=\hat{H}-\sum_{i}t_{ii}n_i^{\phi}-\frac{1}{2}\sum_{ij} V_{ijji} n_i^{\phi}n_j^{\phi}\\
&= \sum_{ij} t^{\phi}_{ij}p_{i}^\dagger p_{j}-\sum_{ij} t^{\phi}_{ij}h_{i}^\dagger h_{j}+\sum_{ ij} t^{\phi}_{ij}\left(p_{i}^\dagger h_{j}^\dagger+H.c.\right)\\
&\quad+\frac{1}{4}\sum_{ijkl} V_{ijkl} p_i^\dagger p_j^\dagger p_k p_l +\frac{1}{4}\sum_{ijkl} V_{ijkl}h_i^\dagger h_j^\dagger h_k h_l\\
&\quad+\frac{1}{4}\sum_{ ijkl} V_{ijkl}(p_i^\dagger p_j^\dagger h_k^\dagger h_l^\dagger+H.c.)\\
&\quad+\frac{1}{2}\sum_{ ijkl} V_{ijkl}(p_i^\dagger p_j^\dagger h_k^\dagger p_l+H.c.)\\
&\quad+\frac{1}{2}\sum_{ ijkl} V_{ijkl}(h_i^\dagger h_j^\dagger p_k^\dagger h_l+H.c.)- \sum_{ijkl} V_{ikjl}p_i^\dagger h_j^\dagger h_k p_l\,.
\end{split}
\end{equation}
Although this representation is lengthier than \eqref{eq:H_K_V}, it is convenient when working with the expansion of the wave function in numbers of particle-holes excitations, Eq. \eqref{eq:psi_sum}, either in CI or CC calculations, as illustrated in Appendix \ref{sec:deriv_SCC_eqns}.

\section{Generalized CC equations for the energy and CI wave function coefficients}\label{sec:eqns_coeffs}

Let us now write the equations \eqref{eq:SCC_gen_equation_lhs} for $T$ given by \eqref{eq:T_SCC}, in terms of the CI wave function coefficients and Hamiltonian parameters, and projection SD's $\bra{\phi_{j_1,j_2\ldots j_l}^{i_1i_2\ldots i_l}}$ with $0\leq l\leq 2$. To derive the equations, we have used the excited particle and hole operators, Eqs. \eqref{eq:def_p_i_h_i}, and the representation \eqref{eq:H_phi_r_fin} of the Hamiltonian. However, although that representation is convenient for that task, it remains a quite lengthy derivation. We therefore only provide the equations here. The derivation of the equations with projection on $\bra{\phi}$ and $\bra{\phi_j^i}$ are given in Appendix \ref{sec:deriv_SCC_eqns}, while the derivation of the equations with projection on $\bra{\phi_{j_1j_2}^{i_1i_2}}$ is provided as \hyperlink{suppl_mat}{supplemental material}.

Using the shifted Hamiltonian, \eqref{eq:H_phi_r_fin}, the CC equations are
\begin{equation}\label{eq:CC_equations_H_phi}
\begin{split}
\bra{\phi}e^{-T}\hat{H}^{\phi}e^{T}\ket{\phi}&=\Delta E\,,\\
\bra{\phi_{j_1,j_2\ldots j_l}^{i_1i_2\ldots i_l}}e^{-T}\hat{H}^{\phi}e^{T}\ket{\phi}&=0\,,\quad 1\leq l\leq n\,,
\end{split}
\end{equation}
where $\Delta E=E-\bra{\phi}\hat{H}\ket{\phi}$, and we will take $n=2$.


First, because $T$, Eq. \eqref{eq:T_SCC}, is an excitation operator, $\bra{\phi}e^{-T}=\bra{\phi}$. Then, taking into account the fact that $\hat{H}^{\phi}$ can destroy at most two particle-hole pairs and that $\bra{\phi}\hat{H}^{\phi}\ket{\phi}=0$, the equation for the energy is
\begin{equation}\label{eq:eqn_Delta_E}
\begin{split}
\Delta E&=\bra{\phi}e^{-T}\hat{H}^{\phi}e^T\ket{\phi}\\
&=\bra{\phi}\hat{H}^{\phi}\left(1+T+\frac{1}{2}T^2\right)\ket{\phi}\\
\Delta E&=\bra{\phi}\hat{H}^{\phi}T\ket{\phi}+\frac{1}{2}\bra{\phi}\hat{H}^{\phi}T^2\ket{\phi}\,.
\end{split}
\end{equation}
which, in terms of CI coefficients and Hamiltonian parameters, is
\begin{equation}\label{eq:CC_eq_0}
\Delta E=\sum_{ij} t^{\phi}_{ij}c^i_j-\frac{1}{4}\sum_{ijkl} c^{ij}_{kl}V_{ijkl}\,.
\end{equation}

Then, the equations with projection on $\bra{\phi_j^i}$ is
\begin{equation}\label{eq:eqn_phi_1}
\begin{split}
0&=\bra{\phi_j^i}e^{-T}\hat{H}^{\phi} e^T\ket{\phi}\\
&=\bra{\phi_j^i}\left(1-T\right)\hat{H}^{\phi}\left(1+T+\frac{1}{2}T^2+\frac{1}{3!}T^3\right)\ket{\phi}\\
0&=\bra{\phi_j^i}\hat{H}^{\phi}\ket{\phi}+ \bra{\phi_j^i}\hat{H}^{\phi}T\ket{\phi}-\bra{\phi_j^i}T\hat{H}^{\phi}T\ket{\phi}+\frac{1}{2}\bra{\phi_j^i}\hat{H}^{\phi}T^2\ket{\phi}\\
&-\frac{1}{2}\bra{\phi_j^i}T\hat{H}^{\phi}T^2\ket{\phi}+\frac{1}{3!}\bra{\phi_j^i}\hat{H}^{\phi}T^3\ket{\phi}\,,
\end{split}
\end{equation}
which yields
\begin{equation}\label{eq:CC_eq_1}
\begin{split}
0&=t^{\phi}_{ij}+ \sum_l t^{\phi}_{il}c^l_j - \sum_l t^{\phi}_{jl} c^i_l-\sum_{mn} V_{imjn} c^n_m\\
&\qquad+\sum_{kl}t^{\phi}_{kl}\,\left(c^{ik}_{jl}-c^i_jc^k_l\right)+\frac{1}{4}\sum_{klm} V_{klmi}c_{jm}^{kl}\\
&\qquad-\frac{1}{4}\sum_{klm} V_{klmj}c_{kl}^{im}-\frac{1}{4}\sum_{klmn} V_{klmn} (c^{kli}_{mnj}-c^i_jc^{kl}_{mn})\,.
\end{split}
\end{equation}

\begin{widetext}
Finally, the equations with projection on $\bra{\phi_{j_1j_2}^{i_1i_2}}$ are
\begin{equation}\label{eq:eqn_phi_2}
\begin{split}
0&=\bra{\phi_{j_1j_2}^{i_1i_2}}e^{-T}\hat{H}^{\phi}e^T\ket{\phi}\\
&=\bra{\phi_{j_1j_2}^{i_1i_2}}\left(1-T+\frac{1}{2}T^2\right)\hat{H}^{\phi}\Bigg(1+T+\frac{1}{2}T^2+\frac{1}{3!}T^3+\frac{1}{4!}T^4\Bigg)\ket{\phi}\\
&=\bra{\phi_{j_1j_2}^{i_1i_2}}\hat{H}^{\phi}\ket{\phi}+\bra{\phi_{j_1j_2}^{i_1i_2}}\hat{H}^{\phi}T\ket{\phi}-\bra{\phi_{j_1j_2}^{i_1i_2}}T\hat{H}^{\phi}\ket{\phi}+\frac{1}{2}\bra{\phi_{j_1j_2}^{i_1i_2}}\hat{H}^{\phi}T^2\ket{\phi}-\bra{\phi_{j_1j_2}^{i_1i_2}}T\hat{H}^{\phi}T\ket{\phi}+\frac{1}{3!}\bra{\phi_{j_1j_2}^{i_1i_2}}\hat{H}^{\phi}T^3\ket{\phi}\\
&-\frac{1}{2}\bra{\phi_{j_1j_2}^{i_1i_2}}T\hat{H}^{\phi}T^2\ket{\phi}+\frac{1}{2}\bra{\phi_{j_1j_2}^{i_1i_2}}T^2\hat{H}^{\phi}T\ket{\phi}+\frac{1}{4!}\bra{\phi_{j_1j_2}^{i_1i_2}}\hat{H}^{\phi}T^4\ket{\phi}-\frac{1}{3!}\bra{\phi_{j_1j_2}^{i_1i_2}}T\hat{H}^{\phi}T^3\ket{\phi}+\frac{1}{4}\bra{\phi_{j_1j_2}^{i_1i_2}}T^2\hat{H}^{\phi}T^2\ket{\phi}
\end{split}
\end{equation}
which yields (details in \hyperlink{suppl_mat}{supplemental material})
\begin{equation}\label{eq:CC_eq_2}
\begin{split}
0&=-V_{i_1i_2j_1j_2}+t^{\phi}_{i_1j_1}c_{j_2}^{i_2}-t^{\phi}_{i_1j_2}c_{j_1}^{i_2}-t^{\phi}_{i_2j_1}c_{j_2}^{i_1}+t^{\phi}_{i_2j_2}c_{j_1}^{i_1}-\sum_{k} \left(V_{i_1i_2j_1k}c_{j_2}^{k}-V_{i_1i_2j_2k}c_{j_1}^{k}\right)\\
&+\sum_{k} \left(V_{j_1j_2i_1k}c_{k}^{i_2}-V_{j_1j_2i_2k}c_{k}^{i_1}\right)+\sum_{k}\left(t^{\phi}_{i_1k}c_{j_1j_2}^{ki_2} +t^{\phi}_{i_2k}c_{j_1j_2}^{i_1k}\right)-\sum_{k}\left(t^{\phi}_{j_1k}c_{kj_2}^{i_1i_2}+t^{\phi}_{j_2k}c_{j_1k}^{i_1i_2}\right)\\
&-\frac{1}{2}\sum_{kl}V_{i_1i_2kl}c_{j_1j_2}^{kl}-\frac{1}{2}\sum_{kl}V_{j_1j_2kl}c_{kl}^{i_1i_2}-\sum_{kl} \left(V_{i_1kj_1l}c_{kj_2}^{li_2}+V_{i_2kj_1l}c_{kj_2}^{i_1l}+V_{i_1kj_2l}c_{j_1k}^{li_2}+V_{i_2kj_2l}c_{j_1k}^{i_1l}\right)\\
&+\sum_{kl} t^{\phi}_{kl}c_{lj_1j_2}^{ki_1i_2}-\frac{1}{2}\sum_{klm} \left(V_{klmi_1}c_{mj_1j_2}^{kli_2}-V_{klmi_2}c_{mj_1j_2}^{kli_1}\right)+\frac{1}{2}\sum_{klm} \left(V_{klmj_1}c_{klj_2}^{mi_1i_2}-V_{klmj_2}c_{klj_1}^{mi_1i_2}\right)\\
&+c_{j_1j_2}^{i_1i_2}\sum_{kl} t^{\phi}_{kl}c_{l}^{k}-\frac{1}{4}\sum_{klmn}V_{klmn} c_{mnj_1j_2}^{kli_1i_2}-\frac{1}{4}c_{j_1j_2}^{i_1i_2} \sum_{klmn} V_{klmn}c^{kl}_{mn}\\
&+c_{j_1j_2}^{i_1i_2}\Bigg[-\frac{1}{2}\left(\frac{t^{\phi}_{i_2j_2}}{c_{j_2}^{i_2}}+\frac{t^{\phi}_{i_1j_2}}{c_{j_2}^{i_1}}+\frac{t^{\phi}_{i_2j_1}}{c_{j_1}^{i_2}}+\frac{t^{\phi}_{i_1j_1}}{c_{j_1}^{i_1}}\right)-\frac{1}{2}\sum_{k}\left(\frac{t^{\phi}_{i_2k}c_{j_2}^{k}}{c_{j_2}^{i_2}}+\frac{t^{\phi}_{i_1k}c_{j_2}^{k}}{c_{j_2}^{i_1}}+\frac{t^{\phi}_{i_2k}c_{j_1}^{k}}{c_{j_1}^{i_2}}+\frac{t^{\phi}_{i_1k}c_{j_1}^{k}}{c_{j_1}^{i_1}}\right)\\
&+\frac{1}{2}\sum_{k}\left(\frac{t^{\phi}_{j_2k}c_{k}^{i_2}}{c_{j_2}^{i_2}}+\frac{t^{\phi}_{j_2k}c_{k}^{i_1}}{c_{j_2}^{i_1}}+\frac{t^{\phi}_{j_1k}c_{k}^{i_2}}{c_{j_1}^{i_2}}+\frac{t^{\phi}_{j_1k}c_{k}^{i_1}}{c_{j_1}^{i_1}}\right)+\frac{1}{2}\sum_{kl}\left(\frac{V_{i_2lj_2k}c_{l}^{k}}{c_{j_2}^{i_2}}+\frac{V_{i_1lj_2k}c_{l}^{k}}{c_{j_2}^{i_1}}+\frac{V_{i_2lj_1k}c_{l}^{k}}{c_{j_1}^{i_2}}+\frac{V_{i_1lj_1k}c_{l}^{k}}{c_{j_1}^{i_1}}\right)\\
&-\sum_{kl} t^{\phi}_{kl}\left(\frac{c_{lj_2}^{ki_2}}{c_{j_2}^{i_2}}+\frac{c_{lj_2}^{ki_1}}{c_{j_2}^{i_1}}+\frac{c_{lj_1}^{ki_2}}{c_{j_1}^{i_2}}+\frac{c_{lj_1}^{ki_1}}{c_{j_1}^{i_1}}\right)+\frac{1}{2}\sum_{klm}\left(\frac{V_{klmi_2}c_{mj_2}^{kl}}{c_{j_2}^{i_2}}+\frac{V_{klmi_1}c_{mj_2}^{kl}}{c_{j_2}^{i_1}}+\frac{V_{klmi_2}c_{mj_1}^{kl}}{c_{j_1}^{i_2}}+\frac{V_{klmi_1}c_{mj_1}^{kl}}{c_{j_1}^{i_1}}\right)\\
&-\frac{1}{2}\sum_{klm}\left(\frac{V_{klmj_2}c_{kl}^{mi_2}}{c_{j_2}^{i_2}}+\frac{V_{klmj_2}c_{kl}^{mi_1}}{c_{j_2}^{i_1}}+\frac{V_{klmj_1}c_{kl}^{mi_2}}{c_{j_1}^{i_2}}+\frac{V_{klmj_1}c_{kl}^{mi_1}}{c_{j_1}^{i_1}}\right)+\frac{1}{8}\sum_{klmn}V_{klmn}\left(\frac{c_{mnj_2}^{kli_2}}{c_{l_2}^{i_2}}+\frac{c_{mnj_1}^{kli_2}}{c_{j_1}^{i_2}}+\frac{c_{mnj_2}^{kli_1}}{c_{j_2}^{i_1}}+\frac{c_{mnj_1}^{kli_1}}{c_{j_1}^{i_1}}\right)\Bigg]\,.
\end{split}
\end{equation}

\end{widetext}

In equations \eqref{eq:CC_eq_0}, \eqref{eq:CC_eq_1} and \eqref{eq:CC_eq_2}, the upper indices in the coefficients are summed over unoccupied spin-orbitals of the reference SD $\ket{\phi}$ and the lower indices are summed over occupied spin-orbitals.

Since the form \eqref{eq:CC_ansatz} with $T$ given by \eqref{eq:T_SCC} can represent the exact wave function, Eq. \eqref{eq:psi_CC_exact}, the equations above apply to the exact wave function. However, because they depend on the triple- and quadruple-excitation coefficients, while the number of equations is $N^{CI}_{\leq 2}$, they form an underdetermined system of equations if the coefficients have full rank. As mentioned in the previous sections, the equations can be closed by using LRTD to parametrize the coefficients in Eqs. \eqref{eq:CC_eq_0}, \eqref{eq:CC_eq_1} and \eqref{eq:CC_eq_2}, such that the number of parameters grows polynomially with the number of excitations. In particular, we suggest to use the representations described in section \ref{sec:coeffs_tensor_decomp} that follows, which are designed to parametrize all the involved coefficients in a globally compact way. 

While Eqs. \eqref{eq:CC_eq_0} and \eqref{eq:CC_eq_1} are very similar to CC equations, the difference with CC is clear in Eq. \eqref{eq:CC_eq_2} because of the coefficients appearing as denominators. The degree of the system is therefore higher than in standard CC equations. In practice, one could either use Eq. \eqref{eq:CC_eq_2} as given here, or the version without coefficients at denominators, obtained after multiplication by $c_{j_1}^{i_1}c_{j_2}^{i_1}c_{j_1}^{i_2}c_{j_2}^{i_2}$.

Note also that, for a given number of free parameters $N^{CI}_{\leq n-1}< N^{TD}_{\leq n}< N^{CI}_{\leq n}$, not all the equations with projections on $\bra{\phi_{j_1j_2\ldots j_n}^{i_1i_2\ldots i_n}}$ have to be used, but only a number $N_{n}^{CC}\geq N^{TD}_{\leq n}$. However, if $N_{n}^{CC}<N^{CI}_{\leq n}$, the projection states should be chosen carefully to take into account all the non-vanishing terms of $H$.

Finally, because Eq. \eqref{eq:emT_H_eT} still applies when the operator $T$ in the CC ansatz is given by \eqref{eq:T_SCC}, the energy remains linked and is thus size-extensive, as in standard CC. Note however that Eq. \eqref{eq:simil_H_CC} does not apply anymore, which is clear from the presence of coefficients at denominators in Eq. \eqref{eq:CC_eq_2}, which do not appear in $He^T$.



\section{Tensor representations for cluster operator amplitudes or CI coefficients}\label{sec:coeffs_tensor_decomp}

To implement the tensor extensions of CC discussed in section \ref{sec:CC_LRTD}, practical LRTD-based representations are required for either the CI coefficients or the cluster operator amplitudes. In this section, we describe representations designed to be compact specifically in the context of the proposed CC extensions, which will also allow us to estimate their computational complexity. Note however that those tensor-based representations are provided to complete the discussed CC extensions and are not tested. Other types of polynomially bounded parametrization are possible, and the results provided in the previous sections are valid for any such parametrization.

Let us assume that we use CC equations without truncation of $T$ with respect to excitation number, or the generalized CC equations of section \ref{sec:eqns_coeffs}, where the projection SD's have two excitations at most, namely $n=2$ in Eq. \eqref{eq:CC_equations_H_phi}. According to Eq. \eqref{eq:CC_eq_2}, we need low rank tensor decompositions for the $l\leq 4$ excitation coefficients to close the equations. The simplest approach would be to use generic types of LRTD such as singular value decompositions (SVD), polyadic decompositions and higher-order SVD (HOSVD) \cite{Bergqvist_2010}, to decompose the sets of coefficients, or $T$-amplitudes, at each number of excitations independently. However, because we have many tensors to decompose, of orders up to eight, we can quickly loose control over the number of parameters with such an approach. On way to reduce the number of free parameters would be to use a tensor network (TN) instead. In that case however, it is not clear how to express the different sets of coefficients $\{c_{j_1,j_2\ldots j_{l}}^{i_1i_2\ldots i_{l}}\}$ using a single TN, or even how to construct the TN for a non-local Hamiltonian. In the following, we describe one possible solution to those problems, which involves multiple binary tree tensor networks (TTN) constructed from the same smaller TTN's, and then a representation of a given set of coefficients by a superposition of TTN's (STTN). This produces a globally compact structure that relies, on one hand, on the low order of the tensors, as compared to usual TN approaches, and, on the other hand, on some basic assumptions about the dominant correlations in the system, though not on the specific connections between spin-orbitals in the Hamiltonian. More specifically, the STTN primarily take into account all types of pairwise entanglement based on the charge and the spin of the excited quasiparticles involved, and only some types of entanglement between pairs and larger groups of excitations. In addition to the fact that the same tensors are shared by different decompositions, the tensors are of third order or less, so that the total number of free parameters is kept under control.

In the following examples, we will use a variant of the $c^{i_1i_2\ldots}_{j_1j_2\ldots}$ notation of Eq. \eqref{eq:psi_sum}, as if the decomposition were used to represent CI coefficients, but they can also be applied to cluster operator amplitudes. The spin indices will be labeled explicitly because different combinations of spins require different types of decompositions in order to take the Pauli exclusion principle into account.

\begin{figure}
\includegraphics[width=\columnwidth]{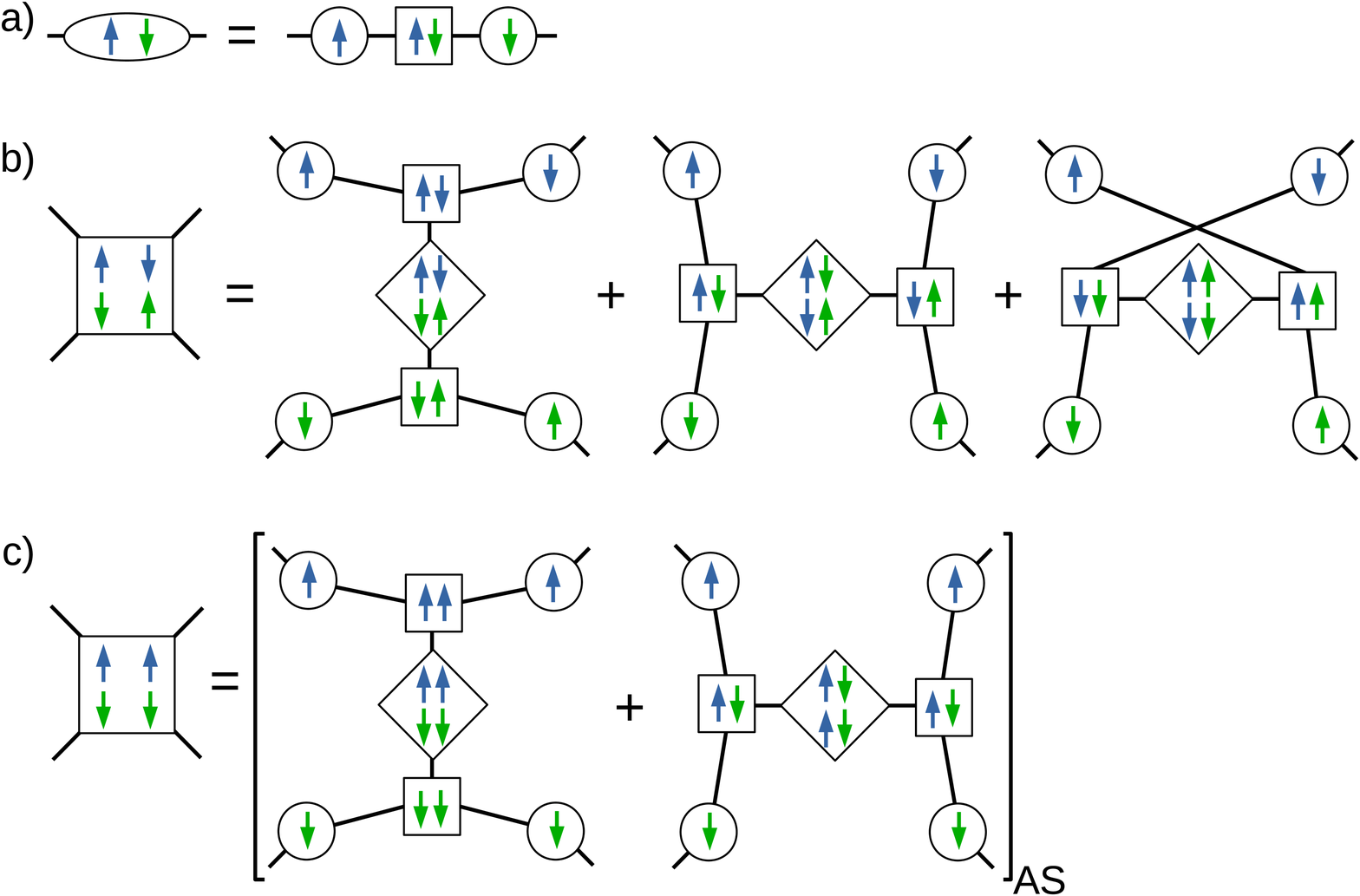}
\caption{\label{fig:decomp_c2} Graphical representation of the STTN for single- and double-excitation coefficients. The blue and green arrows represent particles and holes, respectively, and their spin orientations. Figure (a) is a representation of Eq. \eqref{eq:c1_SVD}, (b) corresponds to Eq. \eqref{eq:c2_ud_sum} and (c), Eq. \eqref{eq:c2_uu_sum}. The circles represent the $u$ and $v$ matrices, the squares with two arrows correspond to the $\kappa$ tensors and the diamonds with four arrows, the $\lambda$ tensors. The brackets with subscript ``AS'' in (c) indicates that the decompositions are antisymmetrized with respect to exchange of identical particle or hole indices.}
\end{figure}

For the single-excitation coefficients, the most compact decomposition is a truncated singular value decomposition (SVD):
\begin{equation}\label{eq:c1_SVD}
c^{i\uparrow}_{j\downarrow}=\sum_{k=1}^{s_{1}} \kappa_{kk1}^{p\bar{h}} u_{i k}^{\uparrow} v_{j k}^{\downarrow}\\
\end{equation}
where the columns of $u^{\uparrow}$ and $v^{\downarrow}$ are orthogonal unit vectors and $s_{1}\leq min(L-N_\sigma,N_\sigma)$. Note that the spins are opposite since the spin carried by a hole is opposite to the spin of the particle in a pair created from a single occupied spin-orbital in $\ket{\phi}$. In the absence of spin-rotational symmetry, $c^{i\downarrow}_{j\uparrow}$ is defined similarly, using the matrices $u^{\downarrow}$ and $v^{\uparrow}$ and the singular values $\kappa_{kk1}^{\bar{p}h}$. Expression \eqref{eq:c1_SVD} is represented graphically in Fig. \ref{fig:decomp_c2}(a). 

In the following, we will label tensors using combination of $p$, $\bar{p}$, $h$ and $\bar{h}$ as superscripts, corresponding respectively to spin up and spin down particle and spin up and spin down hole. The order in which they appear will determine how they are entangled, assuming a binary tree structure, with an additional coma for odd numbers of excitations, as will be seen below.



For two particle-hole excitations with opposite spins, there are three different possible decompositions based on pairwise entanglement: Singlet particle-particle and hole-hole pairing:
\begin{equation}\label{eq:decomp_c2_ud_pp_s}
\begin{split}
\left(c^{i_1\uparrow i_2\downarrow}_{j_1\downarrow j_2\uparrow}\right)_{p\bar{p}\bar{h}h}=&\sum_{k=1}^{s_{p\bar{p}}}\sum_{l=1}^{s_{h\bar{h}}}\sum_{m,n=1}^{s_{p}}\sum_{q,r=1}^{s_{h}} \lambda_{kl1}^{p\bar{p}\bar{h}h} \kappa_{mnk}^{p\bar{p}} \kappa_{qrl}^{\bar{h}h}\\
&\times u^{\uparrow}_{i_1m}u^{\downarrow}_{i_2n} v^{\downarrow}_{j_1q}v^{\uparrow}_{j_2r}\,,
\end{split}
\end{equation}
singlet particle-hole pairing:
\begin{equation}\label{eq:decomp_c2_ud_ph_s}
\begin{split}
\left(c^{i_1\uparrow i_2\downarrow}_{j_1\downarrow j_2\uparrow}\right)_{p\bar{h}\bar{p}h}=&\sum_{k,l=1}^{s_{p\bar{h}}}\sum_{m,q=1}^{s_{p}}\sum_{n,r=1}^{s_{h}} \lambda_{kl1}^{p\bar{h}\bar{p}h} \kappa_{mnk}^{p\bar{h}} \kappa_{qrl}^{\bar{p}h}\\
&\times u^{\uparrow}_{i_1m}v^{\downarrow}_{j_1n}u^{\downarrow}_{i_2q} v^{\uparrow}_{j_2r}\,,
\end{split}
\end{equation}
and triplet particle-hole pairing:
\begin{equation}\label{eq:decomp_c2_ud_ph_t}
\begin{split}
\left(c^{i_1\uparrow i_2\downarrow}_{j_1\downarrow j_2\uparrow}\right)_{ph\bar{p}\bar{h}}=&\sum_{k,l=1}^{s_{ph}}\sum_{m,q=1}^{s_{p}}\sum_{n,r=1}^{s_{h}} \lambda_{kl1}^{ph\bar{p}\bar{h}} \kappa_{mnk}^{ph} \kappa_{qrl}^{\bar{p}\bar{h}}\\
&\times u^{\uparrow}_{i_1m}v^{\uparrow}_{j_2n} u^{\downarrow}_{i_2q}v^{\downarrow}_{j_1r} \,,
\end{split}
\end{equation}
where the first $s_1$ columns of $u^{\sigma}$ and $v^{\sigma}$ and the first matrix slice of $\kappa^{p\bar{h}}$ and $\kappa^{\bar{p}h}$ are the same as in \eqref{eq:c1_SVD}, and the similar decomposition for $c^{i\downarrow}_{j\uparrow}$, where $s_1=min(s_p,s_h)$, and we have assumed $s_{\bar{p}h}=s_{p\bar{h}}$ and $s_{\bar{p}\bar{h}}=s_{ph}$. As shown in Appendix \ref{sec:SVD_TTN}, the decompositions \eqref{eq:decomp_c2_ud_pp_s}, \eqref{eq:decomp_c2_ud_ph_s} and \eqref{eq:decomp_c2_ud_ph_t} are related to combinations of SVD's by internal rotations and all become exact when the tensor dimensions are large. They are therefore not orthogonal in general. However, at small tensor dimensions they occupy different regions of Hilbert space since they are based on different composite excitations, and thus each of those decompositions allows a compact representation for its particular type of entanglement, but not for the two other types. Therefore, to obtain a compact representation allowing all those types of entanglement to coexist, we can combine \eqref{eq:decomp_c2_ud_pp_s}, \eqref{eq:decomp_c2_ud_ph_s} and \eqref{eq:decomp_c2_ud_ph_t} and express the coefficients as
\begin{equation}\label{eq:c2_ud_sum}
c^{i_1\uparrow i_2\downarrow}_{j_1\downarrow j_2\uparrow}=\left(c^{i_1\uparrow i_2\downarrow}_{j_1\downarrow j_2\uparrow}\right)_{p\bar{p}\bar{h}h}+\left(c^{i_1\uparrow i_2\downarrow}_{j_1\downarrow j_2\uparrow}\right)_{p\bar{h}\bar{p}h}+\left(c^{i_1\uparrow i_2\downarrow}_{j_1\downarrow j_2\uparrow}\right)_{ph\bar{p}\bar{h}}\,,
\end{equation}
where the dimensions of the $u$, $v$ and $\kappa$ tensors must be large enough to include most of the entanglement in the set of coefficients $\left\{c^{i_1\uparrow i_2\downarrow}_{j_1\downarrow j_2\uparrow}\right\}$ while the overlap between the terms must remain small to avoid redundancy. The graphical representation of Eq. \eqref{eq:c2_ud_sum} is shown in Fig. \ref{fig:decomp_c2}(b).

Now, let us assume $s_p=s_h=s_1$ so that the number of additional parameters in \eqref{eq:c2_ud_sum}, with respect to the single-excitation representations, depends only on $s_{p\bar{p}}$, $s_{h\bar{h}}$, $s_{p\bar{h}}$ and $s_{ph}$. In addition, let us assume that the total number of matrix slices in the $\kappa$ tensors, $S_{\kappa}=s_{p\bar{p}}+s_{h\bar{h}}+2s_{p\bar{h}}+2s_{ph}$, is constant, and thus the total number of parameters for the $\kappa$ tensors is constant. Then, the total number of parameters in \eqref{eq:c2_ud_sum} is determined only by the total numbers of elements in the $\lambda$ matrices, $N_{\lambda}=s_{p\bar{p}}s_{h\bar{h}}+s_{p\bar{h}}^2+s_{ph}^2$. The minimum of $N_\lambda$ under the constraint that $S_{\kappa}$ is constant is at $s_{p\bar{p}}=s_{h\bar{h}}=s_{p\bar{h}}=s_{ph}=S_{\kappa}/6$. Therefore, for $s_p=s_h=s_1$, if the optimal representation depended only on the value of $S_{\kappa}$, the most compact representation would be the one in which all the terms have the same number of parameters in \eqref{eq:c2_ud_sum}. In practice, the optimal values of $s_p$ and $s_h$ are different and so are the optimal $s_{p\bar{p}}$, $s_{h\bar{h}}$, $s_{p\bar{h}}$ and $s_{ph}$. It is clear however that the inclusion of different terms tends to reduce the number of parameters. In addition, including the decompositions associated with the types of correlations actually present in the set of coefficients also minimizes $S_{\kappa}$ because each type is represented in the most compact way possible. Therefore, combining the relevant decompositions in \eqref{eq:c2_ud_sum} tends to minimize the numbers of parameters in both the $\lambda$ and the $\kappa$ tensors and produces a very compact representation. As will become more clear below, the compactness of the STTN structure also results from the sharing of tensors amongst different decompositions. 

For spin-orbital with same spin, the decompositions must respect the Pauli exclusion principle. The decompositions based on the two possible types of pairing are
\begin{equation}
\begin{split}
\left(c^{i_1\uparrow i_2\uparrow}_{j_1\downarrow j_2\downarrow}\right)_{pp\bar{h}\bar{h}}=&\sum_{k=1}^{s_{pp}}\sum_{l=1}^{s_{hh}}\sum_{n>m=1}^{s_{p}} \sum_{r>q=1}^{s_{h}} \lambda_{kl1}^{pp\bar{h}\bar{h}} \kappa_{mnk}^{pp} \kappa_{qrl}^{\bar{h}\bar{h}}\\
&\times 
\begin{vmatrix}
u^{\uparrow}_{i_1m} & u^{\uparrow}_{i_1n}\\
u^{\uparrow}_{i_2m} & u^{\uparrow}_{i_2n}
\end{vmatrix}
\begin{vmatrix}
v^{\downarrow}_{j_1q}& v^{\downarrow}_{j_1r}\\ 
v^{\downarrow}_{j_2q} & v^{\downarrow}_{j_2r}
\end{vmatrix}\,,
\end{split}
\end{equation}
\begin{equation}\label{eq:decomp_c2_uu_ph}
\begin{split}
\left(c^{i_1\uparrow i_2\uparrow}_{j_1\downarrow j_2\downarrow}\right)_{p\bar{h}p\bar{h}}=&\sum_{k,l=1}^{s_{p\bar{h}}}\sum_{m,q=1}^{s_{p}}\sum_{n,r=1}^{s_{h}} \lambda_{kl1}^{p\bar{h}p\bar{h}} \kappa_{mnk}^{p\bar{h}} \kappa_{qrl}^{p\bar{h}}\\
&\times u^{\uparrow}_{i_1m}u^{\uparrow}_{i_2q}
\begin{vmatrix}
v^{\downarrow}_{j_1n}& v^{\downarrow}_{j_1r}\\ 
v^{\downarrow}_{j_2n} & v^{\downarrow}_{j_2r}
\end{vmatrix}\,,
\end{split}
\end{equation}
where $\lambda_{kl}^{p\bar{h}p\bar{h}}$ is symmetric, and
\begin{equation}\label{eq:c2_uu_sum}
c^{i_1\uparrow i_2\uparrow}_{j_1\downarrow j_2\downarrow}=\left(c^{i_1\uparrow i_2\uparrow}_{j_1\downarrow j_2\downarrow}\right)_{pp\bar{h}\bar{h}}+\left(c^{i_1\uparrow i_2\uparrow}_{j_1\downarrow j_2\downarrow}\right)_{p\bar{h}p\bar{h}}\,.
\end{equation}
In the absence of spin-rotational symmetry, the similar decompositions for $c^{i_1\downarrow i_2\downarrow}_{j_1\uparrow j_2\uparrow}$ involve the different tensors $u^{\downarrow}$, $v^{\uparrow}$, $\kappa^{\bar{p}\bar{p}}$, $\kappa^{hh}$, $\lambda^{\bar{p}\bar{p}hh}$, $\kappa^{\bar{p}h}$ and $\lambda^{\bar{p}h\bar{p}h}$. In \eqref{eq:decomp_c2_uu_ph}, the exchange of $i_1$ and $i_2$ is equivalent to exchanging the columns in the determinant, so that $\left(c^{i_1\uparrow i_2\uparrow}_{j_1\downarrow j_2\downarrow}\right)_{p\bar{h}p\bar{h}}$ is also antisymmetric with respect to those indices. Here, because $\kappa^{p\bar{h}}$ in \eqref{eq:decomp_c2_uu_ph} is the same as in \eqref{eq:decomp_c2_ud_ph_s}, if we assumed that only $s_{pp}$ and $s_{hh}$ can be optimized in \eqref{eq:c2_uu_sum}, the presence of the term $\left(c^{i_1\uparrow i_2\uparrow}_{j_1\downarrow j_2\downarrow}\right)_{p\bar{h}p\bar{h}}$ would allow $s_{pp}$ and $s_{hh}$ to be as small as possible. Expression \eqref{eq:c2_uu_sum} is depicted in Fig. \ref{fig:decomp_c2}(c), where the antisymmetrization is indicated by braquets, instead of displaying explicitly the similar decompositions corresponding to all combinations of permutations of identical particle or hole lines. 

At two excitations, we can easily include all the types of entanglement between the different pairs. In addition, all the $\kappa$ tensors connecting the pairs are also involved in the higher order decompositions described below, and are therefore required. Furthermore, if the LRTD are used to represent sets of cluster operator amplitudes, it is particularly important for the double-excitation ones to be well represented since higher order coefficients depend on them. When the number of particle-hole excitations $l$ increases, since the number of possible binary TTN increases exponentially with $l$, only an exponentially small fraction of them can be included. This is not a problem however when assuming a polynomially bounded parametrization, since it means that there necessarily exists subsets with a polynomially bounded number of decompositions that can produce accurate representations. To ensure that the strongest correlations are taken into account while keeping the number of parameters as small as possible, it is however important to include all the types of pairing in which the two quasiparticles in a pair can occupy the same spatial orbital. Those are the triplet particle-hole pairing and singlet particle-particle and hole-hole pairing. This is another reason to use superpositions of TTN instead of a single TTN representation, which is insufficient to satisfy that condition in the present context. In the following, we will keep including decompositions based on all possible types of pairing in the representations, which satisfy that condition, but also accounts for other, strictly non-local (in the space of the chosen orbital basis), correlations as well.

\begin{figure}
\includegraphics[width=\columnwidth]{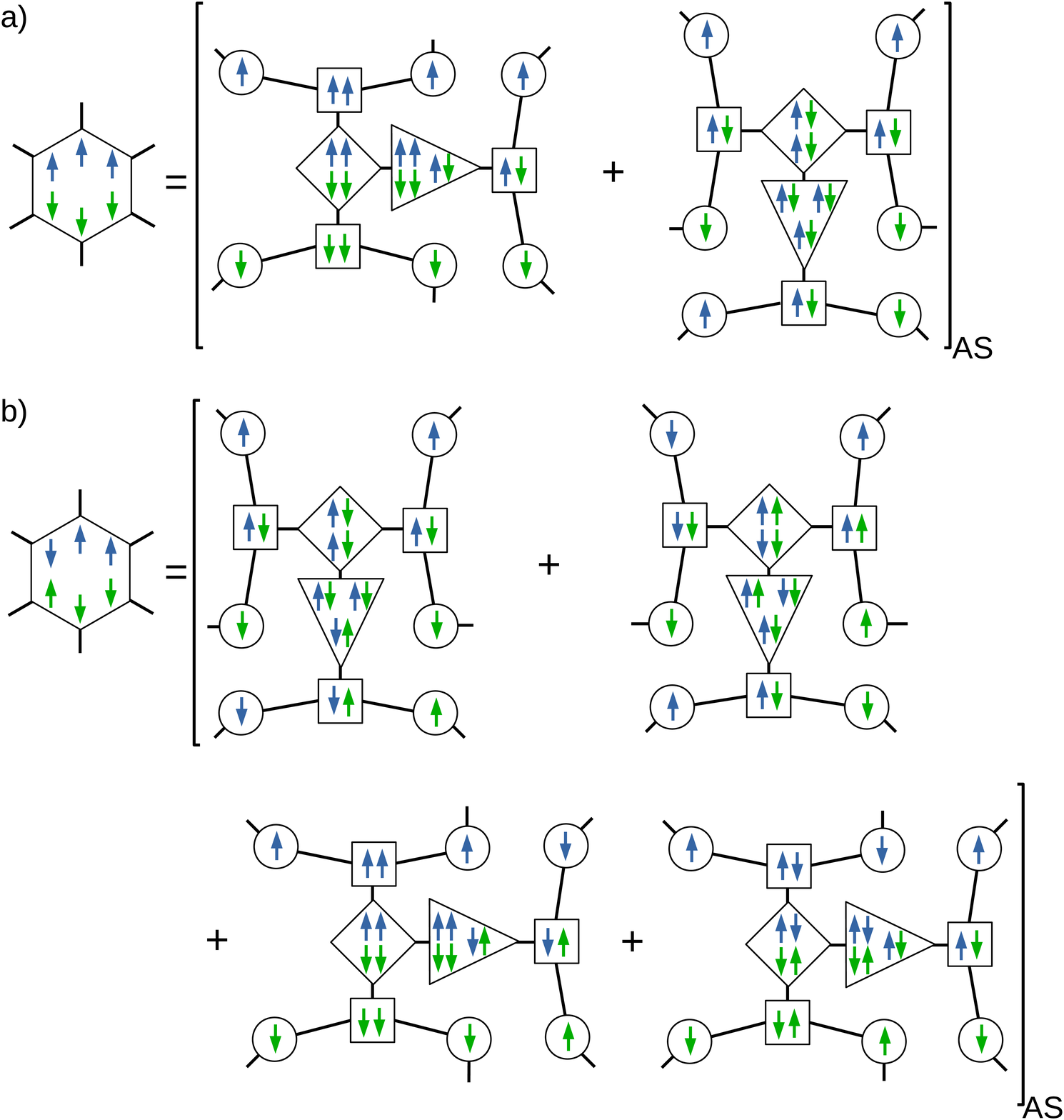}
\caption{\label{fig:decomp_c3} Graphical representation of the STTN for triple-excitation coefficients. Figure (a) is a representation of Eq. \eqref{eq:c3_u_sum} and (b) corresponds to Eq. \eqref{eq:c3_duu_sum}. Here, the triangles represent the $\mu$ matrices. See Fig. \ref{fig:decomp_c2} for other details on the notation. Note that, to avoid line crossings, the relative position of the circles representing the $u$ and $v$ matrices is different for different decompositions, unlike in Fig. \ref{fig:decomp_c2}, which is of no consequence since the antisymmetrization indicated by the brackets implies a summation with permutation of identical quasiparticle indices.}
\end{figure}

At three excitations, with particle-hole entanglement only at the lowest level of the tree, we have
\begin{equation}\label{eq:decomp_c3_uuu_ph}
\begin{split}
&\left(c^{i_1\uparrow i_2\uparrow i_3 \uparrow}_{j_1\downarrow j_2\downarrow j_3 \downarrow}\right)_{p\bar{h}p\bar{h}p\bar{h}}=\sum_{m=1}^{s_{p\bar{h}p\bar{h}}}\sum_{l_1,l_2,l_3=1}^{s_{p\bar{h}}} \sum_{k_1,k_3,k_5=1}^{s_{p}}\sum_{k_2,k_4,k_6=1}^{s_{h}}\\
&\sum_{\pi\in S_3}\left(\mu_{l_1m}^{p\bar{h}p\bar{h}p\bar{h}} \lambda_{l_2l_3m}^{p\bar{h}p\bar{h}}+\mu_{l_2m}^{p\bar{h}p\bar{h}p\bar{h}} \lambda_{l_1l_3m}^{p\bar{h}p\bar{h}}+\mu_{l_3m}^{p\bar{h}p\bar{h}p\bar{h}} \lambda_{l_1l_2m}^{p\bar{h}p\bar{h}}\right)\\
&\qquad\times \kappa_{k_1k_2l_1}^{p\bar{h}}\kappa_{k_3k_4l_2}^{p\bar{h}}\kappa_{k_5k_6l_3}^{p\bar{h}}\epsilon_{\pi_1\pi_2\pi_3}\\
&\qquad\times u_{i_1k_1}^{\uparrow}u_{i_2k_3}^{\uparrow}u_{i_3k_5}^{\uparrow}v_{j_{\pi_1}k_2}^{\downarrow}v_{j_{\pi_2}k_4}^{\downarrow}v_{j_{\pi_3}k_6}^{\downarrow}\,,
\end{split}
\end{equation}
where $\epsilon_{\pi_1\pi_2\pi_3}$ is the Levi-Civita symbol, $S_3$ is the permutation group for the set $\{1,2,3\}$, and the symmetrization of the product $\mu\lambda$ ensures the antisymmetry of the coefficient with respect to exchange of the particle ($i$) indices. 
Then, we can also have
\begin{equation}\label{eq:decomp_c3_uuu_phpphh}
\begin{split}
&\left(c^{i_1\uparrow i_2\uparrow i_3 \uparrow}_{j_1\downarrow j_2\downarrow j_3 \downarrow}\right)_{p\bar{h},pp\bar{h}\bar{h}}=\sum_{m=1}^{s_{pp\bar{h}\bar{h}}}\sum_{l_1=1}^{s_{p\bar{h}}}\sum_{l_2=1}^{s_{pp}}\sum_{l_3=1}^{s_{hh}}\\
&\sum_{k_1,k_3,k_4=1}^{s_{p}} \sum_{k_2,k_5,k_6=1}^{s_{h}} \sum_{\pi,\chi\in S_3}\mu_{l_1m}^{p\bar{h},pp\bar{h}\bar{h}} \lambda_{l_2l_3m}^{pp\bar{h}\bar{h}}\\
&\qquad \times\kappa_{k_1k_2l_1}^{p\bar{h}}\kappa_{k_3k_4l_2}^{pp}\kappa_{k_5k_6l_3}^{\bar{h}\bar{h}} \epsilon_{\pi_1\pi_2\pi_3}\epsilon_{\chi_1\chi_2\chi_3}\\
&\qquad\times u_{i_{\pi_1}k_1}^{\uparrow}u_{i_{\pi_2}k_3}^{\uparrow}u_{i_{\pi_1}k_4}^{\uparrow}v_{j_{\chi_1}k_2}^{\downarrow}v_{j_{\chi_2}k_5}^{\downarrow}v_{j_{\chi_3}k_6}^{\downarrow}\,.
\end{split}
\end{equation}
Here, we could also have entangled first the particle-hole pair with either of the two other pairs. However, to limit the number of terms we choose only one of those three possibilities, namely, the only one that respect the symmetry $(p,\bar{p})\leftrightarrow (\bar{h},h)$. We can then use the representation
\begin{equation}\label{eq:c3_u_sum}
c^{i_1\uparrow i_2\uparrow i_3 \uparrow}_{j_1\downarrow j_2\downarrow j_3 \downarrow}=\left(c^{i_1\uparrow i_2\uparrow i_3 \uparrow}_{j_1\downarrow j_2\downarrow j_3 \downarrow}\right)_{p\bar{h}p\bar{h}p\bar{h}}+\left(c^{i_1\uparrow i_2\uparrow i_3 \uparrow}_{j_1\downarrow j_2\downarrow j_3 \downarrow}\right)_{p\bar{h},pp\bar{h}\bar{h}}\,,
\end{equation}
which is represented graphically in Fig. \ref{fig:decomp_c3}(a).

In the other representations described below, we will also use only decompositions that respect the $(p,\bar{p})\leftrightarrow (\bar{h},h)$ symmetry to simplify the representations. Otherwise, the decompositions that do not respect that symmetry would have to be included in pairs to avoid artificially breaking particle-hole symmetry. On the other hand, including only the decompositions satisfying that symmetry does not constrain the resulting wave function to be particle-hole symmetric, since that would also require that the reference and all the tensors themselves be particle-hole symmetric.


When one spin is different at three excitations, there are four combinations of pairs that respect the $(p,\bar{p})\leftrightarrow (\bar{h},h)$ symmetry, namely, two combinations of particle-hole pairs only and two combinations with particle-hole, particle-particle and hole-hole pairs. First there is a combination of singlet particle-hole pairs:
\begin{equation}
\begin{split}
\left(c^{i_1\downarrow i_2\uparrow i_3 \uparrow}_{j_1\uparrow j_2\downarrow j_3 \downarrow}\right)_{\bar{p}h,p\bar{h}p\bar{h}}=&\sum_{m=1}^{s_{p\bar{h}p\bar{h}}}\sum_{l_1,l_2,l_3=1}^{s_{p\bar{h}}}\sum_{k_1,k_3,k_5=1}^{s_{p}}\sum_{k_2,k_4,k_6=1}^{s_{h}}\\
&\mu_{l_1m}^{\bar{p}h,p\bar{h}p\bar{h}} \lambda_{l_2l_3m}^{p\bar{h}p\bar{h}} \kappa_{k_1k_2l_1}^{\bar{p}h}\kappa_{k_3k_4l_2}^{p\bar{h}}\kappa_{k_5k_6l_3}^{p\bar{h}}\\
&\times u_{i_1k_1}^{\downarrow}v_{j_1k_2}^{\uparrow}
u^{\uparrow}_{i_2k_3}u^{\uparrow}_{i_3k_5}
\begin{vmatrix}
v^{\downarrow}_{j_2k_4}& v^{\downarrow}_{j_2k_6}\\ 
v^{\downarrow}_{j_3k_4} & v^{\downarrow}_{j_3k_6}
\end{vmatrix}\,,
\end{split}
\end{equation}
then a combination of singlet and triplet particle-hole pairs:
\begin{equation}
\begin{split}
&\left(c^{i_1\downarrow i_2\uparrow i_3 \uparrow}_{j_1\uparrow j_2\downarrow j_3 \downarrow}\right)_{p\bar{h},\bar{p}\bar{h}ph}=\sum_{m=1}^{s_{\bar{p}\bar{h}ph}}\sum_{l_1=1}^{s_{p\bar{h}}}\sum_{l_2,l_3=1}^{s_{ph}}\sum_{k_1,k_3,k_5=1}^{s_{p}}\sum_{k_2,k_4,k_6=1}^{s_{h}}\\
&\qquad\mu_{l_1m}^{p\bar{h},\bar{p}\bar{h}ph} \lambda_{l_2l_3m}^{\bar{p}\bar{h}ph} \kappa_{k_1k_2l_1}^{p\bar{h}}\kappa_{k_3k_4l_2}^{\bar{p}\bar{h}}\kappa_{k_5k_6l_3}^{ph}\\
&\qquad\times u_{i_1k_3}^{\downarrow}v_{j_1k_6}^{\uparrow}
\begin{vmatrix}
u^{\uparrow}_{i_2k_5} & u^{\uparrow}_{i_2k_1}\\
u^{\uparrow}_{i_3k_5} & u^{\uparrow}_{i_3k_1}
\end{vmatrix}
\begin{vmatrix}
v^{\downarrow}_{j_2k_4}& v^{\downarrow}_{j_2k_2}\\ 
v^{\downarrow}_{j_3k_4} & v^{\downarrow}_{j_3k_2}
\end{vmatrix}\,,
\end{split}
\end{equation}
a combination with triplet particle-particle and hole-hole pairs:
\begin{equation}
\begin{split}
&\left(c^{i_1\downarrow i_2\uparrow i_3 \uparrow}_{j_1\uparrow j_2\downarrow j_3 \downarrow}\right)_{\bar{p}h,pp\bar{h}\bar{h}}=\sum_{m=1}^{s_{pp\bar{h}\bar{h}}}\sum_{l_1=1}^{s_{p\bar{h}}}\sum_{l_2=1}^{s_{pp}}\sum_{l_3=1}^{s_{hh}} \sum_{k_1,k_3,k_4=1}^{s_{p}}\\
&\quad\sum_{k_2,k_5,k_6=1}^{s_{h}}\mu_{l_1m}^{\bar{p}h,pp\bar{h}\bar{h}} \lambda_{l_2l_3m}^{pp\bar{h}\bar{h}} \kappa_{k_1k_2l_1}^{\bar{p}h}\kappa_{k_3k_4l_2}^{pp}\kappa_{k_5k_6l_3}^{\bar{h}\bar{h}}\\
&\qquad\times u_{i_1k_1}^{\downarrow}v_{j_1k_2}^{\uparrow}
\begin{vmatrix}
u^{\uparrow}_{i_2k_3} & u^{\uparrow}_{i_2k_4}\\
u^{\uparrow}_{i_3k_3} & u^{\uparrow}_{i_3k_4}
\end{vmatrix}
\begin{vmatrix}
v^{\downarrow}_{j_2k_5}& v^{\downarrow}_{j_2k_6}\\ 
v^{\downarrow}_{j_3k_5} & v^{\downarrow}_{j_3k_6}
\end{vmatrix}
\end{split}
\end{equation}
and finally, a combination with singlet particle-particle and hole-hole pairs:
\begin{equation}
\begin{split}
&\left(c^{i_1\downarrow i_2\uparrow i_3 \uparrow}_{j_1\uparrow j_2\downarrow j_3 \downarrow}\right)_{p\bar{h},\bar{p}ph\bar{h}}=\sum_{m=1}^{s_{\bar{p}ph\bar{h}}}\sum_{l_1=1}^{s_{p\bar{h}}}\sum_{l_2=1}^{s_{p\bar{p}}}\sum_{l_3=1}^{s_{h\bar{h}}} \sum_{k_1,k_3,k_4=1}^{s_{p}}\\
&\quad\sum_{k_2,k_5,k_6=1}^{s_{h}}\mu_{l_1m}^{p\bar{h},\bar{p}ph\bar{h}} \lambda_{l_2l_3m}^{\bar{p}ph\bar{h}} \kappa_{k_1k_2l_1}^{p\bar{h}}\kappa_{k_3k_4l_2}^{\bar{p}p}\kappa_{k_5k_6l_3}^{h\bar{h}}\\
&\qquad\times u_{i_1k_3}^{\downarrow}v_{j_1k_5}^{\uparrow}
\begin{vmatrix}
u^{\uparrow}_{i_2k_4} & u^{\uparrow}_{i_2k_1}\\
u^{\uparrow}_{i_3k_4} & u^{\uparrow}_{i_3k_1}
\end{vmatrix}
\begin{vmatrix}
v^{\downarrow}_{j_2k_6}& v^{\downarrow}_{j_2k_2}\\ 
v^{\downarrow}_{j_3k_6} & v^{\downarrow}_{j_3k_2}
\end{vmatrix}\,,
\end{split}
\end{equation}
where $\kappa_{k_3k_4l_2}^{\bar{p}p}=\kappa_{k_4k_3l_2}^{p\bar{p}}$ and $\kappa_{k_5k_6l_3}^{h\bar{h}}=\kappa_{k_6k_5l_3}^{\bar{h}h}$.
We then use
\begin{equation}\label{eq:c3_duu_sum}
\begin{split}
c^{i_1\downarrow i_2\uparrow i_3 \uparrow}_{j_1\uparrow j_2\downarrow j_3 \downarrow}=&\left(c^{i_1\downarrow i_2\uparrow i_3 \uparrow}_{j_1\uparrow j_2\downarrow j_3 \downarrow}\right)_{\bar{p}h,p\bar{h}p\bar{h}}+\left(c^{i_1\downarrow i_2\uparrow i_3 \uparrow}_{j_1\uparrow j_2\downarrow j_3 \downarrow}\right)_{p\bar{h},\bar{p}\bar{h}ph}\\
&+\left(c^{i_1\downarrow i_2\uparrow i_3 \uparrow}_{j_1\uparrow j_2\downarrow j_3 \downarrow}\right)_{\bar{p}h,pp\bar{h}\bar{h}}+\left(c^{i_1\downarrow i_2\uparrow i_3 \uparrow}_{j_1\uparrow j_2\downarrow j_3 \downarrow}\right)_{p\bar{h},\bar{p}ph\bar{h}}\,,
\end{split}
\end{equation}
which is depicted in Fig. \ref{fig:decomp_c3}(b).

By comparing Fig. \ref{fig:decomp_c3} with Fig. \ref{fig:decomp_c2}, it is clear how the decompositions for the triple-excitation coefficients are constructed from the decompositions defining single- and double-excitation coefficients, using the other matrix slices of the $\lambda$ tensors. Note that other $\lambda$ tensors, not present in the representations above, are also used below in the quadruple-excitation coefficients.

For quadruple-excitation coefficients, we will only provide here the short STTN expressions and the graphical representation for the coefficients. The algebraic expressions are provided in Appendix \ref{sec:STTN_c4}. As for triple-excitation coefficients, we will include decompositions with all types of pairings in the STTN representations and use only decompositions respecting the $(p,\bar{p})\leftrightarrow (\bar{h},h)$ symmetry.

\begin{figure}
\includegraphics[width=\columnwidth]{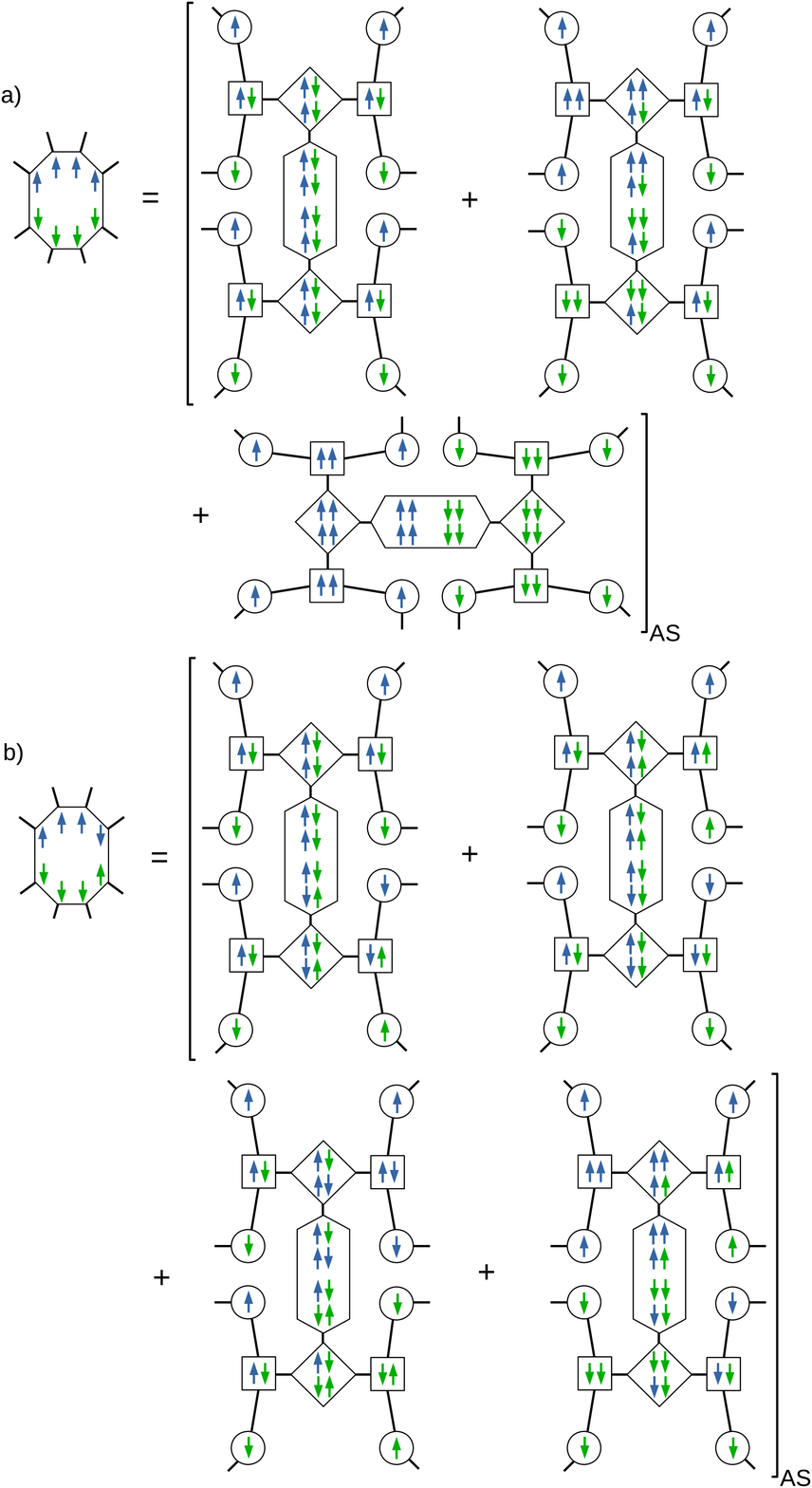}
\caption{\label{fig:decomp_c4_u} Graphical representation of the quadruple-excitation coefficients \eqref{eq:c4_u_sum} in (a) and \eqref{eq:c4_uuud_sum} in (b). See Fig. \ref{fig:decomp_c2} for details on the notation.}
\end{figure}

\begin{figure}
\includegraphics[width=\columnwidth]{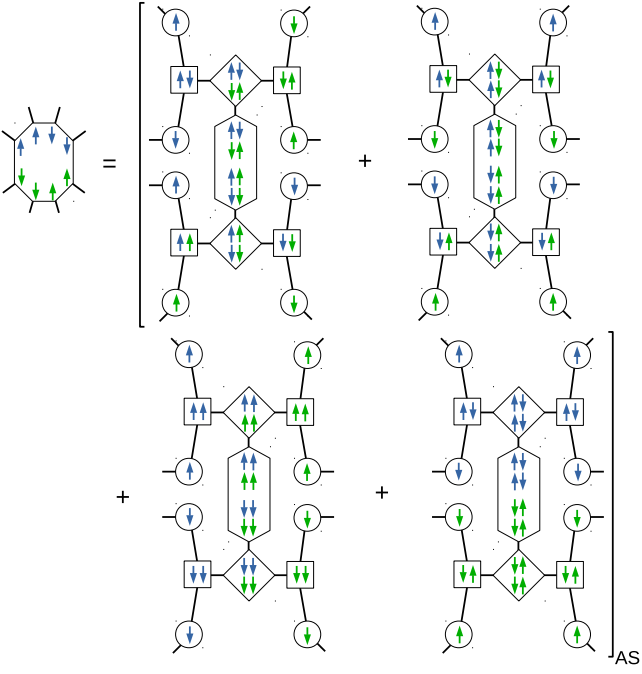}
\caption{\label{fig:decomp_c4_ud} Graphical representation of \eqref{eq:c4_uudd_sum}. See Fig. \ref{fig:decomp_c2} for details on the notation.}
\end{figure}

First when all the spins are equal, we can use
\begin{equation}\label{eq:c4_u_sum}
\begin{split}
&c^{i_1\uparrow i_2\uparrow i_3\uparrow i_4\uparrow}_{j_1\downarrow j_2\downarrow j_3\downarrow j_4\downarrow}=\left(c^{i_1\uparrow i_2\uparrow i_3\uparrow i_4\uparrow}_{j_1\downarrow j_2\downarrow j_3\downarrow j_4\downarrow}\right)_{p\bar{h}p\bar{h}p\bar{h}p\bar{h}}\\
&\qquad+\left(c^{i_1\uparrow i_2\uparrow i_3\uparrow i_4\uparrow}_{j_1\downarrow j_2\downarrow j_3\downarrow j_4\downarrow}\right)_{ppp\bar{h}\bar{h}\bar{h}p\bar{h}}+\left(c^{i_1\uparrow i_2\uparrow i_3\uparrow i_4\uparrow}_{j_1\downarrow j_2\downarrow j_3\downarrow j_4\downarrow}\right)_{pppp\bar{h}\bar{h}\bar{h}\bar{h}}\,,
\end{split}
\end{equation}
depicted in Fig. \ref{fig:decomp_c4_u}(a).
Then, when one spin is different, we use
\begin{equation}\label{eq:c4_uuud_sum}
\begin{split}
&c^{i_1\uparrow i_2\uparrow i_3\uparrow i_4\downarrow}_{j_1\downarrow j_2\downarrow j_3\downarrow j_4\uparrow}=\left(c^{i_1\uparrow i_2\uparrow i_3\uparrow i_4\downarrow}_{j_1\downarrow j_2\downarrow j_3\downarrow j_4\uparrow}\right)_{p\bar{h}p\bar{h}p\bar{h}\bar{p}h}\\
&\quad+\left(c^{i_1\uparrow i_2\uparrow i_3\uparrow i_4\downarrow}_{j_1\downarrow j_2\downarrow j_3\downarrow j_4\uparrow}\right)_{p\bar{h}php\bar{h}\bar{p}\bar{h}}+\left(c^{i_1\uparrow i_2\uparrow i_3\uparrow i_4\downarrow}_{j_1\downarrow j_2\downarrow j_3\downarrow j_4\uparrow}\right)_{p\bar{h}p\bar{p}p\bar{h}\bar{h}h}\\
&\quad+\left(c^{i_1\uparrow i_2\uparrow i_3\uparrow i_4\downarrow}_{j_1\downarrow j_2\downarrow j_3\downarrow j_4\uparrow}\right)_{ppph\bar{p}\bar{h}\bar{h}\bar{h}}\,,
\end{split}
\end{equation}
shown in Fig. \ref{fig:decomp_c4_u}(b). 
Finally, for two up spins and two down spins, we can use
\begin{equation}\label{eq:c4_uudd_sum}
\begin{split}
&c^{i_1\uparrow i_2\uparrow i_3\downarrow i_4\downarrow}_{j_1\downarrow j_2\downarrow j_3\uparrow j_4\uparrow}=\left(c^{i_1\uparrow i_2\uparrow i_3\downarrow i_4\downarrow}_{j_1\downarrow j_2\downarrow j_3\uparrow j_4\uparrow}\right)_{p\bar{p}\bar{h}hph\bar{p}\bar{h}}\\
&\quad+\left(c^{i_1\uparrow i_2\uparrow i_3\downarrow i_4\downarrow}_{j_1\downarrow j_2\downarrow j_3\uparrow j_4\uparrow}\right)_{p\bar{h}p\bar{h}\bar{p}h\bar{p}h}+\left(c^{i_1\uparrow i_2\uparrow i_3\downarrow i_4\downarrow}_{j_1\downarrow j_2\downarrow j_3\uparrow j_4\uparrow}\right)_{pphh\bar{p}\bar{p}\bar{h}\bar{h}}\\
&\qquad+\left(c^{i_1\uparrow i_2\uparrow i_3\downarrow i_4\downarrow}_{j_1\downarrow j_2\downarrow j_3\uparrow j_4\uparrow}\right)_{p\bar{p}p\bar{p}\bar{h}h\bar{h}h}\,,
\end{split}
\end{equation}
which is depicted in Fig. \ref{fig:decomp_c4_ud}.

Note that, in expressions \eqref{eq:c4_u_sum}, \eqref{eq:c4_uuud_sum} and \eqref{eq:c4_uudd_sum}, depicted in Figs. \ref{fig:decomp_c4_u} and \ref{fig:decomp_c4_ud}, the combinations of particles and holes in the two mains branches of the trees are different between different decompositions. In other words, in the representation of a given set of coefficients, say in \eqref{eq:c4_uuud_sum} depicted in Fig. \ref{fig:decomp_c4_u}(b), one cannot obtain a decomposition from another by applying permutations of quasiparticle matrices (circles) within the main branches. One instead has to apply permutations between the two main branches. The purpose of this choice is to reduce the possible overlap between the decompositions.

The above representations of coefficients are provided as examples to illustrate the STTN structure and as suggestions of representation for the sets of coefficients, or $T$-amplitudes, involved in the (generalized) CC equations. However, terms representing other types of entanglement can easily be added, or some included terms removed. There is in principle an optimal combination of TTN's that minimizes the total number of free parameters required to obtain an accurate representation of all the sets of coefficients. Given that the total number of free parameters is also controlled by the adjustable tensor dimensions, there is plenty of degrees of freedom for optimization, and any scaling with the number of particle-hole excitations can be obtained below the scaling of the number of SD's. Since the number of different sets of coefficients grows linearly with the number of excitations, if the number of terms for each set remains constant or grows only slowly, the total number of different decompositions and tensors remain bounded by low order polynomials and are thus computationally tractable, while the total number of free parameters should be reduced if the included terms are relevant.

Since the tensors always have even order, using decompositions based on pairwise entanglement is quite natural. However, from the point of view of collective excitations, this type of decomposition is based only on bosonic ones, while there can also be fermionic collective excitations in the system. For instance, the quasiparticles in a Fermi liquid are in fact collective fermionic excitations and can be approximated as single particle excitations only in an effective low-energy model. Although the representations used can also account for such collective excitations since they become exact at large tensor dimensions, to allow for a more compact representation of this kind of excitation, one could also add decompositions in which the $u$ and $v$ matrices are entangled with pairs already included. Then, for instance, those three-particles groups can be entangled with $u$ or $v$ in the double-excitation case, or together in the triple-excitation case, and so on.

The large number of adjustable tensor dimensions could be seen as a disadvantage of the STTN structure, as they requires additional optimization algorithms. However, it also offers the possibility to explore different combinations of decompositions in the STTN representation of a given set of coefficients and, assuming that the accuracy of the results can be assessed, the types of decomposition yielding the best results and the relative norms of the different terms can provide useful information about the correlations in the system.

Let us end this section by discussing the computational complexity of the proposed CC extensions, as we must also ensure that they are worth the efforts required for their implementation. The complexity in the evaluation of the generalized CC equations of section \ref{sec:eqns_coeffs} depends on the term
\begin{equation}
\sum_{klmn}V_{klmn} c_{mnj_1j_2}^{kli_1i_2}
\end{equation}
of equation \eqref{eq:CC_eq_2}. To simplify the complexity analysis if we use the quadruple-excitation decompositions provided in Appendix \ref{sec:STTN_c4} in that term, let us set $s_p=s_h=s_1$, all the third dimensions of the $\kappa$ tensors to $s_2$ and all the third dimensions of the $\lambda$ tensors to $s_4$.  Then, starting with the sum over the indices of $V_{klmn}$, it requires $O\left(N^4s_1\right)$ operations, then the sum over the first two indices of the $\kappa$ tensors has $O\left(N^4s_1s_2^4\right)$ complexity, the sums over the first two indices of the $\lambda$ tensors have $O\left(N^4s_2^4s_4\right)$ or $O\left(N^4s_2^2s_4^2\right)$ complexity, and finally, the sum over the $\mu$ tensor indices have $O\left(N^4s_4^2\right)$ complexity. The overall complexity is therefore either $O\left(N^4s_1s_2^4\right)$, $O\left(N^4s_2^4s_4\right)$, or $O\left(N^4s_2^2s_4^2\right)$, depending on the scaling of $s_1$, $s_2$ and $s_4$ with $N$. The more detailed complexity analysis is provided in Appendix \ref{sec:comp_complexity}. For the corresponding CC equations with no truncation of $T$ with respect to the number of excitations, the limiting term has the same form, with the wave function coefficients replaced by the $T$-amplitudes, and thus the complexity is the same. The two proposed approaches are therefore tractable if the tensor dimensions are small and only weakly dependant on $N$, and their scaling can be comparable to CCSD ($O(N^6)$) \cite{Bartlett_Musial_2007}, or better, in a certain range of tensor dimensions.

\section{Discussion}\label{sec:discussion}


We have seen that only low-order coefficients are relevant to the ground state energy when there exist a converging low entanglement single-reference expansion of the wave function. A remarkable aspect of that result is that it applies even if the convergence is slow, and thus the irrelevant coefficients are not vanishingly small, but only smaller than the relevant ones. This is unusual since, on one hand, most approximations based on an expansion are valid only when that expansion converges rapidly and, on the other hand, the energy of an eigenstate is usually assumed to depend on all the coefficients larger than some small threshold magnitude. In the present case, those common assumptions do not apply because of both the low entanglement property of the wave function and the locality of $H$ on the number of excitations axis. Computationally, the fact that the number of relevant parameters is much smaller than the total number of wave function parameters is an interesting advantage of using equations of the CC form, compared to a variational method that requires computing all the parameters. In particular, because only low-order tensors are involved, and the maximum order does not depend on system size, one can also afford tensor product representations which are not based on the specific connections between spin-orbitals in the Hamiltonian, which is very convenient for non-local Hamiltonians. Consequently, if one suspects that there exist a single-reference expansion of the wave function that converges, even if the convergence is slow, namely at strong coupling, not only the proposed CC extensions are theoretically applicable, but they also possess important advantages compared to other tensor network methods.

If it turns out that the convergence of the CI form of the wave function is fast, then the increasing rate of the number of free parameters $N^{TD}_l$ with the number of excitations $l$ is necessarily slow since more components of the wave function can be neglected as $l$ increases. The low entanglement assumption is thus always valid at weak coupling. On the other hand, a slow increasing rate of $N^{TD}_l$ is also possible if the convergence is slow. For instance, this is the case of a strongly correlated system where many SD's are nearly degenerate in energy, hence the slow convergence of the CI coefficients with $l$, while the correlations are only local, hence the low entanglement and slow scaling of $N^{TD}_l$ with $l$. In fact, when the Hamiltonian is local, the low entanglement assumption is essentially always valid. It has indeed been proven recently that physically realizable ground states of such systems can only occupy an exponentially small volume of Hilbert space \cite{Poulin_Qarry_2011}, which implies that such states have low entanglement. Another exact result for local Hamiltonians is that correlations are short range, and thus entanglement is low, if the ground state is gapped \cite{Hastings_2004}. From first principles, the Hamiltonian is not local because of the long-range Coulomb interaction. However, in metallic systems, screening effects yield effective local Hamiltonians at low energy, i.e., in the active space. Therefore, even though the original Hamiltonian is not local, based on the result of Ref. \onlinecite{Poulin_Qarry_2011}, it remains quite reasonable to use a low-entanglement approximation for the ground state of strongly correlated itinerant systems. However, one difficulty for such systems is to determine this effective low-energy Hamiltonian, hence the usefulness of approaches that do not depend on the locality of the Hamiltonian. More generally, in large systems, another argument for assuming a low-entanglement ground state is that the energy range spanned by all the eigenstates grows linearly with the size of the system, while the total number of eigenstates grows exponentially. Many eigenstates therefore become nearly degenerate and can be replaced in practice by a single effective average eigenstate. Since averaging reduces correlations, this effective eigenstate has only low-entanglement. In systems in which this near degeneracy is present at the ground state energy, the low entanglement assumption is therefore valid. On the other hand, in systems with a gapped ground state that breaks the symmetry of the Hamiltonian, whether local or not, the correlation length associated with the order parameter is finite, and thus entanglement is finite as well. In summary, many different types of system have a low-entanglement ground state and can be modelled using tensor networks.


The first tensor extension of the CC method discussed in section \ref{sec:CC_no_trunc}, TCC, in which the LRTD are used to represents sets of amplitudes in the cluster operator $T$, is necessarily valid in the weak coupling limit since it includes the CC method as a special case. However, because $T$ is not truncated with respect to particle-hole excitation, and there are irreducible low-rank corrections to the reducible parts of the wave function coefficients at all orders, TCC remains theoretically valid as long as those corrections can be well represented using LRTD, as the coupling strength increases. As discussed in section \ref{sec:coeffs_tensor_decomp}, not only the local correlations can be accounted for by the STTN representation, but also non-local ones. On the other hand, in the CC method, including its tensor implementations, the wave function coefficients at higher order than the truncation order are always completely expressed as a sum of decoupled terms, a form badly suited for strong correlations, while at the same time the importance of higher order terms increases with the coupling strength. The catastrophic failure of CC at strong coupling \cite{Bulik_2015} is thus inevitable unless the nature of approximation is modified. The TCC approach is a way to do so using LRTD, without projection or correlation operator. Between the two proposed approaches, TCC is the closest one to the CC approach, which should make it the easiest to implement.

The second proposed tensor extension of the CC method,  TCICC, is in principle valid in any situation where the wave function has a converging low-entanglement CI expansion. That includes both the weak and strong coupling regimes. In practice however, at weak coupling, the parametrization would have to essentially reproduce the standard CC approximation, requiring a rather complex parametrization from the point of view of wave function coefficients instead of cluster operator amplitudes, and thus TCC is better suited in that regime. On the other hand, as discussed in section \ref{sec:CI_from_CC}, at some large coupling strength, TCICC could become more compact than TCC, and thus better suited to even stronger couplings. There are also well-known strongly correlated wave functions that have a tensor network representation of the coefficients and are good variational ground states of strongly correlated systems \cite{Changlani_2009, Poilblanc_2014, Zhao_2017}, suggesting that the direct parametrization of wave function coefficients in TCICC is well suited to that regime. Finally, although TCICC can also be seen as a low-entanglement version of the CI method, since the equations \eqref{eq:CC_eq_0}, \eqref{eq:CC_eq_1} and \eqref{eq:CC_eq_2} involve the CI coefficients, it is in fact very different from CI: First, Hilbert space is not truncated in the number of particle-hole excitations, then, the coefficients are not computed variationally or by matrix diagonalization, but instead the tensors defining the coefficients are obtained by solving non-linear equations, and finally, the result is size-extensive because the energy is linked. 

The implementation algorithms and numerical testing of the proposed tensor extensions of the CC method require much more work and are thus not included here. However, at the end of section \ref{sec:coeffs_tensor_decomp}, the obtained scaling of the calculation with system size and tensor dimensions indicate that those CC extensions are applicable in practice. In addition, their complexity could possibly be further improved using a tensor-product representation for the two-particle Coulomb integrals as well \cite{Benedikt_2011, Hohenstein_2012, Benedikt_2013, Hohenstein_2013, Hohenstein_2013a, Parrish_2014, Schutski_2017, Tichai_2019, Parrish_2019}.

\section{Conclusion}\label{sec:conclusion}

We have identified a class of fermionic wave functions which ground state energy can be computed in polynomial time using generalized CC equations. It corresponds to the subclass of the wave functions possessing a converging CI expansion that also possess a low-entanglement representation in which the number of free parameters $N_{\leq k}$ defining the wave function coefficients with $l\leq k$ particle-hole excitations is bounded polynomially in $k$. The CC approximation is the simplest approximation of the class, for which $N_{\leq k}$ is bounded by a constant. The convergence condition only ensures that the energy obtained from the equations is a good approximation to the true ground state energy. There is therefore no lower bound on the convergence rate, which implies that the class contains wave functions of strongly correlated systems which cannot be treated with the standard CC approximations. Based on that result, we have proposed extensions of the CC method to treat such systems using two types of polynomially bounded parametrization different from standard CC, and based on low-rank tensor decompositions (LRTD): a straightforward extension in which the LRTD are used to represent sets of cluster operator ($T$) amplitudes, which involves tensor-adapted standard CC equations, and an extension of CC in which the LRTD are used to represent the CI wave function coefficients directly. For the latter case, we have derived exact generalized CC equations involving the CI coefficients with up to four particle-hole excitations. Finally, although the discussed CC extensions are in principle applicable with any type of polynomially bounded parametrization of the CI coefficients or $T$-amplitudes, to complete the proposals, we have constructed representations of the CI coefficients or cluster operator amplitudes in the form of superpositions of tree tensor networks (STTN), which by design can parametrize all the involved sets of coefficients or $T$-amplitudes in a globally compact way. If the tensor dimensions in the STTN are small and only weakly dependant on system size, the proposed CC extensions are computationally tractable.

\section{Acknowledgements}

Thanks to Andr\'e-Marie Tremblay for useful discussions, comments on the manuscript, and moral support. Thanks to Thomas Baker for comments on the manuscript. 

\appendix

\section{Equal-time correlation functions}\label{sec:eq_time_corr}

In this appendix, we describe how to compute equal-time correlation functions using the source-field method. The derivation also corresponds to a proof of the Hellmann-Feynman theorem. 

If the equations on the second line of Eqs. \eqref{eq:CC_equations}, are satisfied, we have
\begin{equation}
\begin{split}
\frac{\bra{\phi}e^{T^\dagger}\hat{H}^{\phi}e^T\ket{\phi}}{\bra{\phi}e^{T^\dagger}e^T\ket{\phi}}&=\frac{\bra{\phi}e^{T^\dagger}e^Te^{-T}\hat{H}^{\phi}e^T\ket{\phi}}{\bra{\phi}e^{T^\dagger}e^T\ket{\phi}}\\
&=\sum_i\frac{\bra{\phi}e^{T^\dagger}e^T\ket{\phi_i}\bra{\phi_i}e^{-T}\hat{H}^{\phi}e^T\ket{\phi}}{\bra{\phi}e^{T^\dagger}e^T\ket{\phi}}\\
&=\frac{\bra{\phi}e^{T^\dagger}e^T\ket{\phi}\bra{\phi}e^{-T}\hat{H}^{\phi}e^T\ket{\phi}}{\bra{\phi}e^{T^\dagger}e^T\ket{\phi}}\\
&=\bra{\phi}e^{-T}\hat{H}^{\phi}e^T\ket{\phi}\\
&= \Delta E\,.
\end{split}
\end{equation}

Now, if we have a perturbed Hamiltonian
\begin{equation}
H_f=\hat{H}^{\phi}+f\hat{O}
\end{equation}
where $\hat{O}$ is any time-independent operator, and
\begin{equation}
\begin{split}
\Delta E_f &= \frac{\bra{\phi}e^{T_f^\dagger}H_fe^{T_f}\ket{\phi}}{\bra{\phi}e^{T_f^\dagger}e^{T_f}\ket{\phi}}\\
&=\frac{\bra{\psi_f}H_f\ket{\psi_f}}{\bk{\psi_f}{\psi_f}}\\
&=\bra{\phi}e^{-T_f}H_fe^{T_f}\ket{\phi}\,
\end{split}
\end{equation}
then,
\begin{equation}
\begin{split}
&\frac{\partial \Delta E_f}{\partial f}\\
&=\frac{\bra{\psi_f}H_f\ket{\psi_f}}{\bk{\psi_f}{\psi_f}^2}\frac{\partial \bk{\psi_f}{\psi_f}}{\partial f}+\frac{1}{\bk{\psi_f}{\psi_f}}\frac{\partial \bra{\psi_f}H_f\ket{\psi_f}}{\partial f}\\
&=\frac{\Delta E_f}{\bk{\psi_f}{\psi_f}}\frac{\partial \bk{\psi_f}{\psi_f}}{\partial f}+\frac{1}{\bk{\psi_f}{\psi_f}}\left(\frac{\partial }{\partial f}\bra{\psi_f}\right)H_f\ket{\psi_f}\\
&\quad +\frac{1}{\bk{\psi_f}{\psi_f}}\bra{\psi_f}H_f\frac{\partial }{\partial f}\ket{\psi_f} +\frac{1}{\bk{\psi_f}{\psi_f}} \bra{\psi_f}\left(\frac{\partial H_f}{\partial f}\right)\ket{\psi_f}\\
&=\frac{\Delta E_f}{\bk{\psi_f}{\psi_f}}\frac{\partial \bk{\psi_f}{\psi_f}}{\partial f}+\frac{\Delta E_f}{\bk{\psi_f}{\psi_f}}\frac{\partial \bk{\psi_f}{\psi_f}}{\partial f}\\
&\quad +\frac{1}{\bk{\psi_f}{\psi_f}} \bra{\psi_f}\left(\frac{\partial H_f}{\partial f}\right)\ket{\psi_f}\\
&=\frac{2\Delta E_f}{\bk{\psi_f}{\psi_f}}\frac{\partial \bk{\psi_f}{\psi_f}}{\partial f} +\frac{1}{\bk{\psi_f}{\psi_f}} \bra{\psi_f}\left(\frac{\partial H_f}{\partial f}\right)\ket{\psi_f}\,.
\end{split}
\end{equation}
If $f$ is small enough, according to perturbation theory,
\begin{equation}
\ket{\psi_f}\approx \ket{\psi}+f\ket{\mu_1}
\end{equation}
where $\ket{\mu_1}$ is orthogonal to $\ket{\psi}$, and
\begin{equation}
\bk{\psi_f}{\psi_f}=\bk{\psi}{\psi}+f^2\bk{\mu_1}{\mu_1}\,.
\end{equation}
Thus,
\begin{equation}
\frac{\partial \bk{\psi_f}{\psi_f}}{\partial f}=2f\bk{\mu_1}{\mu_1}\,,
\end{equation}
\begin{equation}
\frac{\partial \Delta E_f}{\partial f}=f\frac{4\Delta E_f}{\bk{\psi_f}{\psi_f}}\bk{\mu_1}{\mu_1} +\frac{1}{\bk{\psi_f}{\psi_f}} \bra{\psi_f}\left(\frac{\partial H_f}{\partial f}\right)\ket{\psi_f}\,,
\end{equation}
\begin{equation}
\lim_{f\rightarrow 0}\frac{\partial \Delta E_f}{\partial f}=\lim_{f\rightarrow 0} \frac{1}{\bk{\psi_f}{\psi_f}} \bra{\psi_f}\left(\frac{\partial H_f}{\partial f}\right)\ket{\psi_f}\\
\end{equation}
and finally,
\begin{equation}
\frac{\bra{\psi}\hat{O}\ket{\psi}}{\bk{\psi}{\psi}}=\lim_{f\rightarrow 0}\frac{\partial }{\partial f}\bra{\phi}e^{-T_f}H_fe^{T_f}\ket{\phi}\,.
\end{equation}
We can therefore compute the expectation value of any time-independent operator $\hat{O}$ by computing $\Delta E_f$ for $f=0$ and for $f$ small and obtain
\begin{equation}
\frac{\bra{\psi}\hat{O}\ket{\psi}}{\bk{\psi}{\psi}}\approx \frac{\Delta E_f-\Delta E_0}{f}\,.
\end{equation}

In particular, if
\begin{equation}
\begin{split}
\hat{O}&=a_i^\dagger a_j\\
&=(p_i^\dagger+h_i)(p_j+h_j^\dagger)\\
&=p_i^\dagger p_j + p_i^\dagger h_j^\dagger + h_i p_j +h_i h_j^\dagger\,.
\end{split}
\end{equation}
we can obtain the one particle density-matrix elements
\begin{equation}
\frac{\bra{\psi}a_i^\dagger a_j\ket{\psi}}{\bk{\psi}{\psi}}\,.
\end{equation}
By diagonalizing the one-particle density matrix, we obtain the \textit{natural spin-orbitals} basis, which is the basis that yields the fastest convergence of a configuration interaction series. Although each density matrix elements requires a different calculation, and the number of non-zero elements $2L^2$ can be large, each calculation should converge rapidly if the parameters for the unperturbed Hamiltonian are used as the initial ones in each calculation since the perturbation is very small.

%

\section{Derivation of the generalized CC equations for the CI coefficients}\label{sec:deriv_SCC_eqns}

Here we provide the derivation of the equations \eqref{eq:psi_CC_exact} for $T$ given by Eq. \eqref{eq:T_SCC}, with projection of $e^{-T}H^{\phi}e^T\ket{\phi}$ on $\bra{\phi}$ and $\bra{\phi_j^i}$, using the excited particle and hole operators, Eqs. \eqref{eq:def_p_i_h_i}, and the representation \eqref{eq:H_phi_r_fin} of the Hamiltonian. The derivation of the equations with projection on $\bra{\phi^{i_1i_2}_{j_1j_2}}$ are provided as \hyperlink{suppl_mat}{supplemental material}.

Taking into account the fact that $T$ is an excitation operator, and thus $\bra{\phi}e^{-T}=\bra{\phi}$, that $H$ can destroy at most two particle-hole pairs, and that $\bra{\phi}\hat{H}^{\phi}\ket{\phi}=0$, the equation for the energy is
\begin{equation}\label{eq:eqn_Delta_E_ap}
\begin{split}
\Delta E&=\bra{\phi}e^{-T}\hat{H}^{\phi}e^T\ket{\phi}\\
&=\bra{\phi}\hat{H}^{\phi}\left(1+T+\frac{1}{2}T^2\right)\ket{\phi}\\
\Delta E&=\bra{\phi}\hat{H}^{\phi}T\ket{\phi}+\frac{1}{2}\bra{\phi}\hat{H}^{\phi}T^2\ket{\phi}\,.
\end{split}
\end{equation}
For $\bra{\phi}\hat{H}^{\phi}T\ket{\phi}$, the only term in $H^{\phi}$, Eq. \eqref{eq:H_phi_r_fin}, that contributes is the term annihilating a single particle-hole pair:
\begin{equation}\label{eq:phi_HT_phi}
\begin{split}
\bra{\phi}\hat{H}^{\phi}T\ket{\phi}&=\bra{\phi}\left(\sum_{ij} t^{\phi}_{ij}h_{j}p_{i}\right)\left(\sum_{kl} c^k_l p_k^\dagger h_l^\dagger \right)\ket{\phi}\\
&=\sum_{ijkl} t^{\phi}_{ij}c^k_l \delta_{ik}\delta_{jl}\\
\bra{\phi}\hat{H}^{\phi}T\ket{\phi}&=\sum_{ij} t^{\phi}_{ij}c^i_j\,.
\end{split}
\end{equation}
For $\bra{\phi}\hat{H}^{\phi}T^2\ket{\phi}$, only the part of $H^\phi$ that annihilates two pairs contributes:
\begin{equation}\label{eq:H_T2_vev}
\begin{split}
&\bra{\phi}\hat{H}^{\phi}T^2\ket{\phi}\\
&=\frac{1}{4}\bra{\phi}\left(\sum_{ijkl} V_{ijkl} h_l h_k p_j p_i\right)\frac{1}{2}\sum_{mnqs} c^{mq}_{ns} p^\dagger_m h^\dagger_n p_q^\dagger h_s^\dagger\ket{\phi}\\
&=-\frac{1}{8}\sum_{ijklmnqs} V_{ijkl} c^{mq}_{ns}\bra{\phi}p_j p_i p^\dagger_m p_q^\dagger h_l h_k h^\dagger_n h_s^\dagger\ket{\phi}\\
&=-\frac{1}{8}\sum_{ijklmnqs} V_{ijkl}c^{mq}_{ns}\left(\delta_{im}\delta_{jq}- \delta_{iq}\delta_{jm}\right)\left(\delta_{kn}\delta_{ls}-\delta_{ks}\delta_{ln}\right)\\
&=-\frac{1}{2}\sum_{ijkl} V_{ijkl}c^{ij}_{kl}
\end{split}
\end{equation}
where we have used the result \eqref{eq:Tk_cfs} for $T^2\ket{\phi}$ and where the antisymmetry of $c^{ij}_{kl}$ or $V_{ijkl}$ can be used to obtain the last line. We thus obtain
\begin{equation}\label{eq:CC_eq_0_ap}
\Delta E=\sum_{ij} t^{\phi}_{ij}c^i_j-\frac{1}{4}\sum_{ijkl} c^{ij}_{kl}V_{ijkl}\,.
\end{equation}

Note that, when reordering the products of operators to obtain a group of particle operators times a group of hole operators, the resulting sign can be obtained quickly by counting the number of particle operators on the right-hand side of each group consisting of an odd number of hole operators, and add those numbers. The sign is then negative if that number is odd.

The equations with projection on $\bra{\phi_j^i}$ are
\begin{equation}\label{eq:eqn_phi_1_ap}
\begin{split}
0&=\bra{\phi_j^i}e^{-T}\hat{H}^{\phi} e^T\ket{\phi}\\
&=\bra{\phi_j^i}\left(1-T\right)\hat{H}^{\phi}\left(1+T+\frac{1}{2}T^2+\frac{1}{3!}T^3\right)\ket{\phi}\\
0&=\bra{\phi_j^i}\hat{H}^{\phi}\ket{\phi}+ \bra{\phi_j^i}\hat{H}^{\phi}T\ket{\phi}-\bra{\phi_j^i}T\hat{H}^{\phi}T\ket{\phi}\\
&+\frac{1}{2}\bra{\phi_j^i}\hat{H}^{\phi}T^2\ket{\phi}-\frac{1}{2}\bra{\phi_j^i}T\hat{H}^{\phi}T^2\ket{\phi}+\frac{1}{3!}\bra{\phi_j^i}\hat{H}^{\phi}T^3\ket{\phi}\,,
\end{split}
\end{equation}

For the term $\bra{\phi_j^i}\hat{H}^{\phi}\ket{\phi}$, only the part of $\hat{H}^{\phi}$ that create a single particle-hole pair contributes:
\begin{equation}
\begin{split}
\bra{\phi_j^i}\hat{H}^{\phi}\ket{\phi}&=\bra{\phi}h_jp_i\sum_{kl} t^{\phi}_{kl}p_{k}^\dagger h_{l}^\dagger\ket{\phi}\\
\bra{\phi_j^i}\hat{H}^{\phi}\ket{\phi}&= t^{\phi}_{ij}\,.
\end{split}
\end{equation}
The term $\bra{\phi_j^i}\hat{H}^{\phi}T\ket{\phi}$ includes contribution from all the terms of $\hat{H}^{\phi}$ that act on a single particle-hole pair, without creating or destroying any pair:
\begin{equation}
\begin{split}
&\bra{\phi_j^i}\hat{H}^{\phi}T\ket{\phi}\\
&= \bra{\phi}h_jp_i\Bigg(\sum_{kl} t^{\phi}_{kl}p_{k}^\dagger p_{l} -\sum_{kl} t^{\phi}_{kl}h_{k}^\dagger h_{l}\\
&\qquad\qquad- \sum_{klmn} V_{kmln}p_k^\dagger h_l^\dagger h_m p_n\Bigg)\sum_{qr} c^q_r p_q^\dagger h_r^\dagger \ket{\phi}\\
&=\sum_{klqr}t^{\phi}_{kl}c^q_r\bra{\phi}h_jp_ip_{k}^\dagger p_{l} p_q^\dagger h_r^\dagger \ket{\phi}\\
&\qquad-\sum_{klqr} t^{\phi}_{kl} c^q_r \bra{\phi}h_jp_ih_{k}^\dagger h_{l} p_q^\dagger h_r^\dagger \ket{\phi}\\
&\qquad-\sum_{klmnqr} V_{kmln}c^q_r\bra{\phi}h_jp_ip_k^\dagger h_l^\dagger h_m p_n  p_q^\dagger h_r^\dagger\ket{\phi}\\
&=\sum_{klqr}t^{\phi}_{kl}c^q_r\bra{\phi} p_ip_{k}^\dagger p_{l} p_q^\dagger h_jh_r^\dagger\ket{\phi}\\
&\qquad-\sum_{klqr} t^{\phi}_{kl} c^q_r \bra{\phi}p_ip_q^\dagger h_j h_{k}^\dagger h_{l}  h_r^\dagger \ket{\phi}\\
&\qquad-\sum_{klmnqr} V_{kmln}c^q_r\bra{\phi}p_ip_k^\dagger p_n  p_q^\dagger h_j h_l^\dagger h_m h_r^\dagger\ket{\phi}\\
&=\sum_{klqr}t^{\phi}_{kl}c^q_r \delta_{ik}\delta_{lq}\delta_{jr}-\sum_{klqr} t^{\phi}_{kl} c^q_r \delta_{iq}\delta_{jk}\delta_{lr}\\
&\qquad-\sum_{klmnqr} V_{kmln}c^q_r\delta_{ik}\delta_{nq}\delta_{jl}\delta_{mr}
\end{split}
\end{equation}
and thus
\begin{equation}
\bra{\phi_j^i}\hat{H}^{\phi}T\ket{\phi}=\sum_l t^{\phi}_{il}c^l_j - \sum_l t^{\phi}_{jl} c^i_l-\sum_{mn} V_{imjn} c^n_m\,.
\end{equation}

Now, $\bra{\phi_j^i}T\hat{H}^{\phi}T\ket{\phi}$ factorizes as $\bra{\phi_j^i}T\ket{\phi}\bra{\phi}\hat{H}^{\phi}T\ket{\phi}$. Thus, using the result \eqref{eq:phi_HT_phi},
\begin{equation}
\begin{split}
-\bra{\phi_j^i}T\hat{H}^{\phi}T\ket{\phi}&=-c^i_j \sum_{kl} t^{\phi}_{kl}c^k_l\,.
\end{split}
\end{equation}

In $\bra{\phi_j^i}\hat{H}^{\phi}T^2\ket{\phi}$, all the parts of $\hat{H}^{\phi}$ destroying a single pair contribute:
\begin{widetext}
\begin{equation}
\begin{split}
\bra{\phi_j^i}&\hat{H}^{\phi}T^2\ket{\phi}\\
&=\bra{\phi}h_jp_i\Bigg(\sum_{kl} t^{\phi}_{kl}h_{l}p_{k}+\frac{1}{4}\sum_{klmn} V_{klmn} p_n^\dagger h_m p_l  p_k +\frac{1}{4}\sum_{klmn} V_{klmn} h_n^\dagger p_m h_l  h_k \Bigg)\frac{1}{2}\sum_{qrst} c_{rt}^{qs} p^\dagger_q h^\dagger_r p_s^\dagger h_t^\dagger\ket{\phi}\\
&=\frac{1}{2}\sum_{klqrst}t^{\phi}_{kl}c_{rt}^{qs}\bra{\phi}h_jp_i h_{l}p_{k} p^\dagger_q h^\dagger_r p_s^\dagger h_t^\dagger\ket{\phi}+\frac{1}{8}\sum_{klmnqrst} V_{klmn}c_{rt}^{qs}\bra{\phi}h_jp_i p_n^\dagger h_m p_l  p_k  p^\dagger_q h^\dagger_r  p_s^\dagger h_t^\dagger\ket{\phi}\\
&\quad+\frac{1}{8}\sum_{klmnqrst} V_{klmn}c_{rt}^{qs}\bra{\phi}h_jp_i h_n^\dagger p_m h_l  h_k p^\dagger_q h^\dagger_r  p_s^\dagger h_t^\dagger\ket{\phi}\\
&=\frac{1}{2}\sum_{klqrst}t^{\phi}_{kl}c_{rt}^{qs}\bra{\phi}h_j h_{l} h^\dagger_r  h_t^\dagger p_ip_{k} p^\dagger_q p_s^\dagger \ket{\phi}-\frac{1}{8}\sum_{klmnqrst} V_{klmn}c_{rt}^{qs}\bra{\phi}h_j h_m h^\dagger_r h_t^\dagger p_i p_n^\dagger p_l  p_k  p^\dagger_q p_s^\dagger \ket{\phi}\\
&\quad+\frac{1}{8}\sum_{klmnqrst} V_{klmn}c_{rt}^{qs}\bra{\phi}h_j h_n^\dagger h_l  h_k  h^\dagger_r   h_t^\dagger p_i p_m p^\dagger_q p_s^\dagger \ket{\phi}\\
&=\frac{1}{2}\sum_{klqrst}t^{\phi}_{kl}c_{rt}^{qs}\left(\delta_{lr}\delta_{jt}-\delta_{lt}\delta_{jr}\right)\left(\delta_{kq}\delta_{is}-\delta_{ks}\delta_{iq}\right)-\frac{1}{8}\sum_{klmnqrst} V_{klmn}c_{rt}^{qs}\left(\delta_{mr}\delta_{jt}-\delta_{mt}\delta_{jr}\right)\delta_{in}\left(\delta_{kq}\delta_{ls}-\delta_{ks}\delta_{lq}\right)\\
&\quad+\frac{1}{8}\sum_{klmnqrst} V_{klmn}c_{rt}^{qs}\delta_{jn}\left(\delta_{kr}\delta_{lt}-\delta_{kt}\delta_{lr}\right)\left(\delta_{mq}\delta_{is}-\delta_{ms}\delta_{iq}\right)\\
\end{split}
\end{equation}
which yields, using the antisymmetry of $c_{rt}^{qs}$,
\begin{equation}
\begin{split}
\frac{1}{2}\bra{\phi_j^i}\hat{H}^{\phi}T^2\ket{\phi}=&\sum_{kl}t^{\phi}_{kl}c_{jl}^{ik}+\frac{1}{4}\sum_{klm} V_{klmi}c_{jm}^{kl}-\frac{1}{4}\sum_{klm} V_{klmj}c_{kl}^{im}\,.
\end{split}
\end{equation}

Then, $\bra{\phi_j^i}T\hat{H}^{\phi}T^2\ket{\phi}=\bra{\phi_j^i}T\ket{\phi}\bra{\phi}\hat{H}^{\phi}T^2\ket{\phi}$. Therefore, using the result of \eqref{eq:H_T2_vev},
\begin{equation}
-\frac{1}{2}\bra{\phi_j^i}T\hat{H}^{\phi}T^2\ket{\phi}=\frac{1}{4}c^i_j\sum_{klmn}V_{klmn}c^{kl}_{mn}\,.
\end{equation}

Then, $\bra{\phi_j^i}\hat{H}^{\phi}T^3\ket{\phi}$ involves only the part of $\hat{H}^{\phi}$ destroying two pairs:
\begin{equation}
\begin{split}
\bra{\phi_j^i}\hat{H}^{\phi}T^3\ket{\phi}
&=\frac{1}{4(3!)}\sum_{klmnqrq_1r_1q_2r_2} V_{klmn} c^{qq_1q_2}_{rr_1r_2}\bra{\phi}h_jp_ih_n h_m p_lp_k p_q^\dagger h_r^\dagger  p^\dagger_{q_1} h^\dagger_{r_1} p_{q_2}^\dagger h_{r_2}^\dagger \ket{\phi}\\
&=-\frac{1}{4(3!)}\sum_{klmnqrq_1r_1q_2r_2} V_{klmn} c^{qq_1q_2}_{rr_1r_2}\bra{\phi}h_jh_n h_m h_r^\dagger   h^\dagger_{r_1} h_{r_2}^\dagger p_ip_lp_k p_q^\dagger p^\dagger_{q_1}  p_{q_2}^\dagger  \ket{\phi}\,.
\end{split}
\end{equation}
\end{widetext}
Here, because of the antisymmetry of $c^{qq_1q_2}_{ss_1s_2}$, all $(3!)^2$ terms obtained by normal-ordering the operators are identical, thus
\begin{equation}
\frac{1}{3!}\bra{\phi_j^i}\hat{H}^{\phi}T^3\ket{\phi}=-\frac{1}{4}\sum_{klmn} V_{klmn} c^{kli}_{mnj}\,.
\end{equation}

Therefore, Eq. \eqref{eq:eqn_phi_1_ap} becomes
\begin{equation}
\begin{split}
0&=t^{\phi}_{ij}+ \sum_k t^{\phi}_{ik}c^k_j - \sum_k t^{\phi}_{jk} c^i_k-\sum_{kl} V_{ikjl} c^l_k-c^i_j \sum_{kl} t^{\phi}_{kl}c^k_l\\
&\qquad+\sum_{kl}t^{\phi}_{kl}\,c^{ik}_{jl}+\frac{1}{4}\sum_{klm} V_{klmi}c_{jm}^{kl}-\frac{1}{4}\sum_{klm} V_{klmj}c_{kl}^{im}\\
&\qquad+\frac{1}{4}c^i_j\sum_{klmn}V_{klmn}c^{kl}_{mn}-\frac{1}{4}\sum_{klmn} V_{klmn} c^{kli}_{mnj}\,,
\end{split}
\end{equation}
or, if we group disconnect terms with similar connected ones,
\begin{equation}\label{eq:CC_eq_1_ap}
\begin{split}
0&=t^{\phi}_{ij}+ \sum_k t^{\phi}_{ik}c^k_j - \sum_k t^{\phi}_{jk} c^i_k-\sum_{kl} V_{ikjl} c^l_k\\
&\qquad+\sum_{kl}t^{\phi}_{kl}\,\left(c^{ik}_{jl}-c^i_jc^k_l\right)+\frac{1}{4}\sum_{klm} V_{klmi}c_{jm}^{kl}\\
&\qquad-\frac{1}{4}\sum_{klm} V_{klmj}c_{kl}^{im}-\frac{1}{4}\sum_{klmn} V_{klmn} (c^{kli}_{mnj}-c^i_jc^{kl}_{mn})\,.
\end{split}
\end{equation}

See the \hyperlink{suppl_mat}{supplemental material} for the derivation of the equations with projection on $\bra{\phi^{i_1i_2}_{j_1j_2}}$.

\section{Tensor representation for quadruple-excitation coefficients}\label{sec:STTN_c4}

Provided here are the decompositions used in the STTN representation of the quadruple-excitation CI coefficients or cluster operator amplitudes discussed in section \ref{sec:coeffs_tensor_decomp}.

When all spins are equal, the decomposition for the particle-hole pairing only is
\begin{equation}\label{eq:decomp_c4_uuuu_4ph}
\begin{split}
&\left(c^{i_1\uparrow i_2\uparrow i_3\uparrow i_4\uparrow}_{j_1\downarrow j_2\downarrow j_3\downarrow j_4\downarrow}\right)_{p\bar{h}p\bar{h}p\bar{h}p\bar{h}}=\sum_{m_1,m_2=1}^{s_{p\bar{h}p\bar{h}}}\sum_{l_1,l_2,l_3,l_4=1}^{s_{p\bar{h}}}\\
& \sum_{k_1,k_3,k_5,k_7=1}^{s_{p}}\sum_{k_2,k_4,k_6,k_8=1}^{s_{h}}\sum_{\pi\in S_4} \mu_{m_1m_2}^{p\bar{h}p\bar{h},p\bar{h}p\bar{h}}\\
&\quad \times\left(\lambda_{l_1l_2 m_1}^{p\bar{h}p\bar{h}}\lambda_{l_3l_4 m_2}^{p\bar{h}p\bar{h}}+\lambda_{l_1l_3 m_1}^{p\bar{h}p\bar{h}}\lambda_{l_2l_4 m_2}^{p\bar{h}p\bar{h}}+\lambda_{l_1l_4 m_1}^{p\bar{h}p\bar{h}}\lambda_{l_2l_3 m_2}^{p\bar{h}p\bar{h}}\right)\\
&\quad\times\kappa_{k_1k_2l_1}^{p\bar{h}}\kappa_{k_3k_4l_2}^{p\bar{h}}\kappa_{k_5k_6l_3}^{p\bar{h}}\kappa_{k_7k_8l_4}^{p\bar{h}}\epsilon_{\pi_1\pi_2\pi_3\pi_4}\\
&\quad \times  u_{i_1 k_1}^{\uparrow}u_{i_2 k_3}^{\uparrow}u_{i_3 k_5}^{\uparrow}u_{i_4 k_7}^{\uparrow}v_{j_{\pi_1}k_2}^{\downarrow}v_{j_{\pi_2}k_4}^{\downarrow}v_{j_{\pi_3}k_6}^{\downarrow}v_{j_{\pi_4}k_8}^{\downarrow}\,,
\end{split}
\end{equation}
Then, if we combine particle-hole, particle-particle and hole-hole pairing, we can have
\begin{equation}
\begin{split}
&\left(c^{i_1\uparrow i_2\uparrow i_3\uparrow i_4\uparrow}_{j_1\downarrow j_2\downarrow j_3\downarrow j_4\downarrow}\right)_{ppp\bar{h}\bar{h}\bar{h}p\bar{h}}=\sum_{m_1=1}^{s_{ppp\bar{h}}}\sum_{m_2=1}^{s_{\bar{h}\bar{h}p\bar{h}}}\sum_{l_1=1}^{s_{pp}}\sum_{l_2,l_4=1}^{s_{p\bar{h}}}\sum_{l_3=1}^{s_{hh}}\\
&\sum_{k_1,k_2,k_3,k_7=1}^{s_{p}}\sum_{k_4,k_5,k_6,k_8=1}^{s_{h}}\sum_{\pi,\chi \in S_4} \mu_{m_1m_2}^{ppp\bar{h},\bar{h}\bar{h}p\bar{h}} \lambda_{l_1l_2 m_1}^{ppp\bar{h}}\lambda_{l_3l_4 m_2}^{\bar{h}\bar{h}p\bar{h}}\\
&\quad\times\kappa_{k_1k_2l_1}^{pp}\kappa_{k_3k_4l_2}^{p\bar{h}}\kappa_{k_5k_6l_3}^{\bar{h}\bar{h}}\kappa_{k_7k_8l_4}^{p\bar{h}}\epsilon_{\pi_1\pi_2\pi_3\pi_4}\epsilon_{\chi_1\chi_2\chi_3\chi_4}\\
&\quad \times  u_{i_{\pi_1} k_1}^{\uparrow}u_{i_{\pi_2} k_2}^{\uparrow}u_{i_{\pi_3} k_3}^{\uparrow}u_{i_{\pi_4} k_7}^{\uparrow}v_{j_{\chi_1}k_4}^{\downarrow}v_{j_{\chi_2}k_5}^{\downarrow}v_{j_{\chi_3}k_6}^{\downarrow}v_{j_{\chi_4}k_8}^{\downarrow}\,.
\end{split}
\end{equation}
Then, for particle-particle and hole-hole pairing only, the expression is
\begin{equation}
\begin{split}
&\left(c^{i_1\uparrow i_2\uparrow i_3\uparrow i_4\uparrow}_{j_1\downarrow j_2\downarrow j_3\downarrow j_4\downarrow}\right)_{pppp\bar{h}\bar{h}\bar{h}\bar{h}}=\sum_{m_1=1}^{s_{pppp}}\sum_{m_2=1}^{s_{\bar{h}\bar{h}\bar{h}\bar{h}}}\sum_{l_1,l_2=1}^{s_{pp}}\sum_{l_3,l_4=1}^{s_{\bar{h}\bar{h}}}\\
& \sum_{k_1,k_2,k_3,k_4=1}^{s_{p}}\sum_{k_5,k_6,k_7,k_8=1}^{s_{h}}\sum_{\pi,\chi \in S_4} \mu_{m_1m_2}^{pppp,\bar{h}\bar{h}\bar{h}\bar{h}} \lambda_{l_1l_2 m_1}^{pppp}\lambda_{l_3l_4 m_2}^{\bar{h}\bar{h}\bar{h}\bar{h}}\\
&\quad\times\kappa_{k_1k_2l_1}^{pp}\kappa_{k_3k_4l_2}^{pp}\kappa_{k_5k_6l_3}^{\bar{h}\bar{h}}\kappa_{k_7k_8l_4}^{\bar{h}\bar{h}}\epsilon_{\pi_1\pi_2\pi_3\pi_4}\epsilon_{\chi_1\chi_2\chi_3\chi_4}\\
&\quad \times  u_{i_{\pi_1} k_1}^{\uparrow}u_{i_{\pi_2} k_2}^{\uparrow}u_{i_{\pi_3} k_3}^{\uparrow}u_{i_{\pi_4} k_4}^{\uparrow}v_{j_{\chi_1}k_5}^{\downarrow}v_{j_{\chi_2}k_6}^{\downarrow}v_{j_{\chi_3}k_7}^{\downarrow}v_{j_{\chi_4}k_8}^{\downarrow}\,,
\end{split}
\end{equation}
and finally,
\begin{equation}
\begin{split}
&c^{i_1\uparrow i_2\uparrow i_3\uparrow i_4\uparrow}_{j_1\downarrow j_2\downarrow j_3\downarrow j_4\downarrow}=\left(c^{i_1\uparrow i_2\uparrow i_3\uparrow i_4\uparrow}_{j_1\downarrow j_2\downarrow j_3\downarrow j_4\downarrow}\right)_{p\bar{h}p\bar{h}p\bar{h}p\bar{h}}\\
&\qquad+\left(c^{i_1\uparrow i_2\uparrow i_3\uparrow i_4\uparrow}_{j_1\downarrow j_2\downarrow j_3\downarrow j_4\downarrow}\right)_{ppp\bar{h}\bar{h}\bar{h}p\bar{h}}+\left(c^{i_1\uparrow i_2\uparrow i_3\uparrow i_4\uparrow}_{j_1\downarrow j_2\downarrow j_3\downarrow j_4\downarrow}\right)_{pppp\bar{h}\bar{h}\bar{h}\bar{h}}\,.
\end{split}
\end{equation}

Then, when one spin is different, the decomposition based on singlet particle-hole pairing only is
\begin{equation}\label{eq:decomp_c4_uuud_4ph}
\begin{split}
&\left(c^{i_1\uparrow i_2\uparrow i_3\uparrow i_4\downarrow}_{j_1\downarrow j_2\downarrow j_3\downarrow j_4\uparrow}\right)_{p\bar{h}p\bar{h}p\bar{h}\bar{p}h}=\sum_{m_1=1}^{s_{p\bar{h}p\bar{h}}}\sum_{m_2=1}^{s_{p\bar{h}\bar{p}h}}\sum_{l_1,l_2,l_3,l_4=1}^{s_{p\bar{h}}}\\
&\sum_{k_1,k_3,k_5,k_7=1}^{s_{p}}\sum_{k_2,k_4,k_6,k_8=1}^{s_{h}}\sum_{\pi\in S_3} \mu_{m_1m_2}^{p\bar{h}p\bar{h},p\bar{h}\bar{p}h}\\
&\quad \times\left(\lambda_{l_1l_2 m_1}^{p\bar{h}p\bar{h}}\lambda_{l_3l_4 m_2}^{p\bar{h}\bar{p}h}+\lambda_{l_1l_3 m_1}^{p\bar{h}p\bar{h}}\lambda_{l_2l_4 m_2}^{p\bar{h}\bar{p}h}+\lambda_{l_2l_3 m_1}^{p\bar{h}p\bar{h}}\lambda_{l_1l_4 m_2}^{p\bar{h}\bar{p}h}\right)\\
&\quad\times\kappa_{k_1k_2l_1}^{p\bar{h}}\kappa_{k_3k_4l_2}^{p\bar{h}}\kappa_{k_5k_6l_3}^{p\bar{h}}\kappa_{k_7k_8l_4}^{\bar{p}h}\epsilon_{\pi_1\pi_2\pi_3}\\
&\quad \times  u_{i_1 k_1}^{\uparrow}u_{i_2 k_3}^{\uparrow}u_{i_3 k_5}^{\uparrow}u_{i_4 k_7}^{\downarrow}v_{j_{\pi_1}k_2}^{\downarrow}v_{j_{\pi_2}k_4}^{\downarrow}v_{j_{\pi_3}k_6}^{\downarrow}v_{j_4k_8}^{\uparrow}\,,
\end{split}
\end{equation}
then, we can have both singlet and triplet particle-hole pairing:
\begin{equation}\label{eq:decomp_c4_uuud_phbph}
\begin{split}
&\left(c^{i_1\uparrow i_2\uparrow i_3\uparrow i_4\downarrow}_{j_1\downarrow j_2\downarrow j_3\downarrow j_4\uparrow}\right)_{p\bar{h}php\bar{h}\bar{p}\bar{h}}=\sum_{m_1=1}^{s_{p\bar{h}ph}}\sum_{m_2=1}^{s_{p\bar{h}\bar{p}\bar{h}}}\sum_{l_1,l_3=1}^{s_{p\bar{h}}}\sum_{l_2,l_4=1}^{s_{ph}}\\
& \sum_{k_1,k_3,k_5,k_7=1}^{s_{p}}\sum_{k_2,k_4,k_6,k_8=1}^{s_{h}}\sum_{\pi,\chi \in S_3} \mu_{m_1m_2}^{p\bar{h}ph,p\bar{h}\bar{p}\bar{h}}\lambda_{l_1l_2 m_1}^{p\bar{h}ph}\\
&\quad\times\lambda_{l_3l_4 m_2}^{p\bar{h}\bar{p}\bar{h}}\kappa_{k_1k_2l_1}^{p\bar{h}}\kappa_{k_3k_4l_2}^{ph}\kappa_{k_5k_6l_3}^{p\bar{h}}\kappa_{k_7k_8l_4}^{\bar{p}\bar{h}}\epsilon_{\pi_1\pi_2\pi_3}\epsilon_{\chi_1\chi_2\chi_3}\\
&\quad \times  u_{i_{\pi_1} k_1}^{\uparrow}u_{i_{\pi_2} k_3}^{\uparrow}u_{i_{\pi_3} k_5}^{\uparrow}u_{i_4 k_7}^{\downarrow}v_{j_{\chi_1}k_2}^{\downarrow}v_{j_{\chi_2}k_6}^{\downarrow}v_{j_{\chi_3}k_8}^{\downarrow}v_{j_4k_4}^{\uparrow}\,,
\end{split}
\end{equation}
singlet particle-hole, particle-particle and hole-hole pairing:
\begin{equation}\label{eq:decomp_c4_uuud_phbppb}
\begin{split}
&\left(c^{i_1\uparrow i_2\uparrow i_3\uparrow i_4\downarrow}_{j_1\downarrow j_2\downarrow j_3\downarrow j_4\uparrow}\right)_{p\bar{h}p\bar{p}p\bar{h}\bar{h}h}=\sum_{m_1=1}^{s_{p\bar{h}p\bar{p}}}\sum_{m_2=1}^{s_{p\bar{h}\bar{h}h}}\sum_{l_1,l_3=1}^{s_{p\bar{h}}}\sum_{l_2=1}^{s_{p\bar{p}}}\sum_{l_4=1}^{s_{h\bar{h}}}\\
& \sum_{k_1,k_3,k_4,k_5=1}^{s_{p}}\sum_{k_2,k_6,k_7,k_8=1}^{s_{h}}\sum_{\pi,\chi \in S_3} \mu_{m_1m_2}^{p\bar{h}p\bar{p},p\bar{h}\bar{h}h}\lambda_{l_1l_2 m_1}^{p\bar{h}p\bar{p}}\\
&\quad\times\lambda_{l_3l_4 m_2}^{p\bar{h}\bar{h}h}\kappa_{k_1k_2l_1}^{p\bar{h}}\kappa_{k_3k_4l_2}^{p\bar{p}}\kappa_{k_5k_6l_3}^{p\bar{h}}\kappa_{k_7k_8l_4}^{\bar{h}h}\epsilon_{\pi_1\pi_2\pi_3}\epsilon_{\chi_1\chi_2\chi_3}\\
&\quad \times  u_{i_{\pi_1} k_1}^{\uparrow}u_{i_{\pi_2} k_3}^{\uparrow}u_{i_{\pi_3} k_5}^{\uparrow}u_{i_4 k_4}^{\downarrow}v_{j_{\chi_1}k_2}^{\downarrow}v_{j_{\chi_2}k_6}^{\downarrow}v_{j_{\chi_3}k_7}^{\downarrow}v_{j_4k_8}^{\uparrow}\,,
\end{split}
\end{equation}
or triplet particle-hole, particle-particle and hole-hole pairing:
\begin{equation}
\begin{split}
&\left(c^{i_1\uparrow i_2\uparrow i_3\uparrow i_4\downarrow}_{j_1\downarrow j_2\downarrow j_3\downarrow j_4\uparrow}\right)_{ppph\bar{p}\bar{h}\bar{h}\bar{h}}=\sum_{m_1=1}^{s_{ppph}}\sum_{m_2=1}^{s_{\bar{p}\bar{h}\bar{h}\bar{h}}}\sum_{l_1=1}^{s_{pp}}\sum_{l_2,l_3=1}^{s_{ph}}\sum_{l_4=1}^{s_{hh}}\\
& \sum_{k_1,k_2,k_3,k_5=1}^{s_{p}}\sum_{k_4,k_6,k_7,k_8=1}^{s_{h}}\sum_{\pi,\chi \in S_3} \mu_{m_1m_2}^{ppph,\bar{p}\bar{h}\bar{h}\bar{h}}\lambda_{l_1l_2 m_1}^{ppph}\\
&\quad\times\lambda_{l_3l_4 m_2}^{\bar{p}\bar{h}\bar{h}\bar{h}}\kappa_{k_1k_2l_1}^{pp}\kappa_{k_3k_4l_2}^{ph}\kappa_{k_5k_6l_3}^{\bar{p}\bar{h}}\kappa_{k_7k_8l_4}^{\bar{h}\bar{h}}\epsilon_{\pi_1\pi_2\pi_3}\epsilon_{\chi_1\chi_2\chi_3}\\
&\quad \times  u_{i_{\pi_1} k_1}^{\uparrow}u_{i_{\pi_2} k_2}^{\uparrow}u_{i_{\pi_3} k_3}^{\uparrow}u_{i_4 k_5}^{\downarrow}v_{j_{\chi_1}k_6}^{\downarrow}v_{j_{\chi_2}k_7}^{\downarrow}v_{j_{\chi_3}k_8}^{\downarrow}v_{j_4k_4}^{\uparrow}\,,
\end{split}
\end{equation}
and we can use
\begin{equation}
\begin{split}
&c^{i_1\uparrow i_2\uparrow i_3\uparrow i_4\downarrow}_{j_1\downarrow j_2\downarrow j_3\downarrow j_4\uparrow}=\left(c^{i_1\uparrow i_2\uparrow i_3\uparrow i_4\downarrow}_{j_1\downarrow j_2\downarrow j_3\downarrow j_4\uparrow}\right)_{p\bar{h}p\bar{h}p\bar{h}\bar{p}h}\\
&\quad+\left(c^{i_1\uparrow i_2\uparrow i_3\uparrow i_4\downarrow}_{j_1\downarrow j_2\downarrow j_3\downarrow j_4\uparrow}\right)_{p\bar{h}php\bar{h}\bar{p}\bar{h}}+\left(c^{i_1\uparrow i_2\uparrow i_3\uparrow i_4\downarrow}_{j_1\downarrow j_2\downarrow j_3\downarrow j_4\uparrow}\right)_{p\bar{h}p\bar{p}p\bar{h}\bar{h}h}\\
&\quad+\left(c^{i_1\uparrow i_2\uparrow i_3\uparrow i_4\downarrow}_{j_1\downarrow j_2\downarrow j_3\downarrow j_4\uparrow}\right)_{ppph\bar{p}\bar{h}\bar{h}\bar{h}}\,.
\end{split}
\end{equation}

Finally, for two up spins and two down spins, we can use
\begin{equation}\label{eq:decomp_c4_uudd_ppbhhb}
\begin{split}
&\left(c^{i_1\uparrow i_2\uparrow i_3\downarrow i_4\downarrow}_{j_1\downarrow j_2\downarrow j_3\uparrow j_4\uparrow}\right)_{p\bar{p}\bar{h}hph\bar{p}\bar{h}}=\sum_{m_1=1}^{s_{p\bar{p}\bar{h}h}}\sum_{m_2=1}^{s_{ph\bar{p}\bar{h}}}\sum_{l_1=1}^{s_{p\bar{p}}}\sum_{l_2=1}^{s_{h\bar{h}}}\sum_{l_3,l_4=1}^{s_{ph}}\\
&\sum_{k_1,k_2,k_5,k_7=1}^{s_{p}}\sum_{k_3,k_4,k_6,k_8=1}^{s_{h}}\sum_{\pi,\rho,\tau,\chi\in S_2} \mu_{m_1m_2}^{p\bar{p}\bar{h}h,ph\bar{p}\bar{h}}\lambda_{l_1l_2 m_1}^{p\bar{p}\bar{h}h}\\
&\quad\times\lambda_{l_3l_4 m_2}^{ph\bar{p}\bar{h}}\kappa_{k_1k_2l_1}^{p\bar{p}}\kappa_{k_3k_4l_2}^{\bar{h}h}\kappa_{k_5k_6l_3}^{ph}\kappa_{k_7k_8l_4}^{\bar{p}\bar{h}}\epsilon_{\pi_1\pi_2}\epsilon_{\rho_1\rho_2}\\
&\quad \times \epsilon_{\tau_1\tau_2}\epsilon_{\chi_1\chi_2} u_{i_{\pi_1} k_1}^{\uparrow}u_{i_{\pi_2} k_5}^{\uparrow}u_{i_{\rho_1+2} k_2}^{\downarrow}u_{i_{\rho_2+2} k_7}^{\downarrow}\\
&\quad \times v_{j_{\tau_1}k_3}^{\downarrow}v_{j_{\tau_2}k_8}^{\downarrow}v_{j_{\chi_1+2}k_4}^{\uparrow}v_{j_{\chi_2+2}k_6}^{\uparrow}\,,
\end{split}
\end{equation}
where $\lambda^{p\bar{p}\bar{h}h}=\lambda^{\bar{p}ph\bar{h}}$ and $\lambda_{l_3l_4 m_2}^{ph\bar{p}\bar{h}}=\lambda_{l_4l_3 m_2}^{\bar{p}\bar{h}ph}$, and then,
\begin{equation}\label{eq:decomp_c4_uudd_4ph}
\begin{split}
&\left(c^{i_1\uparrow i_2\uparrow i_3\downarrow i_4\downarrow}_{j_1\downarrow j_2\downarrow j_3\uparrow j_4\uparrow}\right)_{p\bar{h}p\bar{h}\bar{p}h\bar{p}h}=\sum_{m_1,m_2=1}^{s_{p\bar{h}p\bar{h}}}\sum_{l_1,l_2,l_3,l_4=1}^{s_{p\bar{h}}} \sum_{k_1,k_3,k_5,k_7=1}^{s_{p}}\\
&\sum_{k_2,k_4,k_6,k_8=1}^{s_{h}}\sum_{\pi,\rho,\tau,\chi\in S_2} \mu_{m_1m_2}^{p\bar{h}p\bar{h},\bar{p}h\bar{p}h}\lambda_{l_1l_2 m_1}^{p\bar{h}p\bar{h}}\lambda_{l_3l_4 m_2}^{\bar{p}h\bar{p}h}\\
&\quad\times\kappa_{k_1k_2l_1}^{p\bar{h}}\kappa_{k_3k_4l_2}^{p\bar{h}}\kappa_{k_5k_6l_3}^{\bar{p}h}\kappa_{k_7k_8l_4}^{\bar{p}h}\epsilon_{\pi_1\pi_2}\epsilon_{\rho_1\rho_2}\epsilon_{\tau_1\tau_2}\epsilon_{\chi_1\chi_2}\\
&\quad \times  u_{i_{\pi_1} k_1}^{\uparrow}u_{i_{\pi_2} k_3}^{\uparrow}u_{i_{\rho_1+2} k_5}^{\downarrow}u_{i_{\rho_2+2} k_7}^{\downarrow}\\
&\quad \times v_{j_{\tau_1}k_2}^{\downarrow}v_{j_{\tau_2}k_4}^{\downarrow}v_{j_{\chi_1+2}k_6}^{\uparrow}v_{j_{\chi_2+2}k_8}^{\uparrow}\,,
\end{split}
\end{equation}
\begin{equation}
\begin{split}
&\left(c^{i_1\uparrow i_2\uparrow i_3\downarrow i_4\downarrow}_{j_1\downarrow j_2\downarrow j_3\uparrow j_4\uparrow}\right)_{pphh\bar{p}\bar{p}\bar{h}\bar{h}}=\sum_{m_1,m_2=1}^{s_{pphh}}\sum_{l_1,l_3=1}^{s_{pp}}\sum_{l_3,l_4=1}^{s_{hh}}\\
&\sum_{k_1,k_2,k_5,k_7=1}^{s_{p}}\sum_{k_3,k_4,k_6,k_8=1}^{s_{h}}\sum_{\pi,\rho,\tau,\chi\in S_2} \mu_{m_1m_2}^{pphh,\bar{p}\bar{p}\bar{h}\bar{h}}\lambda_{l_1l_2 m_1}^{pphh}\\
&\quad\times\lambda_{l_3l_4 m_2}^{\bar{p}\bar{p}\bar{h}\bar{h}}\kappa_{k_1k_2l_1}^{pp}\kappa_{k_3k_4l_2}^{hh}\kappa_{k_5k_6l_3}^{\bar{p}\bar{p}}\kappa_{k_7k_8l_4}^{\bar{h}\bar{h}}\epsilon_{\pi_1\pi_2}\epsilon_{\rho_1\rho_2}\\
&\quad \times \epsilon_{\tau_1\tau_2}\epsilon_{\chi_1\chi_2} u_{i_{\pi_1} k_1}^{\uparrow}u_{i_{\pi_2} k_2}^{\uparrow}u_{i_{\rho_1+2} k_5}^{\downarrow}u_{i_{\rho_2+2} k_6}^{\downarrow}\\
&\quad \times v_{j_{\tau_1}k_7}^{\downarrow}v_{j_{\tau_2}k_8}^{\downarrow}v_{j_{\chi_1+2}k_3}^{\uparrow}v_{j_{\chi_2+2}k_4}^{\uparrow}\,,
\end{split}
\end{equation}
\begin{equation}
\begin{split}
&\left(c^{i_1\uparrow i_2\uparrow i_3\downarrow i_4\downarrow}_{j_1\downarrow j_2\downarrow j_3\uparrow j_4\uparrow}\right)_{p\bar{p}p\bar{p}\bar{h}h\bar{h}h}=\sum_{m_1=1}^{s_{p\bar{p}p\bar{p}}}\sum_{m_2=1}^{s_{\bar{h}h\bar{h}h}}\sum_{l_1,l_2=1}^{s_{p\bar{p}}}\sum_{l_3,l_4=1}^{s_{\bar{h}h}}\\
&\sum_{k_1,k_2,k_3,k_4=1}^{s_{p}}\sum_{k_5,k_6,k_7,k_8=1}^{s_{h}}\sum_{\pi,\rho,\tau,\chi\in S_2} \mu_{m_1m_2}^{p\bar{p}p\bar{p},\bar{h}h\bar{h}h}\lambda_{l_1l_2 m_1}^{p\bar{p}p\bar{p}}\\
&\quad\times\lambda_{l_3l_4 m_2}^{\bar{h}h\bar{h}h}\kappa_{k_1k_2l_1}^{p\bar{p}}\kappa_{k_3k_4l_2}^{p\bar{p}}\kappa_{k_5k_6l_3}^{\bar{h}h}\kappa_{k_7k_8l_4}^{\bar{h}h}\epsilon_{\pi_1\pi_2}\epsilon_{\rho_1\rho_2}\\
&\quad \times \epsilon_{\tau_1\tau_2}\epsilon_{\chi_1\chi_2} u_{i_{\pi_1} k_1}^{\uparrow}u_{i_{\pi_2} k_3}^{\uparrow}u_{i_{\rho_1+2} k_2}^{\downarrow}u_{i_{\rho_2+2} k_4}^{\downarrow}\\
&\quad \times v_{j_{\tau_1}k_5}^{\downarrow}v_{j_{\tau_2}k_7}^{\downarrow}v_{j_{\chi_1+2}k_6}^{\uparrow}v_{j_{\chi_2+2}k_8}^{\uparrow}\,,
\end{split}
\end{equation}
and we use
\begin{equation}
\begin{split}
&c^{i_1\uparrow i_2\uparrow i_3\downarrow i_4\downarrow}_{j_1\downarrow j_2\downarrow j_3\uparrow j_4\uparrow}=\left(c^{i_1\uparrow i_2\uparrow i_3\downarrow i_4\downarrow}_{j_1\downarrow j_2\downarrow j_3\uparrow j_4\uparrow}\right)_{p\bar{p}\bar{h}hph\bar{p}\bar{h}}\\
&\quad+\left(c^{i_1\uparrow i_2\uparrow i_3\downarrow i_4\downarrow}_{j_1\downarrow j_2\downarrow j_3\uparrow j_4\uparrow}\right)_{p\bar{h}p\bar{h}\bar{p}h\bar{p}h}+\left(c^{i_1\uparrow i_2\uparrow i_3\downarrow i_4\downarrow}_{j_1\downarrow j_2\downarrow j_3\uparrow j_4\uparrow}\right)_{pphh\bar{p}\bar{p}\bar{h}\bar{h}}\\
&\qquad+\left(c^{i_1\uparrow i_2\uparrow i_3\downarrow i_4\downarrow}_{j_1\downarrow j_2\downarrow j_3\uparrow j_4\uparrow}\right)_{p\bar{p}p\bar{p}\bar{h}h\bar{h}h}\,.
\end{split}
\end{equation}

\section{SVD form of the tree tensor networks}\label{sec:SVD_TTN}

Let us see how the decomposition \eqref{eq:decomp_c2_ud_pp_s},
\begin{equation}\label{eq:decomp_c2_ud_pp_s_ap}
\begin{split}
\left(c^{i_1\uparrow i_2\downarrow}_{j_1\downarrow j_2\uparrow}\right)_{p\bar{p}\bar{h}h}=&\sum_{k=1}^{s_{p\bar{p}}}\sum_{l=1}^{s_{h\bar{h}}}\sum_{m,n=1}^{s_{p}}\sum_{q,r=1}^{s_{h}} \lambda_{kl1}^{p\bar{p}\bar{h}h} \kappa_{mnk}^{p\bar{p}} \kappa_{qrl}^{\bar{h}h}\\
&\times u^{\uparrow}_{i_1m}u^{\downarrow}_{i_2n} v^{\downarrow}_{j_1q}v^{\uparrow}_{j_2r}\,,
\end{split}
\end{equation}
can be written as a combination of SVD's if the matrix slices of the $\kappa$ tensors, labeled by their third index, are orthogonal. First, we define
\begin{equation}\label{eq:decomp_w_ppb}
w_{(i_1i_2),k}^{p\bar{p}}=\sum_{m,n=1}^{s_{p}} \kappa_{mnk}^{p\bar{p}} u^{\uparrow}_{i_1m}u^{\downarrow}_{i_2n}
\end{equation}
and
\begin{equation}\label{eq:decomp_w_hhb}
w_{(j_1,j_2),l}^{\bar{h}h}=\sum_{q,r=1}^{s_{h}} \kappa_{qrl}^{\bar{h}h} v^{\downarrow}_{j_1q}v^{\uparrow}_{j_2r}\,.
\end{equation}
Then \eqref{eq:decomp_c2_ud_pp_s_ap} becomes
\begin{equation}
\left(c^{i_1\uparrow i_2\downarrow}_{j_1\downarrow j_2\uparrow}\right)_{p\bar{p}\bar{h}h}=\sum_{k=1}^{s_{p\bar{p}}}\sum_{l=1}^{s_{h\bar{h}}} \lambda_{kl1}^{p\bar{p}\bar{h}h} w_{(i_1i_2),k} w_{(j_1,j_2),l}
\end{equation}
which can be written in matrix form as
\begin{equation}\label{eq:decomp_c2_ud_pp_s_svd}
\begin{split}
\left(C^{\uparrow \downarrow}_{\downarrow \uparrow}\right)_{p\bar{p}\bar{h}h}&=W_{p\bar{p}} \Lambda^{p\bar{p}\bar{h}h} W^T_{\bar{h}h}\\
&=W_{p\bar{p}} X\bar{\Lambda}^{p\bar{p}\bar{h}h}Y^T W^T_{\bar{h}h}\\
&=\bar{W}_{p\bar{p}} \bar{\Lambda}^{p\bar{p}\bar{h}h}\bar{W}^T_{\bar{h}h}\\
\end{split}
\end{equation}
where $\bar{\Lambda}^{p\bar{p}\bar{h}h}$ is diagonal, $X$ and $Y$ are unitary, $\bar{W}_{p\bar{p}}=W_{p\bar{p}} X$, $\bar{W}_{\bar{h}h}=W_{\bar{h}h}Y$ and the row indices of $W_{p\bar{p}}$ and $W_{\bar{h}h}$ are $(i_1i_2)$ and $(j_1j_2)$ pairs, respectively. Now, expression \eqref{eq:decomp_c2_ud_pp_s_svd} has SVD form, but it is an actual SVD only if the columns of $\bar{W}_{p\bar{p}}$ and $\bar{W}_{\bar{h}h}$ are orthogonal. Since $X$ and $Y$ are unitary, this is the case if the columns of $W_{p\bar{p}}$ and $W_{\bar{h}h}$, corresponding to matrix slices of tensors $w^{p\bar{p}}$ and $w^{\bar{h}h}$, are orthogonal. Now, using the fact the columns of matrices $u^\sigma$ are orthogonal, the matrix slices of $w^{p\bar{p}}$ are orthogonal if the matrix slices of $\kappa^{p\bar{p}}$ are, and similarly for $v^\sigma$, $w^{\bar{h}h}$ and $\kappa_{qrl}^{\bar{h}h}$.

Furthermore, the decompositions \eqref{eq:decomp_w_ppb} and \eqref{eq:decomp_w_hhb} are also closely related to SVD's. For instance, for a fixed $k$, \eqref{eq:decomp_w_ppb} has the matrix form
\begin{equation}
\begin{split}
\tilde{W}_k^{p\bar{p}}&=U^{\uparrow} K_k (U^{\downarrow})^T\\
&=U^{\uparrow} R_k\bar{K}_kS_k^T (U^{\downarrow})^T\\
&=U^{\uparrow}_k \bar{K}_k(U^{\downarrow}_k)^T
\end{split}
\end{equation}
where $\bar{K}_k$ is diagonal, $U^{\uparrow}_k=U^{\uparrow} R_k$, $U^{\downarrow}_k=U^{\downarrow}S_k$ and $R_k$ and $S_k$ are unitary.

Therefore, if the matrix slices of the $\kappa$ tensors are orthogonal, \eqref{eq:decomp_c2_ud_pp_s_ap} is related to a combination of SVD's only by a few rotations and thus, for that partition of the indices, that combination of SVD is the optimal decomposition. However, since $s_{p\bar{p}}$ and $s_{h\bar{h}}$ are not equal in general, $\lambda$ is not diagonal, and the orthogonality condition for the $\kappa$ tensors is not necessarily satisfied, the decomposition \eqref{eq:decomp_c2_ud_pp_s_ap} is less compact than its corresponding SVD form. The other decompositions of sections \ref{sec:coeffs_tensor_decomp} all have the same structure and can be written similarly as combinations of SVD's using internal rotations. Although, because of those rotations, they are not individually as compact as possible, their more general form allows tensors to be shared between decompositions, and therefore the STTN structure to be globally compact.

\section{Computational complexity}\label{sec:comp_complexity}

The computational complexity in evaluating the generalized CC equations of section \ref{sec:eqns_coeffs} depends on the term
\begin{equation}
\sum_{klmn}V_{klmn} c_{mnj_1j_2}^{kli_1i_2}
\end{equation}
in Eq. \eqref{eq:CC_eq_2}. More specifically, assuming the decompositions provided in Appendix \ref{sec:STTN_c4} are used, the complexity depends on the contributions in which the summed indices have the largest entanglement with the external ones, namely, when each index in $\{k,l,m,n\}$ is connected by a $\kappa$ tensor to an index in $\{i_1,i_2,j_1,j_2\}$. Those contributions have the form
\begin{equation}
\begin{split}
R_{j_1j_2}^{i_1i_2}=&V_{n_1n_2n_3n_4}\mu_{m_1m_2}\lambda_{l_1l_2m_1}\lambda_{l_3l_4m_2}\kappa_{k_1k_2l_1}\kappa_{k_3k_4l_2}\\
&\times\kappa_{k_5k_6l_3}\kappa_{k_7k_8l_4}u_{n_1k_1}u_{n_2k_3}u_{i_1k_5}u_{i_2k_7}\\
&\times v_{n_3k_6}v_{n_4k_8}v_{j_1k_2}v_{j_2k_4}\,,
\end{split}
\end{equation}
where we have used Einstein's notation for the sums. To simplify the scaling analysis, we will assume that each $n_i$ index takes $N$ values, the $k_i$ indices take $s_1$ values, the $l_i$ indices, $s_2$ values and the $m_i$ indices, $s_4$ values. We also assume no memory limitation. Now, after summing over $n_1$, we obtain
\begin{equation}
\begin{split}
R_{j_1j_2}^{i_1i_2}=&(Vu)_{n_2n_3n_4}^{k_1}\mu_{m_1m_2}\lambda_{l_1l_2m_1}\lambda_{l_3l_4m_2}\kappa_{k_1k_2l_1}\kappa_{k_3k_4l_2}\\
&\times\kappa_{k_5k_6l_3}\kappa_{k_7k_8l_4}u_{n_2k_3}u_{i_1k_5}u_{i_2k_7}\\
&\times v_{n_3k_6}v_{n_4k_8}v_{j_1k_2}v_{j_2k_4}\,,
\end{split}
\end{equation}
where each of the $N^3s_1$ elements of $(Vu)$ takes $O(N)$ operations to compute, and thus $(Vu)$ takes $O(N^4s_1)$ operations to compute. Then, after summing over $n_2$, we obtain
\begin{equation}
\begin{split}
R_{j_1j_2}^{i_1i_2}=&(Vu^2)_{n_3n_4}^{k_1k_3}\mu_{m_1m_2}\lambda_{l_1l_2m_1}\lambda_{l_3l_4m_2}\kappa_{k_1k_2l_1}\kappa_{k_3k_4l_2}\\
&\times\kappa_{k_5k_6l_3}\kappa_{k_7k_8l_4}u_{i_1k_5}u_{i_2k_7}\\
&\times v_{n_3k_6}v_{n_4k_8}v_{j_1k_2}v_{j_2k_4}\,,
\end{split}
\end{equation}
where $(Vu^2)$ takes $O(N^3s_1^2)$ additional operations to compute. Similarly, the summation over $n_3$ and $n_4$ take respectively $O(N^2s_1^3)$ and $O(Ns_1^4)$ additional operations. We then obtain
\begin{equation}
\begin{split}
R_{j_1j_2}^{i_1i_2}=&F_{k_6k_8}^{k_1k_3}\mu_{m_1m_2}\lambda_{l_1l_2m_1}\lambda_{l_3l_4m_2}\kappa_{k_1k_2l_1}\kappa_{k_3k_4l_2}\\
&\times\kappa_{k_5k_6l_3}\kappa_{k_7k_8l_4}u_{i_1k_5}u_{i_2k_7}v_{j_1k_2}v_{j_2k_4}\,,
\end{split}
\end{equation}
where $F$ thus takes $O(N^4s_1)$ operations to compute, since $N\geq s_1$. Then, summing over $k_1$, we obtain
\begin{equation}
\begin{split}
R_{j_1j_2}^{i_1i_2}=&(F\kappa)_{k_6k_8}^{k_2l_1,k_3}\mu_{m_1m_2}\lambda_{l_1l_2m_1}\lambda_{l_3l_4m_2}\kappa_{k_3k_4l_2}\\
&\times\kappa_{k_5k_6l_3}\kappa_{k_7k_8l_4}u_{i_1k_5}u_{i_2k_7}v_{j_1k_2}v_{j_2k_4}\,,
\end{split}
\end{equation}
where $(F\kappa)$ takes $O(s_1^5s_2)$ additional operations to compute. After the summation over $k_3$, we obtain
\begin{equation}
\begin{split}
R_{j_1j_2}^{i_1i_2}=&(F\kappa^2)_{k_6k_8}^{k_2l_1,k_4l_2}\mu_{m_1m_2}\lambda_{l_1l_2m_1}\lambda_{l_3l_4m_2}\\
&\times\kappa_{k_5k_6l_3}\kappa_{k_7k_8l_4}u_{i_1k_5}u_{i_2k_7}v_{j_1k_2}v_{j_2k_4}\,,
\end{split}
\end{equation}
where $(F\kappa^2)$ takes $O(s_1^5s_2^2)$ additional operations to compute. Similarly, the sums over $k_6$ and $k_8$ take respectively $O(s_1^5s_2^3)$ and $O(s_1^5s_2^4)$ additional operations to compute. We then obtain
\begin{equation}
\begin{split}
R_{j_1j_2}^{i_1i_2}=&G_{k_5l_3,k_7l_4}^{k_2l_1,k_4l_2}\mu_{m_1m_2}\lambda_{l_1l_2m_1}\lambda_{l_3l_4m_2}u_{i_1k_5}u_{i_2k_7}\\
&\times v_{j_1k_2}v_{j_2k_4}\,,
\end{split}
\end{equation}
where $G$ takes $O(s_1^5s_2^4)$ additional operations to compute. Now, if we the sum over $k_2$, we obtain
\begin{equation}
\begin{split}
R_{j_1j_2}^{i_1i_2}=&(Gv)_{k_5l_3,k_7l_4}^{j_1l_1,k_4l_2}\mu_{m_1m_2}\lambda_{l_1l_2m_1}\lambda_{l_3l_4m_2}u_{i_1k_5}u_{i_2k_7}v_{j_2k_4}\,,
\end{split}
\end{equation}
where $(Gv)$ takes $O(Ns_1^4s_2^4)$ additional operations to compute. Then, after summing over $k_4$, we obtain
\begin{equation}
\begin{split}
R_{j_1j_2}^{i_1i_2}=&(Gv^2)_{k_5l_3,k_7l_4}^{j_1l_1,j_2l_2}\mu_{m_1m_2}\lambda_{l_1l_2m_1}\lambda_{l_3l_4m_2}u_{i_1k_5}u_{i_2k_7}\,,
\end{split}
\end{equation}
where $(Gv^2)$ takes $O(N^2s_1^3s_2^4)$ additional operations to compute. Similarly, the sums over $k_5$ and $k_7$ take respectively $O(N^3s_1^2s_2^4)$ and $O(N^4s_1s_2^4)$ additional operations. We thus obtain
\begin{equation}
\begin{split}
R_{j_1j_2}^{i_1i_2}=&J_{i_1l_3,i_2l_4}^{j_1l_1,j_2l_2}\mu_{m_1m_2}\lambda_{l_1l_2m_1}\lambda_{l_3l_4m_2}\,,
\end{split}
\end{equation}
where $J$ takes $O(N^4s_1s_2^4)$ additional operations to compute. Now, after summing over $l_1$ and $l_2$, we obtain
\begin{equation}
\begin{split}
R_{j_1j_2}^{i_1i_2}=&(J\lambda)_{i_1l_3,i_2l_4}^{j_1j_2m_1}\mu_{m_1m_2}\lambda_{l_3l_4m_2}\,,
\end{split}
\end{equation}
where $(J\lambda)$ takes $O(N^4s_2^4s_4)$ additional operations to compute, and after summing over $l_3$ and $l_4$, we obtain
\begin{equation}
\begin{split}
R_{j_1j_2}^{i_1i_2}=&(J\lambda^2)_{i_1i_2m_2}^{j_1j_2m_1}\mu_{m_1m_2}\,,
\end{split}
\end{equation}
where $(J\lambda^2)$ takes $O(N^4s_2^2s_4^2)$ additional operations to compute. Finally, the sums over $m_1$ and $m_2$ to obtain $R$ take $O(N^4s_4^2)$ additional operations. 

We therefore obtain $O(s_1^5s_2^4)+O(N^4s_1s_2^4)+O(N^4s_2^4s_4)+O(N^4s_2^2s_4^2)=O(N^4s_1s_2^4)+O(N^4s_2^4s_4)+O(N^4s_2^2s_4^2)$ operations in total, where the simplification comes from using $N\geq s_1$, and we cannot simplify further without knowing the complexity of $s_1$, $s_2$ and $s_4$.

%

\clearpage
\setcounter{section}{0}
\setcounter{equation}{0}

\begin{widetext}
\begin{center}
\hypertarget{suppl_mat}{\Large\bfseries Supplemental material}
\end{center}

\renewcommand{\appendixname}{}
\renewcommand{\thesection}{\arabic{section}}
\renewcommand{\theequation}{\thesection.\arabic{equation}}

\titleformat*{\section}{\bfseries\raggedright}

\section{Derivation of the generalized CC equations for the double-excitation projection SD}\label{sec:suppl_mat}

Here we derive the generalized CC equations for the projection SD's $\bra{\phi_{j_1j_2}^{i_1i_2}}$ using the operator
\begin{subequations}\label{eq:T_SCC_suppl}
\begin{equation}\label{eq:T_SCC_D_op_suppl}
T=\sum_{i,j} a_i^\dagger a_j \hat{D}_j^i
\end{equation}
where $\hat{D}_j^i$ is defined as
\begin{equation}\label{eq:eig_Eqn_D_op_suppl}
\begin{split}
\hat{D}_{j_k}^{i_k}\, \ket{\phi_{j_1j_2\ldots j_{k-1}}^{i_1i_2\ldots i_{k-1}}}=\frac{1}{k}\frac{c_{j_1j_2\ldots j_k}^{i_1i_2\ldots i_k}}{c_{j_1j_2\ldots j_{k-1}}^{i_1i_2\ldots i_{k-1}}}\,  \ket{\phi_{j_1j_2\ldots j_{k-1}}^{i_1i_2\ldots i_{k-1}}}\,,
\end{split}
\end{equation}
\end{subequations}
the excited particle and hole operators
\begin{subequations}\label{eq:def_p_i_h_i_suppl}
\begin{equation}\label{eq:def_p_i_suppl}
p_i^\dagger =(1-n_i^{\phi})a_i^\dagger\,,
\end{equation}
\begin{equation}\label{eq:def_h_i_suppl}
h_i^\dagger=n_i^{\phi} a_i\,,
\end{equation}
\end{subequations}
respectively, and the hamiltonian
\begin{equation}
\begin{split}
\hat{H}^{\phi}&=\hat{H}-\bra{\phi}\hat{H}\ket{\phi}\\
&=\sum_{ij} t^{\phi}_{ij}p_{i}^\dagger p_{j}-\sum_{ij} t^{\phi}_{ij}h_{i}^\dagger h_{j}+\sum_{ ij} t^{\phi}_{ij}\left(p_{i}^\dagger h_{j}^\dagger+h.c.\right)\\
&+\frac{1}{4}\sum_{ijkl} V_{ijkl} p_i^\dagger p_j^\dagger p_k p_l +\frac{1}{4}\sum_{ijkl} V_{ijkl}h_i^\dagger h_j^\dagger h_k h_l+\frac{1}{4}\sum_{ ijkl} V_{ijkl}(p_i^\dagger p_j^\dagger h_k^\dagger h_l^\dagger+h.c.)\\
& +\frac{1}{2}\sum_{ ijkl} V_{ijkl}(p_i^\dagger p_j^\dagger h_k^\dagger p_l+h.c.)+\frac{1}{2}\sum_{ ijkl} V_{ijkl}(h_i^\dagger h_j^\dagger p_k^\dagger h_l+h.c.)- \sum_{ijkl} V_{ikjl}p_i^\dagger h_j^\dagger h_k p_l\,,
\end{split}
\end{equation}
where $V_{ijkl}$ is the antisymmetrized Coulomb interaction $V_{ijkl}=V_{ijkl}^c-V_{ijlk}^c$, where $V_{ijkl}^c$ are Coulomb integrals.

Taking into account the facts that $T$ creates a single particle-hole pair, $H$ can create or destroy at most two pairs, and $\bra{\phi}\hat{H}^{\phi}\ket{\phi}=0$, the equations written in terms of $H$ and $T$ are
\begin{equation}\label{eq:eqn_phi_2_suppl}
\begin{split}
0&=\bra{\phi_{j_1j_2}^{i_1i_2}}e^{-T}\hat{H}^{\phi}e^T\ket{\phi}\\
&=\bra{\phi_{j_1j_2}^{i_1i_2}}\left(1-T+\frac{1}{2}T^2\right)\hat{H}^{\phi}\left(1+T+\frac{1}{2}T^2+\frac{1}{3!}T^3+\frac{1}{4!}T^4\right)\ket{\phi}\\
0&=\bra{\phi_{j_1j_2}^{i_1i_2}}\hat{H}^{\phi}\ket{\phi}+\bra{\phi_{j_1j_2}^{i_1i_2}}\hat{H}^{\phi}T\ket{\phi}-\bra{\phi_{j_1j_2}^{i_1i_2}}T\hat{H}^{\phi}\ket{\phi}\\
&+\frac{1}{2}\bra{\phi_{j_1j_2}^{i_1i_2}}\hat{H}^{\phi}T^2\ket{\phi}-\bra{\phi_{j_1j_2}^{i_1i_2}}T\hat{H}^{\phi}T\ket{\phi}+\frac{1}{3!}\bra{\phi_{j_1j_2}^{i_1i_2}}\hat{H}^{\phi}T^3\ket{\phi}\\
&-\frac{1}{2}\bra{\phi_{j_1j_2}^{i_1i_2}}T\hat{H}^{\phi}T^2\ket{\phi}+\frac{1}{2}\bra{\phi_{j_1j_2}^{i_1i_2}}T^2\hat{H}^{\phi}T\ket{\phi}+\frac{1}{4!}\bra{\phi_{j_1j_2}^{i_1i_2}}\hat{H}^{\phi}T^4\ket{\phi}\\
&-\frac{1}{3!}\bra{\phi_{j_1j_2}^{i_1i_2}}T\hat{H}^{\phi}T^3\ket{\phi}+\frac{1}{4}\bra{\phi_{j_1j_2}^{i_1i_2}}T^2\hat{H}^{\phi}T^2\ket{\phi}\,.
\end{split}
\end{equation}

The term $\bra{\phi_{j_1j_2}^{i_1i_2}}\hat{H}^{\phi}\ket{\phi}$ involves the part of $H$ that creates two pairs:
\begin{equation}
\begin{split}
\bra{\phi_{j_1j_2}^{i_1i_2}}\hat{H}^{\phi}\ket{\phi}&=\frac{1}{4}\sum_{ klmn} V_{klmn}\bra{\phi}h_{j_2}p_{i_2}h_{j_1}p_{i_1}p_k^\dagger p_l^\dagger h_m^\dagger h_n^\dagger\ket{\phi}\\
&=-\frac{1}{4}\sum_{ klmn} V_{klmn}\bra{\phi}h_{j_2}h_{j_1}h_m^\dagger h_n^\dagger p_{i_2}p_{i_1}p_k^\dagger p_l^\dagger \ket{\phi}\\
&=-\frac{1}{4}\sum_{ klmn} V_{klmn} \left(\delta_{j_1m}\delta_{j_2n}-\delta_{j_1n}\delta_{j_2m}\right)\left(\delta_{i_1k}\delta_{i_2l}-\delta_{i_1l}\delta_{i_2k}\right)\\
\end{split}
\end{equation}
which yields, using the antisymmetry of $V_{klmn}$,
\begin{equation}
\boxed{\bra{\phi_{j_1j_2}^{i_1i_2}}\hat{H}^{\phi}\ket{\phi}=-V_{i_1i_2j_1j_2}\,.}
\end{equation}

The term $\bra{\phi_{j_1j_2}^{i_1i_2}}\hat{H}^{\phi}T\ket{\phi}$ involves all the terms of $H$ creating one pair:
\begin{equation}
\begin{split}
\bra{\phi_{j_1j_2}^{i_1i_2}}&\hat{H}^{\phi}T\ket{\phi}\\
&=\bra{\phi}h_{j_2}p_{i_2}h_{j_1}p_{i_1}\left(\sum_{kl} t^{\phi}_{kl}p_{k}^\dagger h_{l}^\dagger + \frac{1}{2}\sum_{klmn} V_{klmn}p_k^\dagger p_l^\dagger h_m^\dagger p_n+\frac{1}{2}\sum_{klmn} V_{klmn}h_k^\dagger h_l^\dagger p_m^\dagger h_n\right)\sum_{qr} c_{r}^{q} p_q^\dagger h_r^\dagger \ket{\phi}\\
&=\sum_{klqr} t^{\phi}_{kl}c_{r}^{q}\bra{\phi}h_{j_2}p_{i_2}h_{j_1}p_{i_1}p_{k}^\dagger h_{l}^\dagger p_q^\dagger h_r^\dagger\ket{\phi}+\frac{1}{2}\sum_{klmnqr} V_{klmn}c_{r}^{q}\bra{\phi}h_{j_2}p_{i_2}h_{j_1}p_{i_1}p_k^\dagger p_l^\dagger h_m^\dagger p_n p_q^\dagger h_r^\dagger \ket{\phi}\\
&+\frac{1}{2}\sum_{klmnqr} V_{klmn}c_{r}^{q}\bra{\phi}h_{j_2}p_{i_2}h_{j_1}p_{i_1}h_k^\dagger h_l^\dagger p_m^\dagger h_n p_q^\dagger h_r^\dagger \ket{\phi}\\
&=\sum_{klqr} t^{\phi}_{kl}c_{r}^{q}\bra{\phi}h_{j_2}h_{j_1}h_{l}^\dagger h_r^\dagger p_{i_2}p_{i_1}p_{k}^\dagger  p_q^\dagger \ket{\phi}-\frac{1}{2}\sum_{klmnqr} V_{klmn}c_{r}^{q}\bra{\phi}h_{j_2}h_{j_1} h_m^\dagger h_r^\dagger p_{i_2}p_{i_1}p_k^\dagger p_l^\dagger  p_n p_q^\dagger  \ket{\phi}\\
&+\frac{1}{2}\sum_{klmnqr} V_{klmn}c_{r}^{q}\bra{\phi}h_{j_2}h_{j_1} h_k^\dagger h_l^\dagger h_n h_r^\dagger p_{i_2}p_{i_1} p_m^\dagger  p_q^\dagger  \ket{\phi}\\
&=\sum_{klqr} t^{\phi}_{kl}c_{r}^{q}\left(\delta_{j_1l}\delta_{j_2r}-\delta_{j_1r}\delta_{j_2l}\right)\left(\delta_{i_1k}\delta_{i_2q}-\delta_{i_1q}\delta_{i_2k}\right)-\frac{1}{2}\sum_{klmqr} V_{klmq}c_{r}^{q}\bra{\phi}h_{j_2}h_{j_1} h_m^\dagger h_r^\dagger p_{i_2}p_{i_1}p_k^\dagger p_l^\dagger    \ket{\phi}\\
&+\frac{1}{2}\sum_{klmqr} V_{klmr}c_{r}^{q}\bra{\phi}h_{j_2}h_{j_1} h_k^\dagger h_l^\dagger  p_{i_2}p_{i_1} p_m^\dagger  p_q^\dagger  \ket{\phi}\\
&=\left(t^{\phi}_{i_1j_1}c_{j_2}^{i_2}-t^{\phi}_{i_1j_2}c_{j_1}^{i_2}-t^{\phi}_{i_2j_1}c_{j_2}^{i_1}+t^{\phi}_{i_2j_2}c_{j_1}^{i_1}\right)\\
&-\frac{1}{2}\sum_{klmqr} V_{klmq}c_{r}^{q}\left(\delta_{j_1m}\delta_{j_2r}-\delta_{j_1r}\delta_{j_2m}\right)\left(\delta_{i_1k}\delta_{i_2l}-\delta_{i_1l}\delta_{i_2k}\right)\\
&+\frac{1}{2}\sum_{klmqr} V_{klmr}c_{r}^{q}\left(\delta_{j_1k}\delta_{j_2l}-\delta_{j_1l}\delta_{j_2k}\right)\left(\delta_{i_1m}\delta_{i_2q}-\delta_{i_1q}\delta_{i_2m}\right)\\
&=\left(t^{\phi}_{i_1j_1}c_{j_2}^{i_2}-t^{\phi}_{i_1j_2}c_{j_1}^{i_2}-t^{\phi}_{i_2j_1}c_{j_2}^{i_1}+t^{\phi}_{i_2j_2}c_{j_1}^{i_1}\right)-\sum_{mqr} V_{i_1i_2mq}c_{r}^{q}\left(\delta_{j_1m}\delta_{j_2r}-\delta_{j_1r}\delta_{j_2m}\right)\\
&+\sum_{mqr} V_{j_1j_2mr}c_{r}^{q}\left(\delta_{i_1m}\delta_{i_2q}-\delta_{i_1q}\delta_{i_2m}\right)\\
\end{split}
\end{equation}
\begin{equation}
\boxed{
\begin{split}
\bra{\phi_{j_1j_2}^{i_1i_2}}\hat{H}^{\phi}T\ket{\phi}&=t^{\phi}_{i_1j_1}c_{j_2}^{i_2}-t^{\phi}_{i_1j_2}c_{j_1}^{i_2}-t^{\phi}_{i_2j_1}c_{j_2}^{i_1}+t^{\phi}_{i_2j_2}c_{j_1}^{i_1}-\sum_{q} \left(V_{i_1i_2j_1q}c_{j_2}^{q}-V_{i_1i_2j_2q}c_{j_1}^{q}\right)\\
&\quad+\sum_{r} \left(V_{j_1j_2i_1r}c_{r}^{i_2}-V_{j_1j_2i_2r}c_{r}^{i_1}\right)\,.
\end{split}}
\end{equation}

The term $\bra{\phi_{j_1j_2}^{i_1i_2}}T\hat{H}^{\phi}\ket{\phi}$ involves only the one-body part of $H$ creating one pair:
\begin{equation}
\begin{split}
\bra{\phi_{j_1j_2}^{i_1i_2}}T\hat{H}^{\phi}\ket{\phi}&=\bra{\phi}h_{j_2}p_{i_2}h_{j_1}p_{i_1}\sum_{kl} p_k^\dagger h_l^\dagger \hat{D}_l^k\sum_{qr} t^{\phi}_{qr}p_{q}^\dagger h_{r}^\dagger\ket{\phi}\\
&=\frac{1}{2}\sum_{klqr}\frac{c_{lr}^{kq}}{c_r^q}t^{\phi}_{qr}\bra{\phi}h_{j_2}p_{i_2}h_{j_1}p_{i_1}p_k^\dagger h_l^\dagger p_{q}^\dagger h_{r}^\dagger\ket{\phi}\\
&=\frac{1}{2}\sum_{klqr}\frac{c_{lr}^{kq}}{c_r^q}t^{\phi}_{qr}\bra{\phi}h_{j_2}h_{j_1}h_l^\dagger h_{r}^\dagger p_{i_2}p_{i_1}p_k^\dagger  p_{q}^\dagger \ket{\phi}\\
&=\frac{1}{2}\sum_{klqr}\frac{c_{lr}^{kq}}{c_r^q}t^{\phi}_{qr}\left(\delta_{j_1l}\delta_{j_2r}-\delta_{j_1r}\delta_{j_2l}\right)\left(\delta_{i_1k}\delta_{i_2q}-\delta_{i_1q}\delta_{i_2k}\right)\\
&=\frac{1}{2}\sum_{klqr}\frac{c_{lr}^{kq}}{c_r^q}t^{\phi}_{qr}\left(\delta_{j_1l}\delta_{j_2r}\delta_{i_1k}\delta_{i_2q}-\delta_{j_1l}\delta_{j_2r}\delta_{i_1q}\delta_{i_2k}-\delta_{j_1r}\delta_{j_2l}\delta_{i_1k}\delta_{i_2q}+\delta_{j_1r}\delta_{j_2l}\delta_{i_1q}\delta_{i_2k}\right)\\
&=\frac{1}{2}\left(\frac{c_{j_1j_2}^{i_1i_2}}{c_{j_2}^{i_2}}t^{\phi}_{i_2j_2}-\frac{c_{j_1j_2}^{i_2i_1}}{c_{j_2}^{i_1}}t^{\phi}_{i_1j_2}-\frac{c_{j_2j_1}^{i_1i_2}}{c_{j_1}^{i_2}}t^{\phi}_{i_2j_1}+\frac{c_{j_2j_1}^{i_2i_1}}{c_{j_1}^{i_1}}t^{\phi}_{i_1j_1}\right)\\
\end{split}
\end{equation}
and thus, using the antisymmetry of $c_{j_1j_2}^{i_1i_2}$,
\begin{equation}
\boxed{
-\bra{\phi_{j_1j_2}^{i_1i_2}}T\hat{H}^{\phi}\ket{\phi}=-\frac{1}{2}c_{j_1j_2}^{i_1i_2}\left(\frac{t^{\phi}_{i_2j_2}}{c_{j_2}^{i_2}}+\frac{t^{\phi}_{i_1j_2}}{c_{j_2}^{i_1}}+\frac{t^{\phi}_{i_2j_1}}{c_{j_1}^{i_2}}+\frac{t^{\phi}_{i_1j_1}}{c_{j_1}^{i_1}}\right)\,.}
\end{equation}

In $\bra{\phi_{j_1j_2}^{i_1i_2}}\hat{H}^{\phi}T^2\ket{\phi}$, all the terms that do not create or destroy any pair contribute:
\begin{equation}
\begin{split}
\bra{\phi_{j_1j_2}^{i_1i_2}}&\hat{H}^{\phi}T^2\ket{\phi}\\
&=\bra{\phi}h_{j_2}p_{i_2}h_{j_1}p_{i_1}\Bigg( \sum_{kl} t^{\phi}_{kl}p_{k}^\dagger p_{l}-\sum_{kl} t^{\phi}_{kl}h_{k}^\dagger h_{l}+\frac{1}{4}\sum_{klmn} V_{klmn} p_k^\dagger p_l^\dagger p_m p_n \\
&+\frac{1}{4}\sum_{klmn} V_{klmn} h_k^\dagger h_l^\dagger h_m h_n- \sum_{klmn} V_{kmln} p_k^\dagger h_{l}^\dagger h_{m} p_{n}\Bigg)\frac{1}{2}\sum_{qrst} c_{rt}^{qs} p_q^\dagger h_r^\dagger p_s^\dagger h_t^\dagger\ket{\phi}\\
&=\frac{1}{2}\sum_{klqrst} t^{\phi}_{kl}c_{rt}^{qs}\bra{\phi}h_{j_2}p_{i_2}h_{j_1}p_{i_1} p_{k}^\dagger p_{l}  p_q^\dagger h_r^\dagger p_s^\dagger h_t^\dagger\ket{\phi}-\frac{1}{2}\sum_{klqrst}t^{\phi}_{kl} c_{rt}^{qs}\bra{\phi}h_{j_2}p_{i_2}h_{j_1}p_{i_1} h_{k}^\dagger h_{l} p_q^\dagger h_r^\dagger p_s^\dagger h_t^\dagger\ket{\phi}\\
&+\frac{1}{8}\sum_{klmnqrst} V_{klmn} c_{rt}^{qs}\bra{\phi}h_{j_2}p_{i_2}h_{j_1}p_{i_1} p_k^\dagger p_l^\dagger p_m p_n p_q^\dagger h_r^\dagger p_s^\dagger h_t^\dagger\ket{\phi}\\
&+\frac{1}{8}\sum_{klmnqrst} V_{klmn} c_{rt}^{qs}\bra{\phi}h_{j_2}p_{i_2}h_{j_1}p_{i_1} h_k^\dagger h_l^\dagger h_m h_n p_q^\dagger h_r^\dagger p_s^\dagger h_t^\dagger\ket{\phi}\\
&-\frac{1}{2}\sum_{klmnqrst} V_{kmln} c_{rt}^{qs}\bra{\phi}h_{j_2}p_{i_2}h_{j_1}p_{i_1} p_k^\dagger h_{l}^\dagger h_{m} p_{n} p_q^\dagger h_r^\dagger p_s^\dagger h_t^\dagger\ket{\phi}\\
\end{split}
\end{equation}
\begin{equation}
\begin{split}
\bra{\phi_{j_1j_2}^{i_1i_2}}&\hat{H}^{\phi}T^2\ket{\phi}\\
&=\frac{1}{2}\sum_{klqrst}t^{\phi}_{kl}c_{rt}^{qs}\bra{\phi}h_{j_2}h_{j_1}h_r^\dagger h_t^\dagger p_{i_2}p_{i_1}  p_{k}^\dagger p_{l} p_q^\dagger   p_s^\dagger \ket{\phi}-\frac{1}{2}\sum_{klqrst}t^{\phi}_{kl}c_{rt}^{qs}\bra{\phi}h_{j_2}h_{j_1} h_{k}^\dagger h_{l}  h_r^\dagger h_t^\dagger p_{i_2}p_{i_1} p_q^\dagger p_s^\dagger \ket{\phi}\\
&+\frac{1}{8}\sum_{klmnqrst}V_{klmn}c_{rt}^{qs}\bra{\phi}h_{j_2}h_{j_1}h_r^\dagger h_t^\dagger p_{i_2}p_{i_1} p_k^\dagger p_l^\dagger p_m p_n p_q^\dagger  p_s^\dagger \ket{\phi}\\
&+\frac{1}{8}\sum_{klmnqrst}V_{klmn}c_{rt}^{qs}\bra{\phi}h_{j_2}h_{j_1}h_k^\dagger h_l^\dagger h_m h_n h_r^\dagger h_t^\dagger p_{i_2}p_{i_1} p_q^\dagger p_s^\dagger \ket{\phi}\\
&-\frac{1}{2}\sum_{klmnqrst} V_{kmln}c_{rt}^{qs} \bra{\phi}h_{j_2}h_{j_1}h_{l}^\dagger h_{m} h_r^\dagger h_t^\dagger p_{i_2} p_{i_1}p_k^\dagger p_{n} p_q^\dagger p_s^\dagger \ket{\phi}\\
\end{split}
\end{equation}
\begin{equation}
\begin{split}
\bra{\phi_{j_1j_2}^{i_1i_2}}&\hat{H}^{\phi}T^2\ket{\phi}\\
&=\frac{1}{2}\sum_{klqrst}t^{\phi}_{kl}c_{rt}^{qs}\left(\delta_{j_1r}\delta_{j_2t}-\delta_{j_1t}\delta_{j_2r}\right)\left[\delta_{i_1k}(\delta_{lq}\delta_{i_2s}-\delta_{ls}\delta_{i_2q})-\delta_{i_2k}(\delta_{lq}\delta_{i_1s}-\delta_{ls}\delta_{i_1q})\right]\\
&-\frac{1}{2}\sum_{klqrst}t^{\phi}_{kl}c_{rt}^{qs}[\delta_{j_1k}(\delta_{lr}\delta_{j_2t}-\delta_{lt}\delta_{j_2r})-\delta_{j_2k}(\delta_{lr}\delta_{j_1t}-\delta_{lt}\delta_{j_1r})](\delta_{i_1q}\delta_{i_2s}-\delta_{i_1s}\delta_{i_2q})\\
&+\frac{1}{8}\sum_{klmnqrst}V_{klmn}c_{rt}^{qs}(\delta_{j_1r}\delta_{j_2t}-\delta_{j_1t}\delta_{j_2r})(\delta_{i_1k}\delta_{i_2l}-\delta_{i_1l}\delta_{i_2k})(\delta_{nq}\delta_{ms}-\delta_{ns}\delta_{mq})\\
&+\frac{1}{8}\sum_{klmnqrst}V_{klmn}c_{rt}^{qs}(\delta_{j_1k}\delta_{j_2l}-\delta_{j_1l}\delta_{j_2k})(\delta_{nr}\delta_{mt}-\delta_{nt}\delta_{mr})(\delta_{i_1q}\delta_{i_2s}-\delta_{i_1s}\delta_{i_2q})\\
&-\frac{1}{2}\sum_{klmnqrst} V_{kmln}c_{rt}^{qs} [\delta_{j_1l}(\delta_{mr}\delta_{j_2t}-\delta_{mt}\delta_{j_2r})-\delta_{j_2l}(\delta_{mr}\delta_{j_1t}-\delta_{mt}\delta_{j_1r})]\\
&\hspace{2in}\times[\delta_{i_1k}(\delta_{nq}\delta_{i_2s}-\delta_{ns}\delta_{i_2q})-\delta_{i_2k}(\delta_{nq}\delta_{i_1s}-\delta_{ns}\delta_{i_1q})]
\end{split}
\end{equation}
which becomes, using the antisymmetry of $c_{rt}^{qs}$,
\begin{equation}
\begin{split}
\bra{\phi_{j_1j_2}^{i_1i_2}}\hat{H}^{\phi}T^2\ket{\phi}
&=2\sum_{klqrst}t^{\phi}_{kl}c_{rt}^{qs}\delta_{j_1r}\delta_{j_2t}(\delta_{i_1k}\delta_{lq}\delta_{i_2s}+\delta_{i_2k}\delta_{ls}\delta_{i_1q})\\
&-2\sum_{klqrst}t^{\phi}_{kl}c_{rt}^{qs}\delta_{i_1q}\delta_{i_2s}(\delta_{j_1k}\delta_{lr}\delta_{j_2t}+\delta_{j_2k}\delta_{lt}\delta_{j_1r})\\
&-\sum_{klmnqrst}V_{klmn}c_{rt}^{qs}\delta_{j_1r}\delta_{j_2t}\delta_{i_1k}\delta_{i_2l}\delta_{ns}\delta_{mq}\\
&-\sum_{klmnqrst}V_{klmn}c_{rt}^{qs}\delta_{j_1k}\delta_{j_2l}\delta_{nt}\delta_{mr}\delta_{i_1q}\delta_{i_2s}\\
&-2\sum_{klmnqrst} V_{kmln}c_{rt}^{qs} (\delta_{j_1l}\delta_{mr}\delta_{j_2t}+\delta_{j_2l}\delta_{mt}\delta_{j_1r})(\delta_{i_1k}\delta_{nq}\delta_{i_2s}+\delta_{i_2k}\delta_{ns}\delta_{i_1q})
\end{split}
\end{equation}
and thus,
\begin{equation}
\boxed{
\begin{split}
\frac{1}{2}\bra{\phi_{j_1j_2}^{i_1i_2}}\hat{H}^{\phi}T^2\ket{\phi}
&=\sum_{l}\left(t^{\phi}_{i_1l}c_{j_1j_2}^{li_2} +t^{\phi}_{i_2l}c_{j_1j_2}^{i_1l}\right)-\sum_{l}\left(t^{\phi}_{j_1l}c_{lj_2}^{i_1i_2}+t^{\phi}_{j_2l}c_{j_1l}^{i_1i_2}\right)\\
&-\frac{1}{2}\sum_{mn}V_{i_1i_2mn}c_{j_1j_2}^{mn}-\frac{1}{2}\sum_{mn}V_{j_1j_2mn}c_{mn}^{i_1i_2}\\
&-\sum_{mn} \left(V_{i_1mj_1n}c_{mj_2}^{ni_2}+V_{i_2mj_1n}c_{mj_2}^{i_1n}+V_{i_1mj_2n}c_{j_1m}^{ni_2}+V_{i_2mj_2n}c_{j_1m}^{i_1n}\right)\,.
\end{split}}
\end{equation}

Now, in $\bra{\phi_{j_1j_2}^{i_1i_2}}T\hat{H}^{\phi}T\ket{\phi}$, only the terms acting on a single pair and that do not create or destroy any pair contribute:
\begin{equation}
\begin{split}
\bra{\phi_{j_1j_2}^{i_1i_2}}&T\hat{H}^{\phi}T\ket{\phi}\\
&=\bra{\phi}h_{j_2}p_{i_2}h_{j_1}p_{i_1}\sum_{kl} p_k^\dagger h_l^\dagger \hat{D}_{l}^k\\
&\times \left( \sum_{qr} t^{\phi}_{qr}p_{q}^\dagger p_{r}-\sum_{qr} t^{\phi}_{qr}h_{q}^\dagger h_{r}- \sum_{qq_1rr_1} V_{qr_1rq_1}p_q^\dagger h_{r}^\dagger h_{r_1} p_{q_1}\right)\sum_{st} c_{t}^{s} p_s^\dagger h_t^\dagger \ket{\phi}\\
&=\bra{\phi}h_{j_2}p_{i_2}h_{j_1}p_{i_1}\sum_{kl} p_k^\dagger h_l^\dagger \hat{D}_{l}^k\sum_{qr} t^{\phi}_{qr}p_{q}^\dagger p_{r}\sum_{st} c_{t}^{s} p_s^\dagger h_t^\dagger \ket{\phi}\\
&-\bra{\phi}h_{j_2}p_{i_2}h_{j_1}p_{i_1}\sum_{kl} p_k^\dagger h_l^\dagger \hat{D}_{l}^k\sum_{qr} t^{\phi}_{qr}h_{q}^\dagger h_{r}\sum_{st} c_{t}^{s} p_s^\dagger h_t^\dagger \ket{\phi}\\
&-\bra{\phi}h_{j_2}p_{i_2}h_{j_1}p_{i_1}\sum_{kl} p_k^\dagger h_l^\dagger \hat{D}_{l}^k\sum_{qq_1rr_1} V_{qr_1rq_1}p_q^\dagger h_{r}^\dagger h_{r_1} p_{q_1}\sum_{st} c_{t}^{s} p_s^\dagger h_t^\dagger\ket{\phi}\\
&=\sum_{klqrst} t^{\phi}_{qr}c_{t}^{s}\bra{\phi}h_{j_2}p_{i_2}h_{j_1}p_{i_1} p_k^\dagger h_l^\dagger \hat{D}_{l}^kp_{q}^\dagger p_{r}  p_s^\dagger h_t^\dagger \ket{\phi}\\
&-\sum_{klqrst} t^{\phi}_{qr} c_{t}^{s}\bra{\phi}h_{j_2}p_{i_2}h_{j_1}p_{i_1} p_k^\dagger h_l^\dagger \hat{D}_{l}^kh_{q}^\dagger h_{r} p_s^\dagger h_t^\dagger \ket{\phi}\\
&-\sum_{klqq_1rr_1st} V_{qr_1rq_1} c_{t}^{s}\bra{\phi}h_{j_2}p_{i_2}h_{j_1}p_{i_1} p_k^\dagger h_l^\dagger \hat{D}_{l}^kp_q^\dagger h_{r}^\dagger h_{r_1} p_{q_1} p_s^\dagger h_t^\dagger\ket{\phi}\\
&=\sum_{klqrst} t^{\phi}_{qr}c_{t}^{s}\delta_{rs}\bra{\phi}h_{j_2}p_{i_2}h_{j_1}p_{i_1} p_k^\dagger h_l^\dagger \hat{D}_{l}^kp_{q}^\dagger h_t^\dagger \ket{\phi}\\
&-\sum_{klqrst} t^{\phi}_{qr} c_{t}^{s}\delta_{rt}\bra{\phi}h_{j_2}p_{i_2}h_{j_1}p_{i_1} p_k^\dagger h_l^\dagger \hat{D}_{l}^k p_s^\dagger h_{q}^\dagger \ket{\phi}\\
&-\sum_{klqq_1rr_1st} V_{qr_1rq_1} c_{t}^{s}\delta_{q_1s}\delta_{r_1t}\bra{\phi}h_{j_2}p_{i_2}h_{j_1}p_{i_1} p_k^\dagger h_l^\dagger \hat{D}_{l}^kp_q^\dagger h_{r}^\dagger \ket{\phi}\\
\end{split}
\end{equation}
\begin{equation}
\begin{split}
\bra{\phi_{j_1j_2}^{i_1i_2}}T\hat{H}^{\phi}T\ket{\phi}&=\frac{1}{2}\sum_{klqrt}\frac{c_{lt}^{kq}}{c_t^q}t^{\phi}_{qr}c_{t}^{r}\bra{\phi}h_{j_2}p_{i_2}h_{j_1}p_{i_1} p_k^\dagger h_l^\dagger  p_{q}^\dagger  h_t^\dagger \ket{\phi}\\
&-\frac{1}{2}\sum_{klqrs}\frac{c_{lq}^{ks}}{c_q^s}t^{\phi}_{qr}c_{r}^{s}\bra{\phi}h_{j_2}p_{i_2}h_{j_1}p_{i_1} p_k^\dagger h_l^\dagger p_s^\dagger h_{q}^\dagger\ket{\phi}\\
&-\frac{1}{2}\sum_{klqq_1rr_1}\frac{c_{lr}^{kq}}{c_r^q}V_{qr_1rq_1}c_{r_1}^{q_1}\bra{\phi}h_{j_2}p_{i_2}h_{j_1}p_{i_1}p_k^\dagger h_l^\dagger p_q^\dagger h_{r}^\dagger \ket{\phi}\\
&=\frac{1}{2}\sum_{klqrt}\frac{c_{lt}^{kq}}{c_t^q}t^{\phi}_{qr}c_{t}^{r}\bra{\phi}h_{j_2}h_{j_1}h_l^\dagger h_t^\dagger p_{i_2}p_{i_1} p_k^\dagger p_{q}^\dagger  \ket{\phi}\\
&-\frac{1}{2}\sum_{klqrs}\frac{c_{lq}^{ks}}{c_q^s}t^{\phi}_{qr}c_{r}^{s}\bra{\phi}h_{j_2}h_{j_1}h_l^\dagger h_{q}^\dagger p_{i_2}p_{i_1} p_k^\dagger p_s^\dagger \ket{\phi}\\
&-\frac{1}{2}\sum_{klqq_1rr_1}\frac{c_{lr}^{kq}}{c_r^q}V_{qr_1rq_1}c_{r_1}^{q_1}\bra{\phi}h_{j_2}h_{j_1}h_l^\dagger h_{r}^\dagger p_{i_2}p_{i_1}p_k^\dagger p_q^\dagger  \ket{\phi}\\
\end{split}
\end{equation}
\begin{equation}
\begin{split}
\bra{\phi_{j_1j_2}^{i_1i_2}}T\hat{H}^{\phi}T\ket{\phi}
&=\frac{1}{2}\sum_{klqrt}\frac{c_{lt}^{kq}}{c_t^q}t^{\phi}_{qr}c_{t}^{r}(\delta_{lj_1}\delta_{tj_2}-\delta_{lj_2}\delta_{tj_1})(\delta_{ki_1}\delta_{qi_2}-\delta_{ki_2}\delta_{qi_1})\\
&-\frac{1}{2}\sum_{klqrs}\frac{c_{lq}^{ks}}{c_q^s}t^{\phi}_{qr}c_{r}^{s}(\delta_{lj_1}\delta_{qj_2}-\delta_{lj_2}\delta_{qj_1})(\delta_{ki_1}\delta_{si_2}-\delta_{ki_2}\delta_{si_1})\\
&-\frac{1}{2}\sum_{klqq_1rr_1}\frac{c_{lr}^{kq}}{c_r^q}V_{qr_1rq_1}c_{r_1}^{q_1}(\delta_{lj_1}\delta_{rj_2}-\delta_{lj_2}\delta_{rj_1})(\delta_{ki_1}\delta_{qi_2}-\delta_{ki_2}\delta_{qi_1})\\
&=\frac{1}{2}\sum_{r}\left(\frac{c_{j_1j_2}^{i_1i_2}}{c_{j_2}^{i_2}}t^{\phi}_{i_2r}c_{j_2}^{r}-\frac{c_{j_1j_2}^{i_2i_1}}{c_{j_2}^{i_1}}t^{\phi}_{i_1r}c_{j_2}^{r}-\frac{c_{j_2j_1}^{i_1i_2}}{c_{j_1}^{i_2}}t^{\phi}_{i_2r}c_{j_1}^{r}+\frac{c_{j_2j_1}^{i_2i_1}}{c_{j_1}^{i_1}}t^{\phi}_{i_1r}c_{j_1}^{r}\right)\\
&-\frac{1}{2}\sum_{r}\left(\frac{c_{j_1j_2}^{i_1i_2}}{c_{j_2}^{i_2}}t^{\phi}_{j_2r}c_{r}^{i_2}-\frac{c_{j_1j_2}^{i_2i_1}}{c_{j_2}^{i_1}}t^{\phi}_{j_2r}c_{r}^{i_1}-\frac{c_{j_2j_1}^{i_1i_2}}{c_{j_1}^{i_2}}t^{\phi}_{j_1r}c_{r}^{i_2}+\frac{c_{j_2j_1}^{i_2i_1}}{c_{j_1}^{i_1}}t^{\phi}_{j_1r}c_{r}^{i_1}\right)\\
&-\frac{1}{2}\sum_{q_1r_1}\left(\frac{c_{j_1j_2}^{i_1i_2}}{c_{j_2}^{i_2}}V_{i_2r_1j_2q_1}c_{r_1}^{q_1}-\frac{c_{j_1j_2}^{i_2i_1}}{c_{j_2}^{i_1}}V_{i_1r_1j_2q_1}c_{r_1}^{q_1}-\frac{c_{j_2j_1}^{i_1i_2}}{c_{j_1}^{i_2}}V_{i_2r_1j_1q_1}c_{r_1}^{q_1}+\frac{c_{j_2j_1}^{i_2i_1}}{c_{j_1}^{i_1}}V_{i_1r_1j_1q_1}c_{r_1}^{q_1}\right)
\end{split}
\end{equation}
which becomes, using the antisymmetry of $c_{j_1j_2}^{i_1i_2}$,
\begin{equation}
\boxed{
\begin{split}
-\bra{\phi_{j_1j_2}^{i_1i_2}}T\hat{H}^{\phi}T\ket{\phi}&=c_{j_1j_2}^{i_1i_2}\Bigg[-\frac{1}{2}\sum_{r}\left(\frac{t^{\phi}_{i_2r}c_{j_2}^{r}}{c_{j_2}^{i_2}}+\frac{t^{\phi}_{i_1r}c_{j_2}^{r}}{c_{j_2}^{i_1}}+\frac{t^{\phi}_{i_2r}c_{j_1}^{r}}{c_{j_1}^{i_2}}+\frac{t^{\phi}_{i_1r}c_{j_1}^{r}}{c_{j_1}^{i_1}}\right)\\
&+\frac{1}{2}\sum_{r}\left(\frac{t^{\phi}_{j_2r}c_{r}^{i_2}}{c_{j_2}^{i_2}}+\frac{t^{\phi}_{j_2r}c_{r}^{i_1}}{c_{j_2}^{i_1}}+\frac{t^{\phi}_{j_1r}c_{r}^{i_2}}{c_{j_1}^{i_2}}+\frac{t^{\phi}_{j_1r}c_{r}^{i_1}}{c_{j_1}^{i_1}}\right)\\
&+\frac{1}{2}\sum_{qr}\left(\frac{V_{i_2rj_2q}c_{r}^{q}}{c_{j_2}^{i_2}}+\frac{V_{i_1rj_2q}c_{r}^{q}}{c_{j_2}^{i_1}}+\frac{V_{i_2rj_1q}c_{r}^{q}}{c_{j_1}^{i_2}}+\frac{V_{i_1rj_1q}c_{r}^{q}}{c_{j_1}^{i_1}}\right)\Bigg]
\end{split}}
\end{equation}

For $\bra{\phi_{j_1j_2}^{i_1i_2}}\hat{H}^{\phi}T^3\ket{\phi}$, all the terms destroying a pair contribute:
\begin{equation}
\begin{split}
\bra{\phi_{j_1j_2}^{i_1i_2}}&\hat{H}^{\phi}T^3\ket{\phi}\\
&=\bra{\phi}h_{j_2}p_{i_2}h_{j_1}p_{i_1}\Bigg(\sum_{mn} t^{\phi}_{mn}h_{n} p_{m}+\frac{1}{2}\sum_{mnqr} V_{mnqr}p_r^\dagger h_q p_n  p_m\\
&+\frac{1}{2}\sum_{mnqr} V_{mnqr}h_r^\dagger p_q h_n h_m\Bigg)\frac{1}{6}\sum_{k_1l_1k_2l_2k_3l_3} c_{l_1l_2l_3}^{k_1k_2k_3} p_{k_1}^\dagger h_{l_1}^\dagger  p_{k_2}^\dagger h_{l_2}^\dagger p_{k_3}^\dagger h_{l_3}^\dagger \ket{\phi}\\
&=\frac{1}{6}\sum_{mnk_1l_1k_2l_2k_3l_3} t^{\phi}_{mn}c_{l_1l_2l_3}^{k_1k_2k_3}\bra{\phi}h_{j_2}p_{i_2}h_{j_1}p_{i_1}h_{n} p_{m} p_{k_1}^\dagger h_{l_1}^\dagger  p_{k_2}^\dagger h_{l_2}^\dagger p_{k_3}^\dagger h_{l_3}^\dagger \ket{\phi}\\
&+\frac{1}{12}\sum_{mnqrk_1l_1k_2l_2k_3l_3} V_{mnqr}c_{l_1l_2l_3}^{k_1k_2k_3}\bra{\phi}h_{j_2}p_{i_2}h_{j_1}p_{i_1}p_r^\dagger h_q p_n  p_m p_{k_1}^\dagger h_{l_1}^\dagger  p_{k_2}^\dagger h_{l_2}^\dagger p_{k_3}^\dagger h_{l_3}^\dagger \ket{\phi}\\
&+\frac{1}{12}\sum_{mnqrk_1l_1k_2l_2k_3l_3} V_{mnqr}c_{l_1l_2l_3}^{k_1k_2k_3}\bra{\phi}h_{j_2}p_{i_2}h_{j_1}p_{i_1}h_r^\dagger p_q h_n h_m p_{k_1}^\dagger h_{l_1}^\dagger p_{k_2}^\dagger h_{l_2}^\dagger p_{k_3}^\dagger h_{l_3}^\dagger \ket{\phi}\\
&=\frac{1}{6}\sum_{mnk_1l_1k_2l_2k_3l_3} t^{\phi}_{mn}c_{l_1l_2l_3}^{k_1k_2k_3}\bra{\phi}h_{j_2}h_{j_1}h_{n} h_{l_1}^\dagger h_{l_2}^\dagger  h_{l_3}^\dagger p_{i_2}p_{i_1}p_{m} p_{k_1}^\dagger p_{k_2}^\dagger p_{k_3}^\dagger  \ket{\phi}\\
&-\frac{1}{12}\sum_{mnqrk_1l_1k_2l_2k_3l_3} V_{mnqr}c_{l_1l_2l_3}^{k_1k_2k_3}\bra{\phi}h_{j_2}h_{j_1}h_q h_{l_1}^\dagger   h_{l_2}^\dagger  h_{l_3}^\dagger p_{i_2}p_{i_1}p_r^\dagger  p_n  p_m p_{k_1}^\dagger p_{k_2}^\dagger p_{k_3}^\dagger  \ket{\phi}\\
&+\frac{1}{12}\sum_{mnqrk_1l_1k_2l_2k_3l_3} V_{mnqr}c_{l_1l_2l_3}^{k_1k_2k_3}\bra{\phi}h_{j_2}h_{j_1}h_r^\dagger h_n h_m h_{l_1}^\dagger h_{l_2}^\dagger h_{l_3}^\dagger p_{i_2}p_{i_1} p_q  p_{k_1}^\dagger  p_{k_2}^\dagger  p_{k_3}^\dagger  \ket{\phi}\\
\end{split}
\end{equation}
Now, because of the antisymmetry of $c_{l_1l_2l_3}^{k_1k_2k_3}$, we obtain factors of $3!$ for each sum over the indices $(k_1,k_2,k_3)$ or $(l_1,l_2,l_3)$:
\begin{equation}
\begin{split}
\bra{\phi_{j_1j_2}^{i_1i_2}}&\hat{H}^{\phi}T^3\ket{\phi}\\
&=6\sum_{mn} t^{\phi}_{mn}c_{nj_1j_2}^{mi_1i_2}-\frac{1}{2}\sum_{mnqrk_1k_2k_3} V_{mnqr}c_{qj_1j_2}^{k_1k_2k_3}\bra{\phi} p_{i_2}p_{i_1}p_r^\dagger  p_n  p_m p_{k_1}^\dagger p_{k_2}^\dagger p_{k_3}^\dagger  \ket{\phi}\\
&+\frac{1}{2}\sum_{mnqrl_1l_2l_3} V_{mnqr}c_{l_1l_2l_3}^{qi_1i_2}\bra{\phi}h_{j_2}h_{j_1}h_r^\dagger h_n h_m h_{l_1}^\dagger h_{l_2}^\dagger h_{l_3}^\dagger \ket{\phi}\\
&=6\sum_{mn} t^{\phi}_{mn}c_{nj_1j_2}^{mi_1i_2}-\frac{1}{2}\sum_{mnqrk_1k_2k_3} V_{mnqr}c_{qj_1j_2}^{k_1k_2k_3}\bigg(\delta_{ri_1}\bra{\phi} p_{i_2}  p_n  p_m p_{k_1}^\dagger p_{k_2}^\dagger p_{k_3}^\dagger  \ket{\phi}-\delta_{ri_2}\bra{\phi} p_{i_1} p_n  p_m p_{k_1}^\dagger p_{k_2}^\dagger p_{k_3}^\dagger  \ket{\phi}\bigg)\\
&+\frac{1}{2}\sum_{mnqrl_1l_2l_3} V_{mnqr}c_{l_1l_2l_3}^{qi_1i_2}\bigg(\delta_{rj_1}\bra{\phi}h_{j_2} h_n h_m h_{l_1}^\dagger h_{l_2}^\dagger h_{l_3}^\dagger \ket{\phi}-\delta_{rj_2}\bra{\phi}h_{j_1} h_n h_m h_{l_1}^\dagger h_{l_2}^\dagger h_{l_3}^\dagger \ket{\phi}\bigg)\\
&=6\sum_{mn} t^{\phi}_{mn}c_{nj_1j_2}^{mi_1i_2}-3\sum_{mnqr} V_{mnqr}\left(\delta_{ri_1}c_{qj_1j_2}^{mni_2}-\delta_{ri_2}c_{qj_1j_2}^{mni_1}\right)+3\sum_{mnqr} V_{mnqr}\left(\delta_{rj_1}c_{mnj_2}^{qi_1i_2}-\delta_{rj_2}c_{mnj_1}^{qi_1i_2}\right)\\
&=6\sum_{mn} t^{\phi}_{mn}c_{nj_1j_2}^{mi_1i_2}-3\sum_{mnq} \left(V_{mnqi_1}c_{qj_1j_2}^{mni_2}-V_{mnqi_2}c_{qj_1j_2}^{mni_1}\right)+3\sum_{mnq} \left(V_{mnqj_1}c_{mnj_2}^{qi_1i_2}-V_{mnqj_2}c_{mnj_1}^{qi_1i_2}\right)
\end{split}
\end{equation}
and thus,
\begin{equation}
\boxed{
\begin{split}
\frac{1}{3!}\bra{\phi_{j_1j_2}^{i_1i_2}}\hat{H}^{\phi}T^3\ket{\phi}&=\sum_{mn} t^{\phi}_{mn}c_{nj_1j_2}^{mi_1i_2}-\frac{1}{2}\sum_{mnq} \left(V_{mnqi_1}c_{qj_1j_2}^{mni_2}-V_{mnqi_2}c_{qj_1j_2}^{mni_1}\right)+\frac{1}{2}\sum_{mnq} \left(V_{mnqj_1}c_{mnj_2}^{qi_1i_2}-V_{mnqj_2}c_{mnj_1}^{qi_1i_2}\right)
\end{split}}
\end{equation}

For $\bra{\phi_{j_1j_2}^{i_1i_2}}T\hat{H}^{\phi}T^2\ket{\phi}$, all the terms of $H$ destroying on pair contribute:
\begin{equation}
\begin{split}
\bra{\phi_{j_1j_2}^{i_1i_2}}&T\hat{H}^{\phi}T^2\ket{\phi}\\
&=\bra{\phi}h_{j_2}p_{i_2}h_{j_1}p_{i_1}\sum_{k_1l_1} p_{k_1}^\dagger h_{l_1}^\dagger \hat{D}_{l_1}^{k_1}\Bigg(\sum_{mn} t^{\phi}_{mn}h_{n} p_{m}+\frac{1}{2}\sum_{mnqr} V_{mnqr}p_r^\dagger h_q p_n  p_m\\
&+\frac{1}{2}\sum_{mnqr} V_{mnqr}h_r^\dagger p_q h_n h_m\Bigg)\frac{1}{2}\sum_{k_2l_2k_3l_3} c_{l_2l_3}^{k_2k_3} p_{k_2}^\dagger h_{l_2}^\dagger p_{k_3}^\dagger h_{l_3}^\dagger \ket{\phi}\\
&=\frac{1}{2}\sum_{k_1l_1k_2l_2k_3l_3mn} t^{\phi}_{mn}c_{l_2l_3}^{k_2k_3}\bra{\phi}h_{j_2}p_{i_2}h_{j_1}p_{i_1} p_{k_1}^\dagger h_{l_1}^\dagger \hat{D}_{l_1}^{k_1} h_{n} p_{m} p_{k_2}^\dagger h_{l_2}^\dagger  p_{k_3}^\dagger h_{l_3}^\dagger \ket{\phi}\\
&+\frac{1}{4}\sum_{k_1l_1k_2l_2k_3l_3mnqr}V_{mnqr}c_{l_2l_3}^{k_2k_3}\bra{\phi}h_{j_2}p_{i_2}h_{j_1}p_{i_1} p_{k_1}^\dagger h_{l_1}^\dagger \hat{D}_{l_1}^{k_1} p_r^\dagger h_q p_n  p_m p_{k_2}^\dagger h_{l_2}^\dagger p_{k_3}^\dagger h_{l_3}^\dagger \ket{\phi}\\
&+\frac{1}{4}\sum_{k_1l_1k_2l_2k_3l_3mnqr}V_{mnqr}c_{l_2l_3}^{k_2k_3}\bra{\phi}h_{j_2}p_{i_2}h_{j_1}p_{i_1} p_{k_1}^\dagger h_{l_1}^\dagger \hat{D}_{l_1}^{k_1}h_r^\dagger p_q h_n h_m p_{k_2}^\dagger h_{l_2}^\dagger  p_{k_3}^\dagger h_{l_3}^\dagger \ket{\phi}\\
&=\frac{1}{2}\sum_{k_1l_1k_2l_2k_3l_3mn} t^{\phi}_{mn}c_{l_2l_3}^{k_2k_3}\bra{\phi}h_{j_2}p_{i_2}h_{j_1}p_{i_1} p_{k_1}^\dagger h_{l_1}^\dagger \hat{D}_{l_1}^{k_1}  p_{m} p_{k_2}^\dagger p_{k_3}^\dagger h_{n} h_{l_2}^\dagger h_{l_3}^\dagger \ket{\phi}\\
&-\frac{1}{4}\sum_{k_1l_1k_2l_2k_3l_3mnqr}V_{mnqr}c_{l_2l_3}^{k_2k_3}\bra{\phi}h_{j_2}p_{i_2}h_{j_1}p_{i_1} p_{k_1}^\dagger h_{l_1}^\dagger \hat{D}_{l_1}^{k_1} p_r^\dagger  p_n  p_m p_{k_2}^\dagger p_{k_3}^\dagger h_q h_{l_2}^\dagger  h_{l_3}^\dagger \ket{\phi}\\
&+\frac{1}{4}\sum_{k_1l_1k_2l_2k_3l_3mnqr}V_{mnqr}c_{l_2l_3}^{k_2k_3}\bra{\phi}h_{j_2}p_{i_2}h_{j_1}p_{i_1} p_{k_1}^\dagger h_{l_1}^\dagger \hat{D}_{l_1}^{k_1} p_q p_{k_2}^\dagger p_{k_3}^\dagger h_r^\dagger h_n h_m  h_{l_2}^\dagger   h_{l_3}^\dagger \ket{\phi}\\
\end{split}
\end{equation}
\begin{equation}
\begin{split}
\bra{\phi_{j_1j_2}^{i_1i_2}}&T\hat{H}^{\phi}T^2\ket{\phi}\\
&=\frac{1}{2}\sum_{k_1l_1k_2l_2k_3l_3mn} t^{\phi}_{mn}c_{l_2l_3}^{k_2k_3}\bra{\phi}h_{j_2}p_{i_2}h_{j_1}p_{i_1} p_{k_1}^\dagger h_{l_1}^\dagger \hat{D}_{l_1}^{k_1}\left(\delta_{mk_2}p_{k_3}^\dagger-\delta_{mk_3}p_{k_2}^\dagger\right)\left(\delta_{nl_2}h_{l_3}^\dagger-\delta_{nl_3}h_{l_2}^\dagger\right)\ket{\phi}\\
&-\frac{1}{4}\sum_{k_1l_1k_2l_2k_3l_3mnqr}V_{mnqr}c_{l_2l_3}^{k_2k_3}\left(\delta_{mk_2}\delta_{nk_3}-\delta_{mk_3}\delta_{nk_2}\right)\bra{\phi}h_{j_2}p_{i_2}h_{j_1}p_{i_1} p_{k_1}^\dagger h_{l_1}^\dagger \hat{D}_{l_1}^{k_1} p_r^\dagger \left(\delta_{ql_2}h_{l_3}^\dagger-\delta_{ql_3}h_{l_2}^\dagger\right) \ket{\phi}\\
&+\frac{1}{4}\sum_{k_1l_1k_2l_2k_3l_3mnqr}V_{mnqr}c_{l_2l_3}^{k_2k_3}\left(\delta_{ml_2}\delta_{nl_3}-\delta_{ml_3}\delta_{nl_2}\right)\bra{\phi}h_{j_2}p_{i_2}h_{j_1}p_{i_1} p_{k_1}^\dagger h_{l_1}^\dagger \hat{D}_{l_1}^{k_1} \left(\delta_{qk_2}p_{k_3}^\dagger-\delta_{qk_3}p_{k_2}^\dagger\right)  h_r^\dagger \ket{\phi}\\
&=\frac{1}{2}\sum_{k_1l_1k_2l_2k_3l_3mn} t^{\phi}_{mn}c_{l_2l_3}^{k_2k_3}\Bigg(\delta_{mk_2}\delta_{nl_2}\frac{c_{l_1l_3}^{k_1k_3}}{c_{l_3}^{k_3}}\bra{\phi}h_{j_2}p_{i_2}h_{j_1}p_{i_1} p_{k_1}^\dagger h_{l_1}^\dagger p_{k_3}^\dagger h_{l_3}^\dagger\ket{\phi}\\
&\hspace{1.7in}-\delta_{mk_2}\delta_{nl_3}\frac{c_{l_1l_2}^{k_1k_3}}{c_{l_2}^{k_3}}\bra{\phi}h_{j_2}p_{i_2}h_{j_1}p_{i_1} p_{k_1}^\dagger h_{l_1}^\dagger p_{k_3}^\dagger h_{l_2}^\dagger\ket{\phi}\\
&\hspace{1.7in}-\delta_{mk_3}\delta_{nl_2}\frac{c_{l_1l_3}^{k_1k_2}}{c_{l_3}^{k_2}}\bra{\phi}h_{j_2}p_{i_2}h_{j_1}p_{i_1} p_{k_1}^\dagger h_{l_1}^\dagger p_{k_2}^\dagger h_{l_3}^\dagger\ket{\phi}\\
&\hspace{1.7in}+\delta_{mk_3}\delta_{nl_3}\frac{c_{l_1l_2}^{k_1k_2}}{c_{l_2}^{k_2}}\bra{\phi}h_{j_2}p_{i_2}h_{j_1}p_{i_1} p_{k_1}^\dagger h_{l_1}^\dagger p_{k_2}^\dagger h_{l_2}^\dagger\ket{\phi}\Bigg)\\
&-\frac{1}{2}\sum_{k_1l_1l_2l_3mnqr}V_{mnqr}c_{l_2l_3}^{mn}\Bigg(\delta_{ql_2}\frac{c_{l_1l_3}^{k_1r}}{c_{l_3}^{r}}\bra{\phi}h_{j_2}p_{i_2}h_{j_1}p_{i_1} p_{k_1}^\dagger h_{l_1}^\dagger p_r^\dagger h_{l_3}^\dagger\ket{\phi}-\delta_{ql_3}\frac{c_{l_1l_2}^{k_1r}}{c_{l_2}^{r}}\bra{\phi}h_{j_2}p_{i_2}h_{j_1}p_{i_1} p_{k_1}^\dagger h_{l_1}^\dagger p_r^\dagger h_{l_2}^\dagger \ket{\phi}\Bigg)\\
&+\frac{1}{2}\sum_{k_1l_1k_2k_3mnqr}V_{mnqr}c_{mn}^{k_2k_3}\Bigg(\delta_{qk_2}\frac{c_{l_1r}^{k_1k_3}}{c_{r}^{k_3}}\bra{\phi}h_{j_2}p_{i_2}h_{j_1}p_{i_1} p_{k_1}^\dagger h_{l_1}^\dagger p_{k_3}^\dagger h_r^\dagger \ket{\phi}-\delta_{qk_3}\frac{c_{l_1r}^{k_1k_2}}{c_{r}^{k_2}}\bra{\phi}h_{j_2}p_{i_2}h_{j_1}p_{i_1} p_{k_1}^\dagger h_{l_1}^\dagger p_{k_2}^\dagger  h_r^\dagger \ket{\phi}\Bigg)\\
&=\frac{1}{2}\sum_{k_1l_1k_2l_2k_3l_3mn} t^{\phi}_{mn}c_{l_2l_3}^{k_2k_3}\Bigg(\delta_{mk_2}\delta_{nl_2}\frac{c_{l_1l_3}^{k_1k_3}}{c_{l_3}^{k_3}}\bra{\phi}h_{j_2}h_{j_1}h_{l_1}^\dagger  h_{l_3}^\dagger p_{i_2}p_{i_1} p_{k_1}^\dagger p_{k_3}^\dagger \ket{\phi}\\
&\hspace{1.7in}-\delta_{mk_2}\delta_{nl_3}\frac{c_{l_1l_2}^{k_1k_3}}{c_{l_2}^{k_3}}\bra{\phi}h_{j_2}h_{j_1}h_{l_1}^\dagger  h_{l_2}^\dagger p_{i_2}p_{i_1} p_{k_1}^\dagger p_{k_3}^\dagger \ket{\phi}\\
&\hspace{1.7in}-\delta_{mk_3}\delta_{nl_2}\frac{c_{l_1l_3}^{k_1k_2}}{c_{l_3}^{k_2}}\bra{\phi}h_{j_2}h_{j_1}h_{l_1}^\dagger  h_{l_3}^\dagger p_{i_2}p_{i_1} p_{k_1}^\dagger p_{k_2}^\dagger \ket{\phi}\\
&\hspace{1.7in}+\delta_{mk_3}\delta_{nl_3}\frac{c_{l_1l_2}^{k_1k_2}}{c_{l_2}^{k_2}}\bra{\phi}h_{j_2}h_{j_1}h_{l_1}^\dagger h_{l_2}^\dagger p_{i_2}p_{i_1} p_{k_1}^\dagger p_{k_2}^\dagger \ket{\phi}\Bigg)\\
&-\frac{1}{2}\sum_{k_1l_1l_2l_3mnqr}V_{mnqr}c_{l_2l_3}^{mn}\Bigg(\delta_{ql_2}\frac{c_{l_1l_3}^{k_1r}}{c_{l_3}^{r}}\bra{\phi}h_{j_2}h_{j_1}h_{l_1}^\dagger h_{l_3}^\dagger p_{i_2}p_{i_1} p_{k_1}^\dagger p_r^\dagger \ket{\phi}-\delta_{ql_3}\frac{c_{l_1l_2}^{k_1r}}{c_{l_2}^{r}}\bra{\phi}h_{j_2}h_{j_1}h_{l_1}^\dagger  h_{l_2}^\dagger p_{i_2}p_{i_1} p_{k_1}^\dagger p_r^\dagger \ket{\phi}\Bigg)\\
&+\frac{1}{2}\sum_{k_1l_1k_2k_3mnqr}V_{mnqr}c_{mn}^{k_2k_3}\Bigg(\delta_{qk_2}\frac{c_{l_1r}^{k_1k_3}}{c_{r}^{k_3}}\bra{\phi}h_{j_2}h_{j_1}h_{l_1}^\dagger  h_r^\dagger  p_{i_2}p_{i_1} p_{k_1}^\dagger p_{k_3}^\dagger \ket{\phi}-\delta_{qk_3}\frac{c_{l_1r}^{k_1k_2}}{c_{r}^{k_2}}\bra{\phi}h_{j_2}h_{j_1}h_{l_1}^\dagger h_r^\dagger p_{i_2}p_{i_1} p_{k_1}^\dagger p_{k_2}^\dagger \ket{\phi}\Bigg)
\end{split}
\end{equation}
\begin{equation}
\begin{split}
\bra{\phi_{j_1j_2}^{i_1i_2}}&T\hat{H}^{\phi}T^2\ket{\phi}\\
&=\frac{1}{2}\sum_{k_1l_1k_2l_2k_3l_3mn} t^{\phi}_{mn}c_{l_2l_3}^{k_2k_3}\Bigg(\frac{c_{l_1l_3}^{k_1k_3}}{c_{l_3}^{k_3}}\delta_{mk_2}\delta_{nl_2}\left(\delta_{l_1j_1}\delta_{l_3j_2}-\delta_{l_1j_2}\delta_{l_3j_1}\right)\left(\delta_{k_1i_1}\delta_{k_3i_2}-\delta_{k_1i_2}\delta_{k_3i_1}\right)\\
&\hspace{1.7in}-\frac{c_{l_1l_2}^{k_1k_3}}{c_{l_2}^{k_3}}\delta_{mk_2}\delta_{nl_3}\left(\delta_{l_1j_1}\delta_{l_2j_2}-\delta_{l_1j_2}\delta_{l_2j_1}\right)\left(\delta_{k_1i_1}\delta_{k_3i_2}-\delta_{k_1i_2}\delta_{k_3i_1}\right)\\
&\hspace{1.7in}-\frac{c_{l_1l_3}^{k_1k_2}}{c_{l_3}^{k_2}}\delta_{mk_3}\delta_{nl_2}\left(\delta_{l_1j_1}\delta_{l_3j_2}-\delta_{l_1j_2}\delta_{l_3j_1}\right)\left(\delta_{k_1i_1}\delta_{k_2i_2}-\delta_{k_1i_2}\delta_{k_2i_1}\right)\\
&\hspace{1.7in}+\frac{c_{l_1l_2}^{k_1k_2}}{c_{l_2}^{k_2}}\delta_{mk_3}\delta_{nl_3}\left(\delta_{l_1j_1}\delta_{l_2j_2}-\delta_{l_1j_2}\delta_{l_2j_1}\right)\left(\delta_{k_1i_1}\delta_{k_2i_2}-\delta_{k_1i_2}\delta_{k_2i_1}\right)\Bigg)\\
&-\frac{1}{2}\sum_{k_1l_1l_2l_3mnqr}V_{mnqr}c_{l_2l_3}^{mn}\Bigg(\delta_{ql_2}\frac{c_{l_1l_3}^{k_1r}}{c_{l_3}^{r}}\left(\delta_{l_1j_1}\delta_{l_3j_2}-\delta_{l_1j_2}\delta_{l_3j_1}\right)\left(\delta_{k_1i_1}\delta_{ri_2}-\delta_{k_1i_2}\delta_{ri_1}\right)\\
&\hspace{1.6in}-\delta_{ql_3}\frac{c_{l_1l_2}^{k_1r}}{c_{l_2}^{r}}\left(\delta_{l_1j_1}\delta_{l_2j_2}-\delta_{l_1j_2}\delta_{l_2j_1}\right)\left(\delta_{k_1i_1}\delta_{ri_2}-\delta_{k_1i_2}\delta_{ri_1}\right)\Bigg)\\
&+\frac{1}{2}\sum_{k_1l_1k_2k_3mnqr}V_{mnqr}c_{mn}^{k_2k_3}\Bigg(\delta_{qk_2}\frac{c_{l_1r}^{k_1k_3}}{c_{r}^{k_3}}\left(\delta_{l_1j_1}\delta_{rj_2}-\delta_{l_1j_2}\delta_{rj_1}\right)\left(\delta_{k_1i_1}\delta_{k_3i_2}-\delta_{k_1i_2}\delta_{k_3i_1}\right)\\
&\hspace{1.6in}-\delta_{qk_3}\frac{c_{l_1r}^{k_1k_2}}{c_{r}^{k_2}}\left(\delta_{l_1j_1}\delta_{rj_2}-\delta_{l_1j_2}\delta_{rj_1}\right)\left(\delta_{k_1i_1}\delta_{k_2i_2}-\delta_{k_1i_2}\delta_{k_2i_1}\right)\Bigg)\\
&=\frac{1}{2}\sum_{mn} t^{\phi}_{mn}\Bigg[c_{nj_2}^{mi_2}\frac{c_{j_1j_2}^{i_1i_2}}{c_{j_2}^{i_2}}-c_{nj_2}^{mi_1}\frac{c_{j_1j_2}^{i_2i_1}}{c_{j_2}^{i_1}}-c_{nj_1}^{mi_2}\frac{c_{j_2j_1}^{i_1i_2}}{c_{j_1}^{i_2}}+c_{nj_1}^{mi_1}\frac{c_{j_2j_1}^{i_2i_1}}{c_{j_1}^{i_1}}\\
&\hspace{0.6in}-\left(c_{j_2n}^{mi_2}\frac{c_{j_1j_2}^{i_1i_2}}{c_{j_2}^{i_2}}-c_{j_2n}^{mi_1}\frac{c_{j_1j_2}^{i_2i_1}}{c_{j_2}^{i_1}}-c_{j_1n}^{mi_2}\frac{c_{j_2j_1}^{i_1i_2}}{c_{j_1}^{i_2}}+c_{j_1n}^{mi_1}\frac{c_{j_2j_1}^{i_2i_1}}{c_{j_1}^{i_1}}\right)\\
&\hspace{0.6in}-\left(c_{nj_2}^{i_2m}\frac{c_{j_1j_2}^{i_1i_2}}{c_{j_2}^{i_2}}-c_{nj_2}^{i_1m}\frac{c_{j_1j_2}^{i_2i_1}}{c_{j_2}^{i_1}}-c_{nj_1}^{i_2m}\frac{c_{j_2j_1}^{i_1i_2}}{c_{j_1}^{i_2}}+c_{nj_1}^{i_1m}\frac{c_{j_2j_1}^{i_2i_1}}{c_{j_1}^{i_1}}\right)\\
&\hspace{0.7in}+c_{j_2n}^{i_2m}\frac{c_{j_1j_2}^{i_1i_2}}{c_{j_2}^{i_2}}-c_{j_2n}^{i_1m}\frac{c_{j_1j_2}^{i_2i_1}}{c_{j_2}^{i_1}}-c_{j_1n}^{i_2m}\frac{c_{j_2j_1}^{i_1i_2}}{c_{j_1}^{i_2}}+c_{j_1n}^{i_1m}\frac{c_{j_2j_1}^{i_2i_1}}{c_{j_1}^{i_1}}\Bigg]\\
&-\frac{1}{2}\sum_{mnq}\Bigg[V_{mnqi_2}c_{qj_2}^{mn}\frac{c_{j_1j_2}^{i_1i_2}}{c_{j_2}^{i_2}}-V_{mnqi_1}c_{qj_2}^{mn}\frac{c_{j_1j_2}^{i_2i_1}}{c_{j_2}^{i_1}}-V_{mnqi_2}c_{qj_1}^{mn}\frac{c_{j_2j_1}^{i_1i_2}}{c_{j_1}^{i_2}}+V_{mnqi_1}c_{qj_1}^{mn}\frac{c_{j_2j_1}^{i_2i_1}}{c_{j_1}^{i_1}}\\
&\hspace{0.4in}-\left(V_{mnqi_2}c_{j_2q}^{mn}\frac{c_{j_1j_2}^{i_1i_2}}{c_{j_2}^{i_2}}-V_{mnqi_1}c_{j_2q}^{mn}\frac{c_{j_1j_2}^{i_2i_1}}{c_{j_2}^{i_1}}-V_{mnqi_2}c_{j_1q}^{mn}\frac{c_{j_2j_1}^{i_1i_2}}{c_{j_1}^{i_2}}+V_{mnqi_1}c_{j_1q}^{mn}\frac{c_{j_2j_1}^{i_2i_1}}{c_{j_1}^{i_1}}\right)\Bigg]\\
&+\frac{1}{2}\sum_{mnq}\Bigg[V_{mnqj_2}c_{mn}^{qi_2}\frac{c_{j_1j_2}^{i_1i_2}}{c_{j_2}^{i_2}}-V_{mnqj_2}c_{mn}^{qi_1}\frac{c_{j_1j_2}^{i_2i_1}}{c_{j_2}^{i_1}}-V_{mnqj_1}c_{mn}^{qi_2}\frac{c_{j_2j_1}^{i_1i_2}}{c_{j_1}^{i_2}}+V_{mnqj_1}c_{mn}^{qi_1}\frac{c_{j_2j_1}^{i_2i_1}}{c_{j_1}^{i_1}}\\
&\hspace{0.4in}-\left(V_{mnqj_2}c_{mn}^{i_2q}\frac{c_{j_1j_2}^{i_1i_2}}{c_{j_2}^{i_2}}-V_{mnqj_2}c_{mn}^{i_1q}\frac{c_{j_1j_2}^{i_2i_1}}{c_{j_2}^{i_1}}-V_{mnqj_1}c_{mn}^{i_2q}\frac{c_{j_2j_1}^{i_1i_2}}{c_{j_1}^{i_2}}+V_{mnqj_1}c_{mn}^{i_1q}\frac{c_{j_2j_1}^{i_2i_1}}{c_{j_1}^{i_1}}\right)\Bigg]
\end{split}
\end{equation}
and again, using the antisymmetry of $c_{j_1j_2}^{i_1i_2}$,
\begin{equation}
\boxed{
\begin{split}
-\frac{1}{2}\bra{\phi_{j_1j_2}^{i_1i_2}}T\hat{H}^{\phi}T^2\ket{\phi}&=c_{j_1j_2}^{i_1i_2}\Bigg[-\sum_{mn} t^{\phi}_{mn}\left(\frac{c_{nj_2}^{mi_2}}{c_{j_2}^{i_2}}+\frac{c_{nj_2}^{mi_1}}{c_{j_2}^{i_1}}+\frac{c_{nj_1}^{mi_2}}{c_{j_1}^{i_2}}+\frac{c_{nj_1}^{mi_1}}{c_{j_1}^{i_1}}\right)\\
&+\frac{1}{2}\sum_{mnq}\left(\frac{V_{mnqi_2}c_{qj_2}^{mn}}{c_{j_2}^{i_2}}+\frac{V_{mnqi_1}c_{qj_2}^{mn}}{c_{j_2}^{i_1}}+\frac{V_{mnqi_2}c_{qj_1}^{mn}}{c_{j_1}^{i_2}}+\frac{V_{mnqi_1}c_{qj_1}^{mn}}{c_{j_1}^{i_1}}\right)\\
&-\frac{1}{2}\sum_{mnq}\left(\frac{V_{mnqj_2}c_{mn}^{qi_2}}{c_{j_2}^{i_2}}+\frac{V_{mnqj_2}c_{mn}^{qi_1}}{c_{j_2}^{i_1}}+\frac{V_{mnqj_1}c_{mn}^{qi_2}}{c_{j_1}^{i_2}}+\frac{V_{mnqj_1}c_{mn}^{qi_1}}{c_{j_1}^{i_1}}\right)\Bigg]
\end{split}}
\end{equation}


For $\bra{\phi_{j_1j_2}^{i_1i_2}}T^2\hat{H}^{\phi}T\ket{\phi}$ we have
\begin{equation}\label{eq:T2_H_T}
\begin{split}
\bra{\phi_{j_1j_2}^{i_1i_2}}T^2\hat{H}^{\phi}T\ket{\phi}&=\bra{\phi_{j_1j_2}^{i_1i_2}}T^2\ket{\phi}\bra{\phi}\hat{H}^{\phi}T\ket{\phi}\\
&=\frac{1}{2}\sum_{k_1l_1k_2l_2}c_{l_1l_2}^{k_1k_2}\bra{\phi}h_{j_2}p_{i_2}h_{j_1}p_{i_1}p_{k_1}^\dagger h_{l_1}^\dagger p_{k_2}^\dagger h_{l_2}^\dagger\ket{\phi}\sum_{mnk_3l_3} t^{\phi}_{mn}c_{l_3}^{k_3}\bra{\phi}h_{n} p_{m}  p_{k_3}^\dagger h_{l_3}^\dagger \ket{\phi}\\
&=2c_{j_1j_2}^{i_1i_2}\sum_{mn} t^{\phi}_{mn}c_{n}^{m}
\end{split}
\end{equation}
and thus,
\begin{equation}
\boxed{
\frac{1}{2}\bra{\phi_{j_1j_2}^{i_1i_2}}T^2\hat{H}^{\phi}T\ket{\phi}=c_{j_1j_2}^{i_1i_2}\sum_{mn} t^{\phi}_{mn}c_{n}^{m}}
\end{equation}

\begin{equation}
\begin{split}
\bra{\phi_{j_1j_2}^{i_1i_2}}&\hat{H}^{\phi}T^4\ket{\phi}\\
&=\frac{1}{4(4!)}\sum_{mnqrk_1l_1k_2l_2k_3l_3k_4l_4}V_{mnqr} c_{l_1l_2l_3l_4}^{k_1k_2k_3k_4}\bra{\phi}h_{j_2}p_{i_2}h_{j_1}p_{i_1}  h_r h_q p_n p_m p_{k_1}^\dagger h_{l_1}^\dagger p_{k_2}^\dagger h_{l_2}^\dagger  p_{k_3}^\dagger h_{l_3}^\dagger p_{k_4}^\dagger h_{l_4}^\dagger \ket{\phi}\\
&=-\frac{1}{4(4!)}\sum_{mnqrk_1l_1k_2l_2k_3l_3k_4l_4}V_{mnqr} c_{l_1l_2l_3l_4}^{k_1k_2k_3k_4}\bra{\phi}p_{i_2}p_{i_1} p_n p_m p_{k_1}^\dagger p_{k_2}^\dagger p_{k_3}^\dagger p_{k_4}^\dagger h_{j_2}h_{j_1} h_r h_q h_{l_1}^\dagger h_{l_2}^\dagger h_{l_3}^\dagger h_{l_4}^\dagger \ket{\phi}\\
&=-\frac{4!}{4}\sum_{mnqr}V_{mnqr} c_{qrj_1j_2}^{mni_1i_2}
\end{split}
\end{equation}
and
\begin{equation}
\boxed{
\frac{1}{4!}\bra{\phi_{j_1j_2}^{i_1i_2}}\hat{H}^{\phi}T^4\ket{\phi}=-\frac{1}{4}\sum_{mnqr}V_{mnqr} c_{qrj_1j_2}^{mni_1i_2}}
\end{equation}

\begin{equation}
\begin{split}
\bra{\phi_{j_1j_2}^{i_1i_2}}&T\hat{H}^{\phi}T^3\ket{\phi}\\
&=\bra{\phi}h_{j_2}p_{i_2}h_{j_1}p_{i_1}\sum_{k_1l_1} p_{k_1}^\dagger h_{l_1}^\dagger \hat{D}_{l_1}^{k_1} \left(\frac{1}{4}\right)\sum_{mnqr} V_{mnqr} h_r h_q p_n p_m \left(\frac{1}{3!}\right)\sum_{k_2l_2k_3l_3k_4l_4} c_{l_2l_3l_4}^{k_2k_3k_4}p_{k_2}^\dagger h_{l_2}^\dagger p_{k_3}^\dagger h_{l_3}^\dagger p_{k_4}^\dagger h_{l_4}^\dagger \ket{\phi}\\
&=\frac{1}{24}\sum_{mnqrk_1l_1k_2l_2k_3l_3k_4l_4}V_{mnqr}c_{l_2l_3l_4}^{k_2k_3k_4}\bra{\phi}h_{j_2}p_{i_2}h_{j_1}p_{i_1} p_{k_1}^\dagger h_{l_1}^\dagger \hat{D}_{l_1}^{k_1}  h_r h_q p_n p_m p_{k_2}^\dagger h_{l_2}^\dagger p_{k_3}^\dagger h_{l_3}^\dagger p_{k_4}^\dagger h_{l_4}^\dagger \ket{\phi}\\
&=-\frac{1}{24}\sum_{mnqrk_1l_1k_2l_2k_3l_3k_4l_4}V_{mnqr}c_{l_2l_3l_4}^{k_2k_3k_4}\bra{\phi}h_{j_2}p_{i_2}h_{j_1}p_{i_1} p_{k_1}^\dagger h_{l_1}^\dagger \hat{D}_{l_1}^{k_1}  p_n p_m p_{k_2}^\dagger p_{k_3}^\dagger p_{k_4}^\dagger h_r h_q h_{l_2}^\dagger h_{l_3}^\dagger h_{l_4}^\dagger \ket{\phi}\\
&=-\frac{1}{24}\sum_{mnqrk_1l_1k_2l_2k_3l_3k_4l_4}V_{mnqr}c_{l_2l_3l_4}^{k_2k_3k_4}\bra{\phi}h_{j_2}p_{i_2}h_{j_1}p_{i_1} p_{k_1}^\dagger h_{l_1}^\dagger \hat{D}_{l_1}^{k_1}\\
&\times \left(\delta_{k_2m}\delta_{k_3n} p_{k_4}^\dagger - \delta_{k_2n}\delta_{k_3m}p_{k_4}^\dagger - \delta_{k_2m}\delta_{k_4n}p_{k_3}^\dagger   + \delta_{k_2n}\delta_{k_4m}p_{k_3}^\dagger + \delta_{k_3m}\delta_{k_4n}p_{k_2}^\dagger-\delta_{k_3n}\delta_{k_4m}p_{k_2}^\dagger\right)\\
&\times\left(\delta_{l_2q}\delta_{l_3r}h_{l_4}^\dagger- \delta_{l_2r}\delta_{l_3q}h_{l_4}^\dagger - \delta_{l_2q}\delta_{l_4r}h_{l_3}^\dagger+\delta_{l_2r}\delta_{l_4q}h_{l_3}^\dagger +  \delta_{l_3q}\delta_{l_4r} h_{l_2}^\dagger-\delta_{l_3r}\delta_{l_4q} h_{l_2}^\dagger  \right)\ket{\phi}\\
\end{split}
\end{equation}
Now, each pair of terms ending with the same operator gives the same contribution, which yields
\begin{equation}
\begin{split}
\bra{\phi_{j_1j_2}^{i_1i_2}}&T\hat{H}^{\phi}T^3\ket{\phi}\\
&=-\frac{1}{6}\sum_{mnqrk_1l_1k_2l_2k_3l_3k_4l_4}V_{mnqr}c_{l_2l_3l_4}^{k_2k_3k_4}\bra{\phi}h_{j_2}p_{i_2}h_{j_1}p_{i_1} p_{k_1}^\dagger h_{l_1}^\dagger \hat{D}_{l_1}^{k_1}\\
&\times \left(\delta_{k_2m}\delta_{k_3n} p_{k_4}^\dagger - \delta_{k_2m}\delta_{k_4n}p_{k_3}^\dagger  + \delta_{k_3m}\delta_{k_4n}p_{k_2}^\dagger\right)\left(\delta_{l_2q}\delta_{l_3r}h_{l_4}^\dagger - \delta_{l_2q}\delta_{l_4r}h_{l_3}^\dagger +  \delta_{l_3q}\delta_{l_4r} h_{l_2}^\dagger \right)\ket{\phi}\\
&=-\frac{1}{6}\sum_{mnqrk_1l_1k_2l_2k_3l_3k_4l_4}V_{mnqr}c_{l_2l_3l_4}^{k_2k_3k_4}\bra{\phi}h_{j_2}p_{i_2}h_{j_1}p_{i_1} p_{k_1}^\dagger h_{l_1}^\dagger \hat{D}_{l_1}^{k_1}\\
&\times \Bigg(\delta_{k_2m}\delta_{k_3n} \delta_{l_2q}\delta_{l_3r} p_{k_4}^\dagger h_{l_4}^\dagger-\delta_{k_2m}\delta_{k_3n} \delta_{l_2q}\delta_{l_4r} p_{k_4}^\dagger h_{l_3}^\dagger+\delta_{k_2m}\delta_{k_3n}\delta_{l_3q}\delta_{l_4r} p_{k_4}^\dagger h_{l_2}^\dagger\\
&-\delta_{k_2m}\delta_{k_4n}\delta_{l_2q}\delta_{l_3r} p_{k_3}^\dagger h_{l_4}^\dagger+\delta_{k_2m}\delta_{k_4n}\delta_{l_2q}\delta_{l_4r} p_{k_3}^\dagger h_{l_3}^\dagger-\delta_{k_2m}\delta_{k_4n}\delta_{l_3q}\delta_{l_4r}p_{k_3}^\dagger h_{l_2}^\dagger\\
&+\delta_{k_3m}\delta_{k_4n}\delta_{l_2q}\delta_{l_3r} p_{k_2}^\dagger h_{l_4}^\dagger-\delta_{k_3m}\delta_{k_4n}\delta_{l_2q}\delta_{l_4r} p_{k_2}^\dagger h_{l_3}^\dagger+\delta_{k_3m}\delta_{k_4n}\delta_{l_3q}\delta_{l_4r}p_{k_2}^\dagger h_{l_2}^\dagger\Bigg)\ket{\phi}\\
&=-\frac{1}{12}\sum_{mnqrk_1l_1}V_{mnqr}\bra{\phi}h_{j_2}p_{i_2}h_{j_1}p_{i_1} p_{k_1}^\dagger h_{l_1}^\dagger \\
&\times \Bigg(\sum_{k_4l_4}c_{qrl_4}^{mnk_4}\frac{c_{l_1l_4}^{k_1k_4}}{c_{l_4}^{k_4}} p_{k_4}^\dagger h_{l_4}^\dagger-\sum_{l_3k_4}c_{ql_3r}^{mnk_4} \frac{c_{l_1l_3}^{k_1k_4}}{c_{l_3}^{k_4}} p_{k_4}^\dagger h_{l_3}^\dagger+\sum_{k_4l_2}c_{l_2qr}^{mnk_4} \frac{c_{l_1l_2}^{k_1k_4}}{c_{l_2}^{k_4}} p_{k_4}^\dagger h_{l_2}^\dagger\\
&- \sum_{k_3l_4}c_{qrl_4}^{mk_3n}\frac{c_{l_1l_4}^{k_1k_3}}{c_{l_4}^{k_3}} p_{k_3}^\dagger h_{l_4}^\dagger+\sum_{k_3l_3}c_{ql_3r}^{mk_3n} \frac{c_{l_1l_3}^{k_1k_3}}{c_{l_3}^{k_3}} p_{k_3}^\dagger h_{l_3}^\dagger-\sum_{k_3l_2} c_{l_2qr}^{mk_3n} \frac{c_{l_1l_2}^{k_1k_3}}{c_{l_2}^{k_3}}p_{k_3}^\dagger h_{l_2}^\dagger\\
&+\sum_{k_2l_4}c_{qrl_4}^{k_2mn} \frac{c_{l_1l_4}^{k_1k_2}}{c_{l_4}^{k_2}} p_{k_2}^\dagger h_{l_4}^\dagger-\sum_{k_2l_3}c_{ql_3r}^{k_2mn}\frac{c_{l_1l_3}^{k_1k_2}}{c_{l_3}^{k_2}} p_{k_2}^\dagger h_{l_3}^\dagger+\sum_{k_2l_2}c_{l_2qr}^{k_2mn} \frac{c_{l_1l_2}^{k_1k_2}}{c_{l_2}^{k_2}} p_{k_2}^\dagger h_{l_2}^\dagger\Bigg)\ket{\phi}\\
&=-\frac{3}{4}\sum_{mnqrk_1l_1k_2l_2}V_{mnqr}c_{qrl_2}^{mnk_2}\frac{c_{l_1l_2}^{k_1k_2}}{c_{l_2}^{k_2}} \bra{\phi}h_{j_2}p_{i_2}h_{j_1}p_{i_1} p_{k_1}^\dagger h_{l_1}^\dagger p_{k_2}^\dagger h_{l_2}^\dagger\ket{\phi}
\end{split}
\end{equation}
\begin{equation}
\begin{split}
\bra{\phi_{j_1j_2}^{i_1i_2}}T\hat{H}^{\phi}T^3\ket{\phi}
&=-\frac{3}{4}\sum_{mnqrk_1l_1k_2l_2}V_{mnqr}c_{qrl_2}^{mnk_2}\frac{c_{l_1l_2}^{k_1k_2}}{c_{l_2}^{k_2}} \bra{\phi}p_{i_2}p_{i_1} p_{k_1}^\dagger p_{k_2}^\dagger h_{j_2}h_{j_1} h_{l_1}^\dagger h_{l_2}^\dagger\ket{\phi}\\
&=-\frac{3}{4}\sum_{mnqrk_1l_1k_2l_2}V_{mnqr}c_{qrl_2}^{mnk_2}\frac{c_{l_1l_2}^{k_1k_2}}{c_{l_2}^{k_2}} \left(\delta_{k_1i_1}\delta_{k_2i_2}-\delta_{k_1i_2}\delta_{k_2i_1}\right)\left(\delta_{l_1j_1}\delta_{l_2j_2}-\delta_{l_1j_2}\delta_{l_2j_1}\right)\\
&=-\frac{3}{4}\sum_{mnqr}V_{mnqr}\left(c_{qrj_2}^{mni_2}\frac{c_{j_1j_2}^{i_1i_2}}{c_{j_2}^{i_2}}-c_{qrj_1}^{mni_2}\frac{c_{j_2j_1}^{i_1i_2}}{c_{j_1}^{i_2}}-c_{qrj_2}^{mni_1}\frac{c_{j_1j_2}^{i_2i_1}}{c_{j_2}^{i_1}}+c_{qrj_1}^{mni_1}\frac{c_{j_2j_1}^{i_2i_1}}{c_{j_1}^{i_1}}\right)\\
&=-\frac{3}{4}c_{j_1j_2}^{i_1i_2}\sum_{mnqr}V_{mnqr}\left(\frac{c_{qrj_2}^{mni_2}}{c_{l_2}^{i_2}}+\frac{c_{qrj_1}^{mni_2}}{c_{j_1}^{i_2}}+\frac{c_{qrj_2}^{mni_1}}{c_{j_2}^{i_1}}+\frac{c_{qrj_1}^{mni_1}}{c_{j_1}^{i_1}}\right)
\end{split}
\end{equation}
and
\begin{equation}
\boxed{
-\frac{1}{3!}\bra{\phi_{j_1j_2}^{i_1i_2}}T\hat{H}^{\phi}T^3\ket{\phi}=\frac{1}{8}c_{j_1j_2}^{i_1i_2}\sum_{mnqr}V_{mnqr}\left(\frac{c_{qrj_2}^{mni_2}}{c_{l_2}^{i_2}}+\frac{c_{qrj_1}^{mni_2}}{c_{j_1}^{i_2}}+\frac{c_{qrj_2}^{mni_1}}{c_{j_2}^{i_1}}+\frac{c_{qrj_1}^{mni_1}}{c_{j_1}^{i_1}}\right)}
\end{equation}

\begin{equation}
\begin{split}
\bra{\phi_{j_1j_2}^{i_1i_2}}T^2\hat{H}^{\phi}T^2\ket{\phi}&=\bra{\phi_{j_1j_2}^{i_1i_2}}T^2\ket{\phi}\bra{\phi}\hat{H}^{\phi}T^2\ket{\phi}\\
&=2c_{j_1j_2}^{i_1i_2}\bra{\phi}\frac{1}{4}\sum_{klmn} V_{klmn} h_n h_m p_l p_k\left(\frac{1}{2}\right)\sum_{qrst} c^{qs}_{rt} p^\dagger_q h^\dagger_r p_s^\dagger h_t^\dagger\ket{\phi}\\
&=\frac{1}{4}c_{j_1j_2}^{i_1i_2}\sum_{klmnqrst} V_{klmn} c^{qs}_{rt} \bra{\phi}  h_n h_m p_l p_k p^\dagger_q h^\dagger_r p_s^\dagger h_t^\dagger\ket{\phi}\\
&=-\frac{1}{4}c_{j_1j_2}^{i_1i_2}\sum_{klmnqrst} V_{klmn} c^{qs}_{rt}\bra{\phi}p_l p_k p^\dagger_q p_s^\dagger h_n h_m h^\dagger_r h_t^\dagger\ket{\phi}\\
&=-c_{j_1j_2}^{i_1i_2} \sum_{klmn} V_{klmn}c^{kl}_{mn}\,,\\
\end{split}
\end{equation}
where we have used the result of \eqref{eq:T2_H_T} for $\bra{\phi_{j_1j_2}^{i_1i_2}}T^2\ket{\phi}$, thus
\begin{equation}
\boxed{
\frac{1}{4}\bra{\phi_{j_1j_2}^{i_1i_2}}T^2\hat{H}^{\phi}T^2\ket{\phi}=-\frac{1}{4}c_{j_1j_2}^{i_1i_2} \sum_{klmn} V_{klmn}c^{kl}_{mn}\,.}
\end{equation}

Finally, after adding and rearranging the terms, equation \eqref{eq:eqn_phi_2_suppl} becomes
\begin{equation}
\begin{split}
0&=-V_{i_1i_2j_1j_2}+t^{\phi}_{i_1j_1}c_{j_2}^{i_2}-t^{\phi}_{i_1j_2}c_{j_1}^{i_2}-t^{\phi}_{i_2j_1}c_{j_2}^{i_1}+t^{\phi}_{i_2j_2}c_{j_1}^{i_1}-\sum_{k} \left(V_{i_1i_2j_1k}c_{j_2}^{k}-V_{i_1i_2j_2k}c_{j_1}^{k}\right)\\
&+\sum_{k} \left(V_{j_1j_2i_1k}c_{k}^{i_2}-V_{j_1j_2i_2k}c_{k}^{i_1}\right)+\sum_{k}\left(t^{\phi}_{i_1k}c_{j_1j_2}^{ki_2} +t^{\phi}_{i_2k}c_{j_1j_2}^{i_1k}\right)-\sum_{k}\left(t^{\phi}_{j_1k}c_{kj_2}^{i_1i_2}+t^{\phi}_{j_2k}c_{j_1k}^{i_1i_2}\right)\\
&-\frac{1}{2}\sum_{kl}V_{i_1i_2kl}c_{j_1j_2}^{kl}-\frac{1}{2}\sum_{kl}V_{j_1j_2kl}c_{kl}^{i_1i_2}-\sum_{kl} \left(V_{i_1kj_1l}c_{kj_2}^{li_2}+V_{i_2kj_1l}c_{kj_2}^{i_1l}+V_{i_1kj_2l}c_{j_1k}^{li_2}+V_{i_2kj_2l}c_{j_1k}^{i_1l}\right)\\
&+\sum_{kl} t^{\phi}_{kl}c_{lj_1j_2}^{ki_1i_2}-\frac{1}{2}\sum_{klm} \left(V_{klmi_1}c_{mj_1j_2}^{kli_2}-V_{klmi_2}c_{mj_1j_2}^{kli_1}\right)+\frac{1}{2}\sum_{klm} \left(V_{klmj_1}c_{klj_2}^{mi_1i_2}-V_{klmj_2}c_{klj_1}^{mi_1i_2}\right)\\
&+c_{j_1j_2}^{i_1i_2}\sum_{kl} t^{\phi}_{kl}c_{l}^{k}-\frac{1}{4}\sum_{klmn}V_{klmn} c_{mnj_1j_2}^{kli_1i_2}-\frac{1}{4}c_{j_1j_2}^{i_1i_2} \sum_{klmn} V_{klmn}c^{kl}_{mn}\\
&+c_{j_1j_2}^{i_1i_2}\Bigg[-\frac{1}{2}\left(\frac{t^{\phi}_{i_2j_2}}{c_{j_2}^{i_2}}+\frac{t^{\phi}_{i_1j_2}}{c_{j_2}^{i_1}}+\frac{t^{\phi}_{i_2j_1}}{c_{j_1}^{i_2}}+\frac{t^{\phi}_{i_1j_1}}{c_{j_1}^{i_1}}\right)-\frac{1}{2}\sum_{k}\left(\frac{t^{\phi}_{i_2k}c_{j_2}^{k}}{c_{j_2}^{i_2}}+\frac{t^{\phi}_{i_1k}c_{j_2}^{k}}{c_{j_2}^{i_1}}+\frac{t^{\phi}_{i_2k}c_{j_1}^{k}}{c_{j_1}^{i_2}}+\frac{t^{\phi}_{i_1k}c_{j_1}^{k}}{c_{j_1}^{i_1}}\right)\\
&+\frac{1}{2}\sum_{k}\left(\frac{t^{\phi}_{j_2k}c_{k}^{i_2}}{c_{j_2}^{i_2}}+\frac{t^{\phi}_{j_2k}c_{k}^{i_1}}{c_{j_2}^{i_1}}+\frac{t^{\phi}_{j_1k}c_{k}^{i_2}}{c_{j_1}^{i_2}}+\frac{t^{\phi}_{j_1k}c_{k}^{i_1}}{c_{j_1}^{i_1}}\right)+\frac{1}{2}\sum_{kl}\left(\frac{V_{i_2lj_2k}c_{l}^{k}}{c_{j_2}^{i_2}}+\frac{V_{i_1lj_2k}c_{l}^{k}}{c_{j_2}^{i_1}}+\frac{V_{i_2lj_1k}c_{l}^{k}}{c_{j_1}^{i_2}}+\frac{V_{i_1lj_1k}c_{l}^{k}}{c_{j_1}^{i_1}}\right)\\
&-\sum_{kl} t^{\phi}_{kl}\left(\frac{c_{lj_2}^{ki_2}}{c_{j_2}^{i_2}}+\frac{c_{lj_2}^{ki_1}}{c_{j_2}^{i_1}}+\frac{c_{lj_1}^{ki_2}}{c_{j_1}^{i_2}}+\frac{c_{lj_1}^{ki_1}}{c_{j_1}^{i_1}}\right)+\frac{1}{2}\sum_{klm}\left(\frac{V_{klmi_2}c_{mj_2}^{kl}}{c_{j_2}^{i_2}}+\frac{V_{klmi_1}c_{mj_2}^{kl}}{c_{j_2}^{i_1}}+\frac{V_{klmi_2}c_{mj_1}^{kl}}{c_{j_1}^{i_2}}+\frac{V_{klmi_1}c_{mj_1}^{kl}}{c_{j_1}^{i_1}}\right)\\
&-\frac{1}{2}\sum_{klm}\left(\frac{V_{klmj_2}c_{kl}^{mi_2}}{c_{j_2}^{i_2}}+\frac{V_{klmj_2}c_{kl}^{mi_1}}{c_{j_2}^{i_1}}+\frac{V_{klmj_1}c_{kl}^{mi_2}}{c_{j_1}^{i_2}}+\frac{V_{klmj_1}c_{kl}^{mi_1}}{c_{j_1}^{i_1}}\right)+\frac{1}{8}\sum_{klmn}V_{klmn}\left(\frac{c_{mnj_2}^{kli_2}}{c_{l_2}^{i_2}}+\frac{c_{mnj_1}^{kli_2}}{c_{j_1}^{i_2}}+\frac{c_{mnj_2}^{kli_1}}{c_{j_2}^{i_1}}+\frac{c_{mnj_1}^{kli_1}}{c_{j_1}^{i_1}}\right)\Bigg]\,.
\end{split}
\end{equation}
or, after multiplying by $c_{j_1}^{i_1}c_{j_2}^{i_1}c_{j_1}^{i_2}c_{j_2}^{i_2}$,
\begin{equation}
\begin{split}
0&=c_{j_1}^{i_1}c_{j_2}^{i_1}c_{j_1}^{i_2}c_{j_2}^{i_2}\Bigg[-V_{i_1i_2j_1j_2}+t^{\phi}_{i_1j_1}c_{j_2}^{i_2}-t^{\phi}_{i_1j_2}c_{j_1}^{i_2}-t^{\phi}_{i_2j_1}c_{j_2}^{i_1}+t^{\phi}_{i_2j_2}c_{j_1}^{i_1}-\sum_{k} \left(V_{i_1i_2j_1k}c_{j_2}^{k}-V_{i_1i_2j_2k}c_{j_1}^{k}\right)\\
&\quad+\sum_{k} \left(V_{j_1j_2i_1k}c_{k}^{i_2}-V_{j_1j_2i_2k}c_{k}^{i_1}\right)+\sum_{k}\left(t^{\phi}_{i_1k}c_{j_1j_2}^{ki_2} +t^{\phi}_{i_2k}c_{j_1j_2}^{i_1k}\right)-\sum_{k}\left(t^{\phi}_{j_1k}c_{kj_2}^{i_1i_2}+t^{\phi}_{j_2k}c_{j_1k}^{i_1i_2}\right)\\
&-\frac{1}{2}\sum_{kl}V_{i_1i_2kl}c_{j_1j_2}^{kl}-\frac{1}{2}\sum_{kl}V_{j_1j_2kl}c_{kl}^{i_1i_2}-\sum_{kl} \left(V_{i_1kj_1l}c_{kj_2}^{li_2}+V_{i_2kj_1l}c_{kj_2}^{i_1l}+V_{i_1kj_2l}c_{j_1k}^{li_2}+V_{i_2kj_2l}c_{j_1k}^{i_1l}\right)\\
&+\sum_{kl} t^{\phi}_{kl}c_{lj_1j_2}^{ki_1i_2}-\frac{1}{2}\sum_{klm} \left(V_{klmi_1}c_{mj_1j_2}^{kli_2}-V_{klmi_2}c_{mj_1j_2}^{kli_1}\right)+\frac{1}{2}\sum_{klm} \left(V_{klmj_1}c_{klj_2}^{mi_1i_2}-V_{klmj_2}c_{klj_1}^{mi_1i_2}\right)\\
&+c_{j_1j_2}^{i_1i_2}\sum_{kl} t^{\phi}_{kl}c_{l}^{k}-\frac{1}{4}\sum_{klmn}V_{klmn} c_{mnj_1j_2}^{kli_1i_2}-\frac{1}{4}c_{j_1j_2}^{i_1i_2} \sum_{klmn} V_{klmn}c^{kl}_{mn}\Bigg]\\
&+c_{j_1j_2}^{i_1i_2}\Bigg[-\frac{1}{2}\left(t^{\phi}_{i_2j_2}c_{j_1}^{i_1}c_{j_2}^{i_1}c_{j_1}^{i_2}+t^{\phi}_{i_1j_2}c_{j_1}^{i_1}c_{j_1}^{i_2}c_{j_2}^{i_2}+t^{\phi}_{i_2j_1}c_{j_1}^{i_1}c_{j_2}^{i_1}c_{j_2}^{i_2}+t^{\phi}_{i_1j_1}c_{j_2}^{i_1}c_{j_1}^{i_2}c_{j_2}^{i_2}\right)\\
&-\frac{1}{2}\sum_{k}\left(t^{\phi}_{i_2k}c_{j_2}^{k}c_{j_1}^{i_1}c_{j_2}^{i_1}c_{j_1}^{i_2}+t^{\phi}_{i_1k}c_{j_2}^{k}c_{j_1}^{i_1}c_{j_1}^{i_2}c_{j_2}^{i_2}+t^{\phi}_{i_2k}c_{j_1}^{k}c_{j_1}^{i_1}c_{j_2}^{i_1}c_{j_2}^{i_2}+t^{\phi}_{i_1k}c_{j_1}^{k}c_{j_2}^{i_1}c_{j_1}^{i_2}c_{j_2}^{i_2}\right)\\
&+\frac{1}{2}\sum_{k}\left(t^{\phi}_{j_2k}c_{k}^{i_2}c_{j_1}^{i_1}c_{j_2}^{i_1}c_{j_1}^{i_2}+t^{\phi}_{j_2k}c_{k}^{i_1}c_{j_1}^{i_1}c_{j_1}^{i_2}c_{j_2}^{i_2}+t^{\phi}_{j_1k}c_{k}^{i_2}c_{j_1}^{i_1}c_{j_2}^{i_1}c_{j_2}^{i_2}+t^{\phi}_{j_1k}c_{k}^{i_1}c_{j_2}^{i_1}c_{j_1}^{i_2}c_{j_2}^{i_2}\right)\\
&+\frac{1}{2}\sum_{kl}\left(V_{i_2lj_2k}c_{l}^{k}c_{j_1}^{i_1}c_{j_2}^{i_1}c_{j_1}^{i_2}+V_{i_1lj_2k}c_{l}^{k}c_{j_1}^{i_1}c_{j_1}^{i_2}c_{j_2}^{i_2}+V_{i_2lj_1k}c_{l}^{k}c_{j_1}^{i_1}c_{j_2}^{i_1}c_{j_2}^{i_2}+V_{i_1lj_1k}c_{l}^{k}c_{j_2}^{i_1}c_{j_1}^{i_2}c_{j_2}^{i_2}\right)\\
&-\sum_{kl} t^{\phi}_{kl}\left(c_{lj_2}^{ki_2}c_{j_1}^{i_1}c_{j_2}^{i_1}c_{j_1}^{i_2}+c_{lj_2}^{ki_1}c_{j_1}^{i_1}c_{j_1}^{i_2}c_{j_2}^{i_2}+c_{lj_1}^{ki_2}c_{j_1}^{i_1}c_{j_2}^{i_1}c_{j_2}^{i_2}+c_{lj_1}^{ki_1}c_{j_2}^{i_1}c_{j_1}^{i_2}c_{j_2}^{i_2}\right)\\
&+\frac{1}{2}\sum_{klm}\left(V_{klmi_2}c_{mj_2}^{kl}c_{j_1}^{i_1}c_{j_2}^{i_1}c_{j_1}^{i_2}+V_{klmi_1}c_{mj_2}^{kl}c_{j_1}^{i_1}c_{j_1}^{i_2}c_{j_2}^{i_2}+V_{klmi_2}c_{mj_1}^{kl}c_{j_1}^{i_1}c_{j_2}^{i_1}c_{j_2}^{i_2}+V_{klmi_1}c_{mj_1}^{kl}c_{j_2}^{i_1}c_{j_1}^{i_2}c_{j_2}^{i_2}\right)\\
&-\frac{1}{2}\sum_{klm}\left(V_{klmj_2}c_{kl}^{mi_2}c_{j_1}^{i_1}c_{j_2}^{i_1}c_{j_1}^{i_2}+V_{klmj_2}c_{kl}^{mi_1}c_{j_1}^{i_1}c_{j_1}^{i_2}c_{j_2}^{i_2}+V_{klmj_1}c_{kl}^{mi_2}c_{j_1}^{i_1}c_{j_2}^{i_1}c_{j_2}^{i_2}+V_{klmj_1}c_{kl}^{mi_1}c_{j_2}^{i_1}c_{j_1}^{i_2}c_{j_2}^{i_2}\right)\\
&+\frac{1}{8}\sum_{klmn}V_{klmn}\left(c_{mnj_2}^{kli_2}c_{j_1}^{i_1}c_{j_2}^{i_1}c_{j_1}^{i_2}+c_{mnj_1}^{kli_2}c_{j_1}^{i_1}c_{j_2}^{i_1}c_{j_2}^{i_2}+c_{mnj_2}^{kli_1}c_{j_1}^{i_1}c_{j_1}^{i_2}c_{j_2}^{i_2}+c_{mnj_1}^{kli_1}c_{j_2}^{i_1}c_{j_1}^{i_2}c_{j_2}^{i_2}\right)\Bigg]
\end{split}
\end{equation}
\end{widetext}

\end{document}